\documentclass[a4paper,11pt]{article}
\usepackage{amsmath}
\usepackage{amssymb}
\usepackage{graphicx}
\usepackage{color}
\usepackage[square,numbers,comma,sort&compress]{natbib}
\usepackage[colorlinks=true,citecolor=blue,linkcolor=purple,urlcolor=purple]{hyperref}
\usepackage{geometry}
\def\varabstract{ }
\def\varkeywords{ }
\def\vararxivnumber{ }
\def\vartitle{ }
\renewcommand{\title}[1]{\gdef\vartitle{#1}}
\renewcommand{\abstract}[1]{\gdef\varabstract{#1}}
\newcommand{\keywords}[1]{\gdef\varkeywords{#1}}
\newcommand{\arxivnumber}[1]{\gdef\vararxivnumber{#1}}
\newtoks\authtoks
\renewcommand{\author}[2][]{%
	\authtoks=\expandafter{\the\authtoks#2$^{#1}$\ }%
}
\newtoks\affiltoks
\newcommand{\affiliation}[2][]{%
    \affiltoks=\expandafter{\the\affiltoks{\item[$^{#1}$]#2}}%
}
\newtoks\emailtoks\newcounter{emailcounter}%
\newcommand{\emailAdd}[1]{%
\ifnum\theemailcounter>0\emailtoks=\expandafter{\the\emailtoks, \typeemail{#1}}%
\else\emailtoks=\expandafter{\typeemail{#1}}%
\fi
\stepcounter{emailcounter}}
\newcommand{\typeemail}[1]{\href{mailto:#1}{\tt #1}}

\renewcommand\maketitle{
	\newgeometry{margin=2cm}
	\pagestyle{empty}\setcounter{page}{0}
{\huge\flushleft\sffamily\bfseries\vartitle\par}
\vskip6ex
{\bfseries\raggedright\sffamily\the\authtoks\par}
\vskip2ex
\begin{list}{}{%
\setlength{\leftmargin}{0.28cm}%
\setlength{\labelsep}{0pt}%
\setlength{\itemsep}{-3pt}%
\setlength{\topsep}{-\parskip}}
\itshape\small%
\the\affiltoks
\end{list}
\vskip2ex
\noindent\hspace{0.28cm}\begin{minipage}[l]{.9\textwidth}
\begin{flushleft}
\textit{E-mail:} \the\emailtoks
\end{flushleft}
\end{minipage}
\vskip5ex
\noindent{\renewcommand\baselinestretch{.9}\textsc{Abstract:}}\ \varabstract
\vskip5ex 
\if!\varkeywords!\else\noindent{\textsc{Keywords:}}\ \varkeywords \vskip2ex\fi
\if!\vararxivnumber!\else\noindent{\textsc{ArXiv ePrint:}} \href{http://arxiv.org/abs/\vararxivnumber}{\vararxivnumber}\vskip2ex\fi

\newpage
\restoregeometry
\pagestyle{plain}
\hrule
\bigskip\bigskip

{
	\hypersetup{linkcolor=black}
	\tableofcontents
}
\bigskip\medskip
\hrule
\bigskip\bigskip
\setcounter{footnote}{0}
} 

\usepackage{subcaption}
\usepackage{verbatim}
\usepackage{textcomp}
\usepackage{nicefrac}
\usepackage{array}
\usepackage{multirow}
\usepackage{slashed}
\usepackage{diagbox}
\usepackage{cancel}
\usepackage[normalem]{ulem}
\usepackage{textgreek}
\usepackage{lmodern}
\usepackage{xparse}
\usepackage{booktabs}
\usepackage{caption}
\usepackage{tikz}
\usepackage[compat=1.1.0]{tikz-feynman}
\usetikzlibrary{positioning}
\usetikzlibrary{decorations.text}
\usetikzlibrary{decorations.pathmorphing}
\usetikzlibrary{calc}
\usetikzlibrary{shapes.misc}
\tikzset{
	>=latex,
    photon/.style={decorate, decoration={snake}, draw=black, thick},
    fermionnoarrow/.style={draw=black, postaction={decorate}, thick},
    scalar/.style={draw=black, postaction={decorate}, decoration={markings,mark=at position .55 with {\arrow{>}}}, thick, dashed},
    scalarnoarrow/.style={draw=black, postaction={decorate},  thick, dashed},
    fermion/.style={draw=black, postaction={decorate},decoration={markings,mark=at position .55 with {\arrow{>}}}, thick},
    gluon/.style={decorate, draw=black, decoration={coil,amplitude=4pt, segment length=5pt}, thick},
    vertex/.style={draw,shape=circle,fill=black,minimum size=3pt,inner sep=0pt},
    fillvertex/.style={draw,shape=circle,fill=black,minimum size=5pt,inner sep=0pt},
    openvertex/.style={draw,shape=circle,minimum size=5pt,inner sep=0pt},
    blob/.style={draw=red,shape=circle,fill=red,minimum size=6pt,inner sep=0pt},
    redvertex/.style={draw=red,shape=circle,fill=red,minimum size=3pt,inner sep=0pt},
    cross/.style={cross out, draw=black,thick, minimum size=5pt, inner sep=0pt, outer sep=0pt}
}

\def\cc{\text{H.c.}}
\def\rpv{$\slashed{R}_p \:$}

\def\vb#1{\vbox to #1 pt{}}

\newcommand{\braket}[1]{\left\langle#1\right\rangle}

\title{From the trees to the forest:\\ a review of radiative neutrino mass models}

\author[a, b]{Yi Cai,} \emailAdd{caiy36@mail.sysu.edu.cn}
\author[c]{Juan Herrero-Garc\'ia,} \emailAdd{juan.herrero-garcia@adelaide.edu.au}
\author[d]{Michael A.~Schmidt,} \emailAdd{michael.schmidt@sydney.edu.au}
\author[e]{Avelino Vicente} \emailAdd{avelino.vicente@ific.uv.es}
\author[b]{and Raymond R.~Volkas} \emailAdd{raymondv@unimelb.edu.au}

\affiliation[a]{School of Physics, Sun Yat-sen University, Guangzhou, 510275, China}
\affiliation[b]{ARC Centre of Excellence for Particle Physics at the Terascale, School of Physics, The University of Melbourne, VIC 3010, Australia}
\affiliation[c]{ARC Centre of Excellence for Particle Physics at the Terascale, Department of Physics, The University of Adelaide,  SA 5005, Australia}
\affiliation[d]{ARC Centre of Excellence for Particle Physics at the Terascale, School of Physics, The University of Sydney, NSW 2006, Australia}
\affiliation[e]{Instituto de F\'{\i}sica Corpuscular (CSIC-Universitat de Val\`{e}ncia), Apdo. 22085, E-46071 Valencia, Spain}

\abstract{A plausible explanation for the lightness of neutrino masses is that neutrinos are massless at tree level, with their mass (typically Majorana) being
generated radiatively at one or more loops. The new couplings, together
with the suppression coming from the loop factors, imply that the new degrees
of freedom cannot be too heavy (they are typically at the TeV scale). Therefore, in these models
there are no large mass hierarchies and they can be tested using different
searches, making their detailed phenomenological study very appealing. In
particular, the new particles can be searched for at colliders and generically induce signals in lepton-flavor and
lepton-number violating processes (in the case of Majorana neutrinos), which are not independent from reproducing
correctly the neutrino masses and mixings. The main focus of the review is on Majorana neutrinos. We order the allowed
theory space from three different perspectives: (i) using an effective operator
approach to lepton number violation, (ii) by the number of loops at which the
Weinberg operator is generated, (iii) within a given loop order, by the possible
irreducible topologies. We also discuss in more detail some popular radiative
models which
involve qualitatively different features, revisiting their most important
phenomenological implications. Finally, we list some promising avenues to pursue.}

\arxivnumber{1706.08524}
\begin{document}

\maketitle

\newpage
\listoffigures

\listoftables

\newpage
\section{Introduction}

The discovery of neutrino oscillations driven by mass mixing is one of the crowning achievements of experimental high-energy physics in recent decades. From its beginnings as the ``solar neutrino problem'' -- a deficit of electron neutrinos from the Sun compared to the prediction of the standard solar model, an anomaly first discovered by the Homestake experiment -- through the emergence of the ``atmospheric neutrino problem'' and its eventual confirmation by SuperKamiokande, to terrestrial verifications by long baseline and reactor neutrino experiments, the existence of nonzero and non-degenerate neutrino masses is now well established~\cite{Hirata:1990xa,Hirata:1992ku,Fukuda:1998mi,Cleveland:1998nv,Hampel:1998xg,Abdurashitov:1999zd,Fukuda:2001nj,Ahmad:2002jz,Altmann:2005ix,Ahn:2006zza,Michael:2006rx,Abe:2008aa,Abe:2011sj,Abe:2011fz,An:2012eh,Ahn:2012nd,Abe:2014ugx}.  In addition, the existence of oscillations proves that the weak eigenstate neutrinos $\nu_e$, $\nu_\mu$ and $\nu_\tau$ are not states of definite mass themselves, but rather non-trivial, coherent superpositions of mass eigenstate fields called simply $\nu_1$, $\nu_2$ and $\nu_3$, with masses $m_1$, $m_2$ and $m_3$, respectively.\footnote{The possibility of additional neutrino-like states will be discussed below.}  The dynamical origin of neutrino mass is at present unknown, including whether neutrinos are Dirac or Majorana fermions. In the former case, neutrinos and antineutrinos are distinct and have a total of four degrees of freedom, exactly as do the charged leptons and quarks. Majorana fermions, on the other hand, are their own antiparticles, and they have just two degrees of freedom corresponding to left- and right-handed helicity. Dirac neutrinos preserve total lepton number conservation, while Majorana neutrino masses violate lepton number conservation by two units. The purpose of this review is to survey one class of possible models, where neutrino masses arise at loop order and are thus called ``radiative''. Almost all of the models we examine are for the Majorana mass case. Before turning to a discussion of possible models, we should summarize the experimental data the models are trying to understand or at least accommodate.

The Pontecorvo-Maki-Nakagawa-Sakata (PMNS) matrix $(U_{\alpha i})$~\cite{Pontecorvo:1957qd,Maki:1962mu} defines the relationship between the weak and mass eigenstates, through
\begin{equation}
\nu_\alpha = \sum_i U_{\alpha i} \nu_i ,
\label{eq:Sec1:weak-mass}
\end{equation}
where $\alpha = e, \mu, \tau$ and $i=1,2,3$.  The PMNS matrix $U$ is unitary, and may be parameterized by three (Euler) mixing angles $\theta_{12}$, $\theta_{23}$ and $\theta_{13}$, a CP-violating Dirac phase $\delta$ that is analogous to the phase in the Cabibbo-Kobayashi-Maskawa (CKM) quark mixing matrix, and two Majorana phases $\alpha_{2,3}$ if neutrinos are Majorana fermions.  The standard parametrisation is
\begin{equation}
U = \left( \begin{array}{ccc}
c_{12} c_{13} & s_{12} c_{13} & s_{13} e^{-i \delta} \\
- s_{12} c_{23} - c_{12} s_{23} s_{13} e^{i \delta} & c_{12} c_{23} - s_{12} s_{23} s_{13} e^{i \delta} & s_{23} c_{13} \\
s_{12} s_{23} - c_{12} c_{23} s_{13} e^{i \delta} & - c_{12} s_{23} - s_{12} c_{23} s_{13} e^{i \delta} & c_{23} c_{13} 
\end{array} \right)
\left( \begin{array}{ccc}
1 & 0 & 0\\
0 & e^{i \frac{\alpha_2}{2}} & 0 \\
0 & 0 & e^{i \frac{\alpha_3}{2}}
\end{array} \right) ,
\label{eq:Sec1:PMNS}
\end{equation}
where $c_{ij} \equiv \cos \theta_{ij}$ and $s_{ij} \equiv \sin \theta_{ij}$. The neutrino oscillation lengths are set by the ratio of squared-mass differences and energy, while the amplitudes are governed by the PMNS mixing angles and the Dirac phase. The Majorana phases do not contribute to oscillation probabilities. The angles $\theta_{12}$, $\theta_{23}$ and $\theta_{13}$ are sometimes referred to as the solar, atmospheric and reactor angles, respectively, because of how they were originally or primarily measured. The ``solar'' and ``atmospheric'' oscillation length parameters are, respectively, 
\begin{equation}
\Delta m^2_{21} \equiv m_2^2 - m_1^2,\quad \Delta m^2_{32} \equiv m_3^2 - m_2^2 \sim \Delta m^2_{31} \equiv m_3^2 - m_1^2 ,
\end{equation}
where the distinction between the two atmospheric quantities will be discussed below.

A recent global fit~\cite{Esteban:2016qun} obtains the following $3\sigma$ ranges for the mixing angle and $\Delta m^2$ parameters:
\begin{eqnarray}
\sin^2 \theta_{12} & \in & [0.271,\ 0.345],\quad \sin^2 \theta_{23} \in [0.385,\ 0.638],\quad \sin^2 \theta_{13} \in [0.01934,\ 0.02397], \label{eq:Sec1:angle-ranges}\\
\Delta m^2_{21} & \in & [7.03,\ 8.09] \times 10^{-5}\ {\rm eV}^2, \quad \Delta m^2_{3 i} \in [-2.629,\ -2.405] \cup [2.407,\ 2.643] \times 10^{-3}\ {\rm eV}^2,  \nonumber\\
\label{eq:Sec1:Deltamsq-ranges} 
\end{eqnarray}
where $i = 1,2$ depending on the sign of the atmospheric squared-mass difference (see Refs.~\cite{Forero:2014bxa,Capozzi:2013csa} for earlier fits).  The sign of $\Delta m^2_{21}$ has been measured because the Mikheyev-Smirnov-Wolfenstein or MSW effect~\cite{Mikheev:1986wj,Wolfenstein:1977ue} in the Sun depends on it.  The sign of the atmospheric equivalent is, however, not currently known, and is a major target for future neutrino oscillation experiments.  Because of this ambiguity, there are two possible neutrino mass orderings: $m_1 < m_2 < m_3$ which is called either ``normal ordering'' or ``normal hierarchy'', and $m_3 < m_1 < m_2$ which is termed ``inverted''.  The global fit results for the other parameters depend somewhat on which ordering is assumed. In Eqs.~\ref{eq:Sec1:angle-ranges} and \ref{eq:Sec1:Deltamsq-ranges}, we quote results that leave the ordering as undetermined. See Ref.~\cite{Esteban:2016qun} for a discussion of these subtleties, but they will not be important for the rest of this review. Note that the convention is $i=1$ in Eq.~\ref{eq:Sec1:Deltamsq-ranges} for normal ordering and $i=2$ for inverted ordering.

At the $3\sigma$ level, the CP-violating phase $\delta$ can be anything.  However, there is a local minimum in $\chi^2$ at $\delta \sim -\pi/2$, which is tantalizing and very interesting. It hints at large CP-violation in the lepton sector, and the specific value of $-\pi/2$ is suggestive of a group theoretic origin (but beware that the definition of this phase is convention dependent).  As with the mass ordering, the discovery of CP violation in neutrino oscillations is a prime goal for future experiments.  One strong motivation for this is the cosmological scenario of baryogenesis via leptogenesis~\cite{Fukugita:1986hr}, and even if other sources of leptonic CP-violation are involved, it is important to experimentally establish the general phenomenon in the lepton sector.  At present, we do not know if neutrinos are Dirac or Majorana fermions, so there is no information about the possible Majorana phases $\alpha_{2,3}$.  Neutrinoless double-beta decay is sensitive to these parameters, as is standard leptogenesis.

The final parameter to discuss is the absolute neutrino mass scale.  The square root of the magnitude of the atmospheric $\Delta m^2$ provides a lower bound of $0.05$ eV on at least one of the mass eigenvalues.  Laboratory experiments performing precision measurements of the tritium beta-decay end-point spectrum currently place a direct kinematic upper bound of about $2$ eV on the absolute mass scale~\cite{Lobashev:2003kt,Kraus:2004zw,Aseev:2011dq} as quantified by an ``effective electron-neutrino mass'' $m_{\nu_e} \equiv \sqrt{|U_{ei}|^2 m_i^2}$, independent of whether the mass is Dirac or Majorana, and the sensitivity of the currently running KATRIN experiment is expected to be about $0.2$ eV~\cite{Osipowicz:2001sq}.  With appropriate caution because of model dependence, cosmology now places a strong upper bound on the sum of neutrino masses of about $0.2$ eV~\cite{Ade:2015xua}, with the precise number depending on exactly what data are combined.  If the neutrino mass sum was much above this figure, then its effect on large-scale structure formation -- washing out structure on small scales -- would be strong enough to cause disagreement with observations.  For Majorana masses, neutrinoless double beta-decay experiments have determined an upper bound on an effective mass defined by
\begin{equation}
| m_{\beta\beta} | \equiv | \sum_i U^2_{e i} m_i |
\label{eq:Sec1:mbetabeta}
\end{equation}
of $0.15 - 0.33$ at 90\% C.L., depending on nuclear matrix element uncertainties~\cite{Agostini:2017iyd}\footnote{The effective mass $m_{\beta\beta}$ depends on the Majorana phases and thus provides a unique probe for them.}.  We can thus see that experimentally and observationally, we are closing in on a determination of the absolute mass scale.

The fact that the laboratory and cosmological bounds require the absolute neutrino mass scale to be so low strongly motivates the hypothesis that neutrinos obtain their masses in a different manner from the charged leptons and quarks. A number of approaches have been explored in the literature, with one of them being the main topic of this review: radiative neutrino mass generation. Other approaches will also be briefly commented on, to place radiative models into the overall context of possible explanations for why neutrino masses are so small. 

This completes a summary of the neutrino mass and mixing data that any model, including radiative models, must explain or accommodate. As noted above, future experiments and observational programs have
excellent prospects to determine the mass ordering, discover leptonic CP
violation, observe neutrinoless double beta-decay ($0\nu\beta\beta$) and hence
the violation of lepton number by two units, and measure the absolute neutrino
mass scale. In addition, the determination of the $\theta_{23}$ octant --
whether or not $\theta_{23}$ is less than or greater than $\pi/4$ -- is an
important goal of future experiments.  Before turning to a discussion of
neutrino mass models, we should review some interesting experimental anomalies
that may imply the existence of light sterile neutrinos~\footnote{Sterile neutrinos are not charged under the SM gauge group.} in addition to the
active flavors $\nu_{e,\mu,\tau}$ (see
Refs.~\cite{Gariazzo:2017fdh,Kopp:2013vaa} for phenomenological fits).

There are three anomalies. The first is $>3\sigma$ evidence from the
LSND~\cite{Athanassopoulos:1997pv,Aguilar:2001ty} and
MiniBooNE~\cite{AguilarArevalo:2007it,Aguilar-Arevalo:2013pmq} experiments of
$\bar{\nu}_e$ appearance in a $\bar{\nu}_\mu$ beam, with MiniBooNE also
reporting a $\nu_e$ signal in a $\nu_\mu$ beam. Interpreted through a neutrino
oscillation hypothesis, these results indicate an oscillation mode with a
$\Delta m^2$ or order $1$ eV$^2$. This cannot be accommodated with just the
three known active neutrinos simultaneously with the extremely well-established
solar and atmospheric modes that require much smaller $\Delta m^2$ parameters.
This hypothesis thus only works if there are four or more light neutrino
flavors, and the additional state or states must be sterile to accord with the
measured $Z$-boson invisible width.\footnote{MiniBooNE also has a mysterious
excess in their low-energy bins that cannot be explained by any oscillation
hypothesis.} The Icecube neutrino telescope has recently tested the sterile
neutrino oscillation explanation of these anomalies through the zenith angle
dependence of muon track signals and excludes this hypothesis at about the 99\%
C.L.~\cite{Aartsen:2017bap}.

The next two anomalies concern $\nu_e$ and $\bar{\nu}_e$ disappearance. Nuclear
reactors produce a $\bar{\nu}_e$ flux that has been measured by several
experiments. When compared to the most recent computation of the expected
flux~\cite{Mueller:2011nm,Huber:2011wv}, a consistent deficit of a few percent
is observed, a set of results known as the ``reactor
anomaly''~\cite{Mention:2011rk}.  The Gallium anomaly arose from neutrino
calibration source measurements by the Gallex and SAGE radiochemical solar
neutrino experiments, also indicating a
deficit~\cite{Hampel:1997fc,Abdurashitov:1998ne,Abdurashitov:2009tn,Kaether:2010ag}.
Both deficits are consistent with very short baseline transitions driven by
eV-scale sterile neutrinos, and a significant number of experiments are
underway to test the oscillation explanation.  It should be noted that a recent
analysis by the Daya Bay collaboration points to the problem being with the
computation of the reactor $\bar{\nu}_e$ flux rather than being an indication
of very short baseline oscillations~\cite{An:2017osx}. The key point is that if
a sterile neutrino was responsible, one should observe the same deficit for all
neutrinos from the reactor fuel, independent of nuclear species origin, but
this was observed to not be the case. There is also a tension between the
appearance and disappearance anomalies when trying to fit both with a
self-consistent oscillation scheme~\cite{Gariazzo:2017fdh,Kopp:2013vaa}, and
there is a cosmological challenge of devising a mechanism to prevent the
active-sterile transitions from thermalising the sterile neutrino in the early
universe, as thermalisation would violate the $\sim 0.2$ eV bound on the sum of
neutrino masses.

Because the situation with the above anomalies is unclear, and there are challenges to explaining them with oscillations, this review will focus on neutrino mass models that feature just the three known light active neutrinos. If any of the above anomalies is eventually shown to be due to oscillations, then all neutrino mass models will need to be extended to incorporate light sterile neutrinos, including the radiative models that are our subject in this review.

The rest of this review is structured as follows:  Sec.~\ref{sec:schemes} provides a general discussion of schemes for neutrino mass generation and attempts a classification.  The structure of radiative neutrino mass models is then described in Sec.~\ref{sec:models}.  Section~\ref{sec:pheno} covers phenomenological constraints and search strategies, including for cosmological observables. Detailed descriptions of specific models are then given in Sec.~\ref{sec:examples}, with the examples chosen so as to exemplify some of the different possibilities that the radiative mechanisms permit.  We conclude in Sec.~\ref{sec:conc}, where we discuss some research directions for the future. Appendix~\ref{app:PowerCounting} gives further details on the relative contributions of the different operators to neutrino masses.


\section{Schemes for neutrino masses and mixings}
\label{sec:schemes}

In this section, we survey the many different general ways that neutrinos can gain mass, and attempt a classification of at least most of the proposed schemes. As part of this, we place both the tree-level and radiative models in an overarching context -- a systematic approach, if you will, or at least as systematic as we can make it.  The number of different kinds of models can seem bewildering, so there is some value in understanding the broad structure of the neutrino mass ``theory space''.

Under the standard model (SM) gauge group $G_{\rm SM}\equiv SU(3)_{\rm c} \times SU(2)_{\rm L} \times U(1)_{\rm Y}$, the left-handed neutrinos feature as the upper isospin component of
\begin{equation}
L = \left( \begin{array}{c} \nu_{\rm L} \\ e_{\rm L} \end{array} \right) \sim (1,2,-\frac{1}{2}) ,
\label{eq:L-multiplet}
\end{equation}
where on the right-hand (RH) side the first entry denotes the representation with respect to the color group $SU(3)_{\rm c}$, the second $SU(2)_{\rm L}$ (weak-isospin), and the third hypercharge Y, normalized so that electric charge is given by $Q = I_3 + Y$. In the minimal standard model, there is no way to generate nonzero neutrino masses and mixings at the renormalizable level.  Dirac masses are impossible because of the absence of RH neutrinos, 
\begin{equation}
	\nu_{\rm R} \sim (1,1, 0) ,
\label{eq:nuR-multiplet}
\end{equation}
as are Majorana masses because there is no scalar isospin triplet
\begin{equation}
\Delta \sim (1,3, 1)
\label{eq:Type2seesaw-triplet}
\end{equation}
to which the lepton bilinear $\overline{L^c} L$ could have a Yukawa coupling. Thus, the
family-lepton numbers $L_e$, $L_\mu$ and $L_\tau$ are (perturbatively)
conserved because of three accidental global $U(1)$ symmetries.  The discovery
of neutrino oscillations means that the family-lepton number symmetries must be
broken.  If they are broken down to the diagonal subgroup generated by total
lepton number $L \equiv L_e + L_\mu + L_\tau$, then the neutrinos must be Dirac
fermions.  If total lepton number is also broken, then the neutrinos are either
fully Majorana fermions or pseudo-Dirac.\footnote{Pseudo-Dirac neutrinos are a special case of Majorana neutrinos where the masses of two Majorana neutrinos are almost degenerate and the breaking of lepton number is small. However they should not be confused with Dirac neutrinos.}

The question of whether neutrinos are Dirac or Majorana (or possibly
pseudo-Dirac) is one of the great unknowns. The answer is vital for model
building, as well as for some aspects of phenomenology.  If neutrinos are
Majorana, then it is not necessary to add RH neutrinos to the SM particle
content. In fact, many of the radiative models we shall review below do not
feature them. If RH neutrinos do not exist, then a possible deep justification
could be $SU(5)$ grand unification, which is content with a $\bar 5 \oplus 10$
structure per family.\footnote{RH neutrinos could obviously be added as a singlet of $SU(5)$.}  But another logical possibility, motivated by
quark-lepton symmetry and $SO(10)$ grand unification, is that RH neutrinos
exist but have large (SM gauge invariant) Majorana masses, leading to the
extremely well-known type-I seesaw
model~\cite{Minkowski:1977sc,Yanagida:1979as,GellMann:1980vs,Mohapatra:1979ia,Glashow:1979nm}.
On the other hand, if neutrinos are Dirac, then RH neutrinos that are singlets
under the SM gauge group, as per Eq.~\ref{eq:nuR-multiplet}, are mandatory and
they must not have Majorana masses even though such terms are SM gauge
invariant and renormalizable. Thus, at the SM level, something like total
lepton-number conservation must be imposed by hand.  Most of the radiative
models we shall discuss lead to Majorana neutrinos, though we shall also
briefly review the few radiative Dirac models that have been proposed.

The choice of Dirac or Majorana is thus a really important step in model
building.  It is perhaps fair to say that theoretical prejudice, as judged by
number of papers, favors the Majorana possibility. There are a couple of
reasons for this. One is simply that Majorana fermions are permitted by the
Poincar\'{e} group, so it might be puzzling if they were never realized in
nature, and the fact is that they constitute the simplest spinorial
representation. (Recall that a Dirac fermion is equivalent to two CP-conjugate,
degenerate Majorana fermions.)  Another was already discussed above: even if RH
neutrinos exist, at the SM level they can have gauge-invariant Majorana masses,
leading to Majorana mass eigenstates overall.  Yet another reason is a
connection between Majorana masses and an approach to understanding electric
charge quantization using classical constraints and gauge anomaly
cancellation~\cite{Foot:1990uf,Babu:1989tq}.  Nevertheless, theoretical
prejudice or popularity in the literature is not necessarily a reliable guide
to how nature actually is, so the Dirac possibility should be given due
consideration.

\subsection{Dirac neutrino schemes}
\label{subsec:Dirac-nu-schemes}

The simplest way to obtain Dirac neutrinos is by copying the way the charged-fermions gain mass.  Right-handed neutrinos are added to the SM particle content, producing the gauge-invariant, renormalizable Yukawa term
\begin{equation}
	y_\nu \overline{L} \tilde{H} \nu_{\rm R} + {\rm H.c.}\,,
\label{eq:Nu-Yukawa}
\end{equation}
where the Higgs doublet $H$ transforms as $(1,2, 1/2)$ with $\tilde{H} \equiv i\tau_2 H^*$.  The Dirac neutrino mass matrix is then
\begin{equation}
\mathcal{M}_\nu = y_\nu \langle H^0 \rangle = y_\nu \frac{v}{\sqrt{2}} \;,
\label{eq:Nu-Dirac-mass}
\end{equation}
To accommodate the $O(0.1)$ eV neutrino mass scale, one simply takes $y_\nu
\sim 10^{-13}$.  The price to pay for this simple and obvious model is a set of
tiny dimensionless parameters, some six or seven orders of magnitude smaller
than the next smallest Yukawa coupling constant (that for the electron), and
smaller even than the value a fine-tuned $\theta_{\rm QCD}$ needs to be from
the upper bound on the neutron electric-dipole moment.  This is of course
logically possible, and it is also technically natural in the 't Hooft
sense~\cite{tHooft:1979rat} because taking $y_\nu$ to zero increases the
symmetry of the theory.  Nevertheless, it seems unsatisfactory to most people.
The really tiny neutrino masses strongly suggest that the generation of
neutrino mass proceeds in some different, less obvious manner, one that
provides a rationale for why the masses are so small.  As well as the Dirac
versus Majorana question, the explanation of the tiny masses has dominated
model-building efforts in the literature.

So, how may one produce very light Dirac neutrinos? We highlight three possibilities, but there may be others: (i) a Dirac seesaw mechanism, (ii) radiative models, and (iii) extra-dimensional theories.

\subsubsection{Dirac seesaw mechanism}

In addition to the $\nu_{\rm L}$ that resides inside the doublet $L$, and the
standard RH neutrino of Eq.~\ref{eq:nuR-multiplet}, we introduce a vector-like
heavy neutral fermion $N_{\rm L,R} \sim (1,1, 0)$ and impose total lepton-number
conservation with $\nu_{\rm L,R}$ and $N_{\rm L,R}$ assigned lepton numbers of $1$.  In
addition, we impose a $Z_2$ discrete symmetry under which $\nu_{\rm R}$ and a new
gauge-singlet real scalar $S$ are odd, with all other fields even. With these
imposed symmetries, the most general Yukawa and fermion bare mass terms are
\begin{equation}
	y_N \overline{L} \tilde{H} N_{\rm R} + y_R \overline{N_{\rm L}} \nu_{\rm R} S + M_N \overline{N_{\rm L}} N_{\rm R} + {\rm H.c.}
\label{eq:Dirac-seesaw-Lag}
\end{equation}
leading to the neutral-fermion mass matrix
\begin{equation}
\left( \begin{array}{cc} \overline{\nu_{\rm L}} & \overline{N_{\rm L}} \end{array} \right)
\left( \begin{array}{cc} 0 & \ \ m_L \\ m_R & \ \ M_N \end{array} \right)
\left( \begin{array}{c} \nu_{\rm R} \\ N_{\rm R} \end{array} \right) + {\rm H.c.},
\label{eq:Dirac-seesaw-matrix}
\end{equation}
where
\begin{equation}
m_L = y_N \frac{v}{\sqrt{2}} \quad {\rm and}\quad m_R = y_R \langle S \rangle \,.
\end{equation}
We now postulate the hierarchy $m_L \ll m_R \ll M_N$ on the justification that
the bare mass term has no natural scale so could be very high, and that the
symmetry breaking scale of the new, imposed $Z_2$ should be higher than the
electroweak scale.  The light neutrino mass eigenvalue is thus
\begin{equation}
m_\nu \sim m_L \frac{m_R}{M_N}\,,
\label{eq:Dirac-seesaw-light-nu-mass}
\end{equation}
and the eigenvector is dominated by the $\nu_{\rm L}$ admixture so does not violate
weak universality bounds.  The inverse relationship of the light neutrino mass
with the large mass $M_N$ is the seesaw effect, with the postulated small
parameter $m_R/M_N$ causing $m_\nu$ to be much smaller than the
electroweak-scale mass $m_L$.  The above structure is the minimal one necessary
to illustrate the Dirac seesaw mechanism (and has a cosmological domain wall
problem because of the spontaneously broken $Z_2$), but the most elegant
implementation is in the left-right symmetric model~\cite{Roy:1983be}.  Under
the extended electroweak gauge group $SU(2)_{\rm L} \times SU(2)_{\rm R} \times
U(1)_{{\rm B}-{\rm L}}$, the RH neutrino sits in an $SU(2)_{\rm R}$ doublet with B$-$L$=-1$, while
$N_{\rm L,R}$ remains as gauge singlets.  The scalars are a left-right symmetric
pair of doublets $H_{\rm L,R}$ with B$-$L$ = 1$.  The usual scalar bidoublet is not
introduced.  The $Z_2$ symmetry is then a subgroup of $SU(2)_{\rm R}$, and $S$ is
embedded in the RH scalar doublet.  The mass and symmetry breaking hierarchy is
then $\langle H^0_{\rm L} \rangle \ll \langle H^0_{\rm R} \rangle \ll M_N$.  The absence of the
bidoublet ensures the zero in the top-left entry of the mass
matrix.\footnote{If one does not impose left-right discrete symmetry on the
Lagrangian, then there will be no cosmological domain wall problem. The Dirac
seesaw mechanism does not require this discrete symmetry.}
Several tree-level Dirac neutrino mass models have been discussed in Ref.~\cite{Ma:2016mwh}: The SM singlet Dirac fermion $N_L+N_R$ can be obviously replaced by an electroweak triplet. Alternatively a neutrinophilic two Higgs doublet model~\cite{Wang:2006jy,Gabriel:2006ns} is an attractive possibility to obtain small Dirac neutrino masses.

\subsubsection{Radiative Dirac schemes}

A generalisation of the symmetry structure of the $Z_2$ Dirac seesaw model discussed above provides us with one perspective on the construction of radiative Dirac neutrino mass models. A basic structural issue with such models is the
prevention of the tree-level term generated by the renormalizable Yukawa
interaction of Eq.~\ref{eq:Nu-Yukawa}.  Some new symmetry must be
imposed that forbids that term, but that symmetry must also be spontaneously or
softly broken in such a way that an effective $\overline{\nu_{\rm L}} \nu_{\rm R}$ operator
is produced. In the case of radiative models, this must be made to happen at loop order. One obvious possibility is to demand that ``RH neutrino number'' is conserved,
meaning that invariance under
\begin{equation}
	\nu_{\rm R} \to e^{i\theta} \nu_{\rm R} ,
\label{eq:nuR-number-symmetry}
\end{equation}
with all other SM fields as singlets, is imposed.  One may then introduce a complex scalar $\rho$ that transforms, for example, as
\begin{equation}
\rho \to e^{-i\theta/n} \rho ,
\label{eq:rho-transform}
\end{equation}
whose nonzero expectation value spontaneously breaks the symmetry.  The effective operator
\begin{equation}
	\frac{1}{\Lambda^n}\, \overline{L} \tilde{H} \nu_{\rm R} \rho^n ,
\label{eq:Dirac-eff-op}
\end{equation}
produced by integrating out new physics at mass scale $\Lambda$, is both SM gauge invariant and invariant under the imposed symmetry~\footnote{This construction resembles the well-known Froggatt-Nielsen mechanism~\cite{Froggatt:1978nt}.}. It generates a neutrino Dirac mass of order
\begin{equation}
m_\nu \sim v \left( \frac{\langle \rho \rangle}{\Lambda} \right)^n
\label{eq:Mnu-from-rho}
\end{equation}
which will be small compared to the weak scale when $\frac{\langle \rho \rangle}{\Lambda} \ll 1$.  If this operator is ``opened up'' -- derived from an underlying renormalizable or ultraviolet (UV) complete theory -- at loop-level, then a radiative neutrino Dirac-mass model is produced.  Note that in a loop-level completion, the parameter $1/\Lambda^n$ depends on powers of renormalizable coupling constants and a $1/16\pi^2$ per loop as well as the actual masses of new, exotic massive particles. See Ref.~\cite{Ma:2016mwh} for a recent systematic study of $1$-loop models based on this kind of idea. Note that the Dirac seesaw model discussed earlier is obtained as a truncated special case: the $U(1)$ symmetry with $n=1$ is replaced with its $Z_2$ subgroup, the complex scalar field $\rho$ is replaced with the real scalar field $S$, and the effective operator $\overline{L} \tilde{H} \nu_{\rm R} S$ is opened up at tree-level. 

Obviously, the phase part of $\rho$ will be a massless Nambu-Goldstone boson (NGB), but its phenomenology might be acceptable because it only couples to neutrinos. If one wishes to avoid this long range force, one could find a way to make the new $U(1)$ anomaly-free and then gauge it so that the NGB gets eaten, or one may use a discrete subgroup of the $U(1)$ to forbid Eq.~\ref{eq:Nu-Yukawa}. See Ref.~\cite{Wang:2016lve} for a discussion of the $Z_2$ case for $1$-loop models that also include a dark matter candidate.

The above is simply an example of the kind of thinking that has to go into the development of a radiative Dirac neutrino model -- we are not claiming it is the preferred option.  To our knowledge, a thorough analysis of symmetries that can prevent a tree-level Dirac mass and thus guide the construction of complete theories has not yet been undertaken in the literature.  That is one of the reasons this review will discuss Majorana models at greater length than Dirac models.

\subsubsection{Extra-dimensional theories}

One way or another, the effective coefficient in front of $\overline{L}
\tilde{H} \nu_{\rm R}$ must be made small.  Seesaw models achieve this by exploiting
powers of a small parameter given by the ratio of symmetry breaking and/or mass
scales.  Radiative models augment the seesaw feature with $1/16\pi^2$ loop
factors and products of perturbative coupling constants. In warped or
Randall-Sundrum extra-dimensional
theories~\cite{Randall:1999ee,Randall:1999vf}, the geometry of fermion
localisation in the bulk~\cite{Grossman:1999ra,Gherghetta:2000qt} can lead to
the suppression of Dirac neutrino masses through having a tiny overlap integral
between the profile functions for the neutrino chiral components and the Higgs
boson~\cite{Grossman:1999ra,Huber:2001ug,Moreau:2005kz,Agashe:2008fe,Chang:2009mv}. The phenomenological implications of Dirac neutrinos in extra-dimensional set-ups have been studied in Ref.~\cite{DeGouvea:2001mz}, where it is shown that these effects can be encoded in specific dimension-six effective operators.

One can also have a ``clockwork'' mechanism~\cite{Choi:2015fiu,Kaplan:2015fuy} to generate exponentially suppressed Dirac masses. In the same way, it is also useful to have low-scale seesaw~\cite{Hambye:2016qkf}. This mechanism can be implemented with a discrete number of new fields or via an extra spatial dimension~\cite{Giudice:2016yja}.

\subsection{Majorana neutrino schemes}
\label{subsec:Majorana-nu-schemes}

We now come to our main subject: radiative Majorana neutrino mass generation.
We also briefly review tree-level seesaw schemes, both for completeness and for
the purposes of comparison and contrast to the loop-level scenarios.  In the
course of the discussion below, an attempt will be made to classify the
different kinds of radiative models.  This is a multidimensional problem: no
single criterion can be singled out as definitely the most useful discriminator
between models.  Instead, we shall see that several overlapping considerations
emerge, including $\Delta L = 2$ effective operators, number of loops, number
of Higgs doublets, nature of the massive exotic particles, whether or not there
are extended symmetries and gauge bosons, distinctive phenomenology, and
whether or not the models address problems or issues beyond just neutrino mass
(e.g.\ dark matter, grand unification, $\ldots$).

The main distinctive feature of Majorana neutrino mass is, of course, that it
violates lepton-number conservation by two units.  It is thus extremely useful
to view the possibilities for the new physics responsible from a bottom-up
perspective, meaning SM gauge-invariant, $\Delta L = 2$ low-energy effective
operators that are to be derived from integrating out new physics that is
assumed to operate at scales higher than the electroweak. This approach permits
the tree-level
seesaw~\cite{Minkowski:1977sc,Yanagida:1979as,GellMann:1980vs,Mohapatra:1979ia,Glashow:1979nm,Magg:1980ut,Schechter:1980gr,Cheng:1980qt,Lazarides:1980nt,Wetterich:1981bx,Mohapatra:1980yp,Foot:1988aq}
and radiative models to be seen from a unified perspective.

Taking the particle content of the minimal SM, it is interesting that the simplest and lowest mass-dimension effective operator one can produce is directly related to Majorana neutrino mass generation.  This is the famous Weinberg operator~\cite{Weinberg:1979sa}
\begin{equation}
O_1 = LLHH ,
\label{eq:O1}
\end{equation}
where the $SU(2)$ indices and Lorentz structures are suppressed (one can check that there is only one independent invariant even though there are three different ways to contract the $SU(2)$ indices of the four doublets.).  We say the singular ``operator'' for convenience, but it is to be understood that there are also family indices so we really have a set of operators.  This is a mass dimension five operator, so enters the Lagrangian with a $1/\Lambda$ coefficient, where $\Lambda$ is the scale of the new physics that violates lepton number by two units.  Replacing the Higgs doublets with their vacuum expectation values (VEVs), one immediately obtains the familiar Majorana seesaw formula,
\begin{equation}
m_\nu \sim \frac{v^2}{\Lambda} ,
\label{eq:Majorana-seesaw-light-nu-mass}
\end{equation}
displaying the required suppression of $m_\nu$ with respect to the weak scale $v$ when $\epsilon \equiv v/\Lambda \ll 1$, so that the $\Delta L = 2$ new physics operates at a really high scale.

The Weinberg operator can be immediately generalized to the set
\begin{equation}
O_1''{^{\cdots}}' = LLHH(H^\dagger H)^n ,
\label{eq:generalized-Weinberg}
\end{equation}
where the number of primes is equal to $n$. One obtains ever more powerful seesaw suppression,
\begin{equation}
m_\nu \sim v \epsilon^{2n+1} ,
\label{eq:generalized-seesaw}
\end{equation}
as $n$ increases.

The task now is to derive, from an underlying renormalizable or UV complete theory, one of the Weinberg-type operators as the leading contribution to neutrino mass.  This process has come to be termed ``opening up the operator''. The choices one makes about which operator (what value of $n$) is to dominate and how it is to be opened up determine the type of theory one obtains.  Here are some possible choices:
\begin{enumerate}
\item Open up $O_1$ at tree-level using only exotic massive fermions and scalars as the new physics. \label{en:1}
\item Open up $O_1$ at $j$-loop level using heavy exotics only. \label{en:2}
\item Open up $O_1$ at $j$-loop level using both light SM particles and heavy exotics. \label{en:3}
\item Open up $O'{^{\cdots}}'_1$ at tree-level using heavy exotics only. \label{en:4}
\item Open up $O'{^{\cdots}}'_1$ at $j$-loop level using heavy exotics only. \label{en:5}
\item Open up $O'{^{\cdots}}'_1$ at $j$-loop level using both light SM particles and heavy exotics. \label{en:6}
\end{enumerate}
\begin{figure}[htb]\centering
	\begin{tikzpicture}[node distance = 1cm and 0.5cm] 
		\node[fill=blue!20,ellipse] (UV) {UV models};
		\node[below right=of UV,fill=green!20,minimum height=5ex] (deltaL2high)  {$\Delta L=2$ EFT operators};
		\node[below=of deltaL2high,fill=green!20,minimum height=5ex] (deltaL2)  {$\Delta L=2$ EFT operators};
		\node[below=of deltaL2,fill=green!20,minimum height=5ex] (d6effhigh)  {$\bar f f \nu \nu$, \dots};
		\node[below=of d6effhigh,fill=green!20,minimum height=5ex] (d6eff)  {$\bar f f \nu \nu$, \dots};
		\node[below left=of UV,fill=red!20,minimum height=5ex] (Wophigh) {$O_1^{\prime\cdots\prime}=LLHH (H^\dagger H)^n$};
		\node[below=of Wophigh,fill=red!20,minimum height=5ex] (Wop)  {$O_1^{\prime\cdots\prime}=LLHH (H^\dagger H)^n$};
		\node[below=4cm of Wop,fill=cyan!20,ellipse] (mnu)  {Neutrino masses};
		\draw[->] (UV) to node[pos=0.7,right,yshift=1ex] (tree) {(generated at tree-level)}(deltaL2high)  ;
		\draw[->] (UV) to node[pos=0.7,left,yshift=1ex] (loop) {(generated at loop-level)}(Wophigh)  ;
		\node at ($(tree)!0.5!(loop)$) {matching};
		\draw[->] (Wophigh) to node[midway,left] {RGE}(Wop);
		\draw[->] (deltaL2high) to node[midway,right] {RGE}(deltaL2);
		\draw[->] (Wop) to node[pos=0.15,left] {EWSB}(mnu);
		\draw[->] (deltaL2) to node[midway,right] {EWSB}(d6effhigh);
		\draw[->] (deltaL2high) to node[midway,below,yshift=-0.5ex] {mixing}(Wop);
		\draw[->] (Wophigh) to node[midway,above,yshift=0.5ex] {RGE} (deltaL2);
		\draw[->] (d6effhigh) to node[midway,right] {RGE}(d6eff);
		\draw[->] (d6eff) to node[midway,below] {matching}(mnu);

		\node[left=1cm of mnu] (mnuscale) {$m_\nu$};
		\draw[->] ($(mnuscale)+(0.7,-0.5)$) to node[pos=0.95,left] {$E$} ++(0,9.5);
		\draw ($(mnuscale)+(0.55,0)$) to ($(mnuscale)+(0.85,0)$);
		\draw ($(mnuscale)+(0.55,3.8)$) to ($(mnuscale)+(0.85,3.8)$);
		\draw ($(mnuscale)+(0.55,6.6)$) to ($(mnuscale)+(0.85,6.6)$);
		\node at ($(mnuscale)+(0,3.8)$) {$m_W$};
		\node at ($(mnuscale)+(0,6.6)$) {$\Lambda$};
	\end{tikzpicture}
	\caption{Running and matching for (radiative) Majorana neutrino masses. See App.~\ref{app:PowerCounting} for a discussion of the relative contribution of the different operators.}
	\label{fig:run-and-match}
\end{figure}
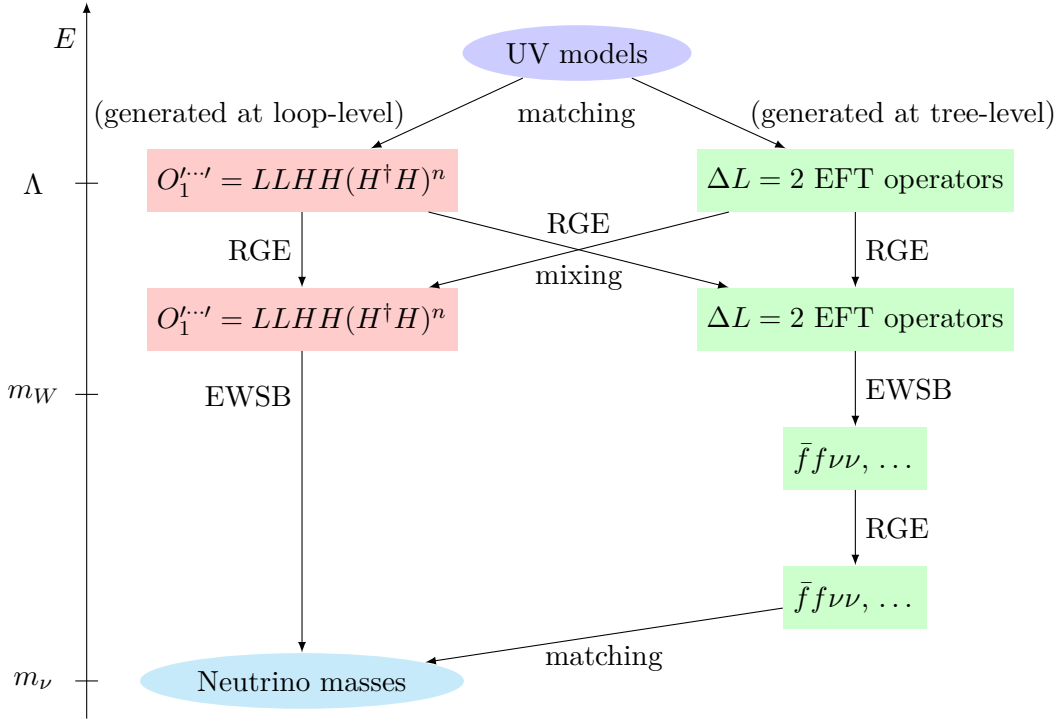

Option~\ref{en:1} leads, in its simplest form, precisely to the familiar
type-I~\cite{Minkowski:1977sc,Yanagida:1979as,GellMann:1980vs,Mohapatra:1979ia,Glashow:1979nm},
type-II~\cite{Magg:1980ut,Schechter:1980gr,Cheng:1980qt,Lazarides:1980nt,Wetterich:1981bx,Mohapatra:1980yp}
and type-III~\cite{Foot:1988aq} seesaw mechanisms, as we review in the next
subsubsection.  Option~\ref{en:2} leads to a certain kind of radiative model,
to be contrasted with that arising from option~\ref{en:3}. The difference
between the two can be expressed in terms of the matching conditions used to
connect an effective theory below the scale $\Lambda$ of the $\Delta L=2$ new
physics to the full theory above that scale, as outlined in Fig.~\ref{fig:run-and-match}. For scenario~\ref{en:2}, the
effective Weinberg operator has a nonzero Wilson coefficient at $\Lambda$, and
for all scales below that. In scenario~\ref{en:3}, on the other hand, the
Weinberg operator has a coefficient at scale $\Lambda$ that is loop-suppressed
compared to the Wilson coefficients of other, non-Weinberg-type $\Delta L = 2$
operators\footnote{The other $\Delta L=2$ operators also play an important role in the classification of radiative neutrino mass models and will be discussed in detail in Sec.~\ref{sec:RadiativeNeutrinoMassClassification}.} at that scale, where these other operators are obtained by
integrating out the heavy fields only.  If the matching is performed at
tree-level approximation, then the coefficient of the Weinberg operator at
$\Lambda$ in fact vanishes. Under renormalization group mixing, the nonzero
$\Delta L=2$ operators will, however, generate an effective Weinberg operator
as the parameters are run to scales below $\Lambda$.  If the matching is
performed at loop-level, then the Weinberg operator will have a nonzero
coefficient at scale $\Lambda$, but it will be loop-suppressed compared to the
coefficients of the relevant non-Weinberg operators.  Below $\Lambda$, the
Weinberg operator coefficient will, once again, receive corrections from the
renormalization group running and operator mixing.  Option~\ref{en:3} will be a
major topic in this review, and it motivates the enumeration of all SM
gauge-invariant $\Delta L = 2$ operators, not just those in the Weinberg class,
since the non-Weinberg operators describe the dominant $\Delta L = 2$ processes
at scale $\Lambda$.  Opening up the non-Weinberg operators at {\em tree-level}
then provides a systematic method of constructing a large class of theories
that generate neutrino masses at {\em loop order}. 

Options~\ref{en:4}-\ref{en:6} obviously repeat the exercise, but with two more
powers of $\epsilon$ which help suppress the neutrino mass.  With these
options, one needs to ensure that $O_1^{\prime\cdots\prime}$ generated from the new physics dominates over $O_1$ and all lower-dimensional operators $O^{\prime\cdots\prime}_1$. Option~\ref{en:6} is similar to \ref{en:3} in that the effective
theory between the weak and new physics scales contains some non-Weinberg type
of $\Delta L = 2$ operator(s) that dominate at scale $\Lambda$.

\subsubsection{Tree-level seesaw mechanisms}

\begin{figure}[t]
\centering
\begin{subfigure}[t]{0.35\textwidth}
\centering
 \begin{tikzpicture}[node distance=1cm and 1cm]
     \coordinate[label=left:$L$] (nu1);
     \coordinate[vertex, above right=of nu1] (v1);
     \coordinate[cross, right=of v1] (lfv);
     \coordinate[ above left=of v1, label=left:$H$] (h1);
     \coordinate[vertex, right=of lfv] (v2);
     \coordinate[below right=of v2, label=right:$L$] (nu2);
     \coordinate[above right=of v2, label=right:$H$] (h2);

     \draw[fermion] (nu1)--(v1);
     \draw[fermion] (v1) -- node[below]{$\nu_{\rm R}$} ++ (lfv);
     \draw[fermion] (v2)-- node[below]{$\nu_{\rm R}$} ++ (lfv);
     \draw[fermion] (nu2)--(v2);
     \draw[scalar] (h1) -- (v1);
     \draw[scalar] (h2) -- (v2);
   \end{tikzpicture}
\caption{}
\end{subfigure}
\begin{subfigure}[t]{0.2\textwidth}
\centering
 \begin{tikzpicture}[node distance=1cm and 1cm]
     \coordinate[label=left:$L$] (nu1);
     \coordinate[vertex, above right=of nu1] (v1);
     \coordinate[below right=of v1, label=right:$L$] (nu2);
     \coordinate[vertex, above=of v1, yshift=0.5cm] (v2);
     \coordinate[ above left=of v2, label=left:$H$] (h1);
     \coordinate[ above right=of v2, label=right:$H$] (h2);

     \draw[fermion] (nu1)--(v1);
     \draw[fermion] (nu2)--(v1);
     \draw[scalar] (v2) -- node[right]{$\Delta$} ++ (v1);
     \draw[scalar] (h1) -- (v2);
     \draw[scalar] (h2) -- (v2);
   \end{tikzpicture}
\caption{}
\end{subfigure}
\begin{subfigure}[t]{0.33\textwidth}
\centering
 \begin{tikzpicture}[node distance=1cm and 1cm]
     \coordinate[label=left:$L$] (nu1); 
     \coordinate[vertex, above right=of nu1] (v1);
     \coordinate[cross, right=of v1] (lfv);
     \coordinate[ above left=of v1, label=left:$H$] (h1);
     \coordinate[vertex, right=of lfv] (v2);
     \coordinate[below right=of v2, label=right:$L$] (nu2);
     \coordinate[above right=of v2, label=right:$H$] (h2);

     \draw[fermion] (nu1)--(v1);
     \draw[fermion] (v1) -- node[below]{$\Sigma$} ++ (lfv);
     \draw[fermion] (v2)-- node[below]{$\Sigma$} ++ (lfv);
     \draw[fermion] (nu2)--(v2);
     \draw[scalar] (h1) -- (v1);
     \draw[scalar] (h2) -- (v2);
   \end{tikzpicture}
\caption{}
\end{subfigure}
\caption[Seesaw topologies.]{Minimally opening up the Weinberg operator at tree-level using either exotic massive fermions or scalars. (a) Type-I seesaw model.  The massive exotic particle integrated out to produce an effective Weinberg operator at low energy is a SM gauge-singlet Majorana fermion, the right-handed neutrino $\nu_{\rm R}$.  (b) Type-II seesaw model. The massive exotic is a $(1,3, 1)$ scalar $\Delta$ coupling to $LL$ and $H^\dagger H^\dagger$.  It gains a small induced VEV from the latter coupling. (c) Type-III seesaw model.  The massive exotic is a $(1,3, 0)$ fermion $\Sigma$ whose middle component mixes with the left-handed neutrino.}%
\label{fig:seesaws}
\end{figure}
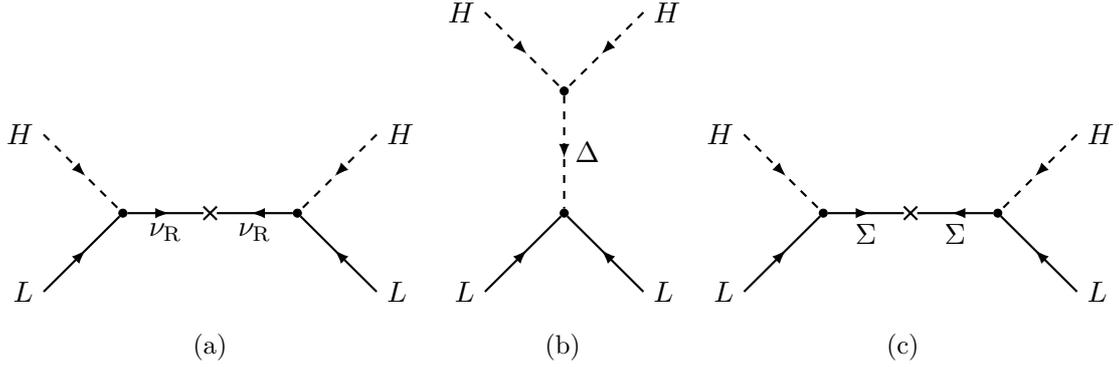

The three familiar seesaw models may be derived in a unified way by opening up
the Weinberg operator $O_1$ at tree level in the simplest possible way, using
as the heavy exotics only scalars or fermions. The available renormalizable
interactions are then just of Yukawa and scalar-scalar type. The opening-up
process is depicted in Fig.~\ref{fig:seesaws}.  The type-I and type-III seesaw
models are obtained by Yukawa coupling $LH$ with the two possible
choices of $(1,1, 0)$ and $(1,3, 0)$ fermions, both of which can have
gauge-invariant bare Majorana masses.  The type-II model is the unique theory
obtained from Yukawa coupling the fermion bilinear $LL \equiv \overline{L^c} L$
to a $(1,3, 1)$ scalar multiplet, which in turn couples to $H^\dagger
H^\dagger$, a cubic interaction term in the scalar potential.\footnote{Note
that the $LL \sim (1,1, -1)$ option is irrelevant for tree level mechanisms because it does not produce
the required $\overline{\nu^c} \nu$ bilinear.} The seesaw effect is obtained in
this case by requiring a positive quadratic term for the triplet in the
scalar potential, that on its own would cause the triplet's VEV to vanish,
but which in combination with the cubic term induces a small VEV for it.

As is clear from Fig.~\ref{fig:seesaws}, there are two interaction vertices for
all three cases, and there is only one type of exotic per case.  An interesting
non-minimal tree-level seesaw model realizing option \ref{en:4} is obtained by allowing four vertices
instead of two, and two exotic multiplets: a $(1,4, -1/2)$ scalar that couples
to $HHH^\dagger$ and a $(1,5, 0)$ massive fermion that Yukawa couples to the
exotic scalar quadruplet and the SM lepton
doublet~\cite{Kumericki:2012bh,Picek:2012ei,Liao:2010cc}.  The resulting model
produces the generalized Weinberg operator $O''_1 = LLHH(H^\dagger H)^2$ which
has mass-dimension nine.  This model is a kind of hybrid of the type-II and
type-III seesaw mechanisms, because it features both a small induced VEV for
the quadruplet and a seesaw suppression from mixing with the fermion
quintuplet.

\subsubsection{Radiative schemes and their classification}
\label{sec:RadiativeNeutrinoMassClassification}

As noted above, there are many different kinds of radiative neutrino mass models and there is probably no single classification scheme that is optimal for all purposes.  We thus discuss a few different perspectives, some much more briefly than others.  Two will be treated at length: (i) the $\Delta L = 2$ effective operator approach, and (ii) classification by loop-order openings of the Weinberg operator.

\vspace{5mm}

\noindent
{\em A.\ Standard model $\Delta L = 2$ effective operators.}  This approach can
be considered as stemming from the observations made about options~\ref{en:3}
and \ref{en:6} in Sec.~\ref{subsec:Majorana-nu-schemes}: when both light SM
particles and heavy exotics appear in the neutrino mass loop graph, it is
useful to first consider integrating out the heavy exotics at tree-level. This
produces effective $\Delta L = 2$ operators that are of non-Weinberg type.
They must be of different type, because if they were not, then the heavy
exotics would produce the Weinberg operator without participation by light SM
particles, leading either to a class~\ref{en:1} model (if $O_1$ is produced at
tree-level) or a class~\ref{en:2} model (if $O_1$ is produced at loop level).
An exhaustive list of gauge-invariant, non-Weinberg $\Delta L = 2$ operators is
thus needed.

Such a list was provided by Babu and Leung (BL)~\cite{Babu:2001ex}, based on
the following assumptions: (i) the gauge group is that of the SM only, (ii) no
internal global symmetries are imposed apart from baryon number, (iii) the
external lines are SM quarks, SM leptons and a single Higgs doublet, and (iv)
no operators of mass dimension higher than $11$ were considered. We first
comment on these assumptions.  Clearly, if the gauge symmetry was extended
beyond that of the SM, then some combination of effective operators might be
restricted to having a single coefficient, and others might be forced to
vanish, compared to the SM-gauge-group-only list.  Similar observations follow
for imposed global symmetries.  It is sensible to impose baryon number
conservation, because otherwise phenomenological constraints will force the new
physics to such high scales that obtaining neutrino masses of the required
magnitude (at least one at $0.05$ eV) will be impossible.  The case of a single
Higgs doublet can readily be generalized to multiple Higgs doublets, given that
the gauge quantum numbers are the same.  This would obviously enrich the
phenomenology of the resulting models, and if additional symmetries were also
admitted, then it would change the model-building options.  The point is simply
that $H^\dagger H$ is invariant under all possible internal symmetries, while
$H_1^\dagger H_2$ is not. (Admitting additional Higgs doublets is also
interesting for generalized-Weinberg-operator models, because then a symmetry
reason can exist for, say, $L L H_{1,2} H_{1,2} (H_1^\dagger H_2)$ being
generated without also generating what would otherwise be dominant $L L H_{1,2}
H_{1,2}$ operators.)  The addition of non-doublet scalar multiplets into the
external lines is a more serious complication.  Some discussion of the possible
roles of additional scalars that gain nonzero VEVs that contribute to neutrino
mass generation will be given in later sections.  Another restriction worth
noting in the BL list is the absence of the gauge-singlet RH neutrinos.  In
assumption (iv), the point to highlight is the absence of SM gauge fields.
Babu and Leung did actually write down the mass-dimension-$7$ operators
containing gauge fields, and Ref.~\cite{Bhattacharya:2015vja} further examined
them.  As far as we know, however, no complete analysis has been undertaken for
the dimension-$9$ and -$11$ cases.  Finally, it is sensible to stop at
dimension $11$ because at any higher order the contribution to neutrino mass
will be insufficiently large.  The BL list, as enumerated from $O_1$ to
$O_{60}$, took operators that could be thought of as products of
lower-dimension operators with the SM invariants $H H^\dagger$ and the three
dimension-4 charged-fermion Yukawa terms as implicit.
Reference~\cite{deGouvea:2007qla} extended their list by explicitly including
the latter cases, thereby augmenting the operator count to $O_{75}$.

Operators meeting all of these requirements exist at all odd mass dimensions~\cite{Babu:2001ex,deGouvea:2014lva,Kobach:2016ami}, starting with the Weinberg operator $O_1$ as the unique dimension-5 case (up to family indices).  The dimension-7 list is as follows:
\begin{align}
 O_2 &= L^i L^j L^k e^c H^l \epsilon_{ij}\epsilon_{kl},  & O_{3a} &= L^i L^j Q^k d^c H^l \epsilon_{ij}\epsilon_{kl},  & O_{3b} &= L^i L^j Q^k d^c H^l \epsilon_{ik}\epsilon_{jl}, & \nonumber\\
 O_{4a} &= L^i L^j \bar{Q}_i \bar{u}^c H^k \epsilon_{jk}, & O_{4b} &= L^i L^j \bar{Q}_k \bar{u}^c H^k \epsilon_{ij},  & O_8 &= L^i \bar{e}^c \bar{u}^c d^c H^j \epsilon_{ij}. &
\label{eq:dim7-BLlist}
\end{align}
We follow the BL numbering scheme, which was based on tracking the number of fermion fields in the operator rather than the mass dimension. The operators are separated in three groups with 2, 4, and 6 fermions.  Some comments now need to be made about the schematic notation and what features are suppressed.  The field-string defining each operator above completely defines the flavor content of that operator.  Thus $L \sim (1,2,-1/2)$ is the lepton doublet, $Q \sim (3,2,1/6)$ is the quark doublet, $e^c \sim (1,1,1)$ is the isosinglet charged anti-lepton, $d^c \sim (\bar{3},1, 1/3)$ is the isosinglet anti-down, 
$u^c \sim (\bar{3},1, -2/3)$ is the isosinglet anti-up, and $H \sim (1,2, 1/2)$ is the Higgs doublet. The color indices and the different possible Lorentz structures are suppressed.  In general, there are a number of independent operators corresponding to each flavor-string.  For the dimension-7 list, operators $O_3$ and $O_4$ each have two independent possibilities for the contraction of the isospin indices, as explicitly defined above, but obviously a unique color contraction.  Babu and Leung specify the independent internal-index contractions, but only make general remarks on the Lorentz structures, and we shall follow suit.  To assist the reader to understand the notation, we write out the above operators more completely in standard 4-component spinor notation, but for scalar and pseudoscalar Lorentz structures only and with isospin indices suppressed:
\begin{eqnarray}
	O_2 & = & L L L e^c H = \left[ \overline{(L_{\rm L})^c} L_{\rm L} \right] \left[ \overline{e_{\rm R}} L_{\rm L} \right] H, \nonumber\\
	O_3 & = & L L Q d^c H = \left[ \overline{(L_{\rm L})^c} L_{\rm L} \right] \left[ \overline{d_{\rm R}} Q_{\rm L} \right] H\ \ {\rm or}\ \ \left[ \overline{(L_{\rm L})^c} Q_{\rm L} \right] \left[ \overline{d_{\rm R}} L_{\rm L} \right] H, \nonumber\\
	O_4 & = & L L \bar{Q} \bar{u}^c H = \left[ \overline{(L_{\rm L})^c} L_{\rm L} \right] \left[ \overline{Q_{\rm L}} u_{\rm R} \right] H, \nonumber\\
	O_8 & = & L \bar{e}^c \bar{u}^c d^c H = \left[\overline{d_{\rm R}} L_{\rm L}\right] \left[ \overline{(e_{\rm R})^c} u_{\rm R} \right] H.
\label{eq:dim7-Lorentz}
\end{eqnarray}
Of course, these operators feature quark and charged-lepton fields in addition
to neutrinos and Higgs bosons, so they do not by themselves produce neutrino
masses.  The charged fermion fields have to be closed off in a loop or loops to
produce a neutrino self-energy graph which then generates a Weinberg-type
operator, as per options~\ref{en:3} and \ref{en:6}. In fact, using this
procedure and naive dimensional analysis one can estimate their matching
contribution to the Weinberg operator, as done in Ref.~\cite{deGouvea:2007qla}.
In addition, every dimension-7 operator in Eq.~\ref{eq:dim7-BLlist} may be multiplied by $H^\dagger H$ to produce a dimension-9 generalisation of that operator, just as $O'_1$ is a generalisation of $O_1$.
At dimension 9, there are many more operators.  Six of the flavor strings feature four fermion fields and three Higgs doublets:
\begin{align}
 O_5 &= L^i L^j Q^k d^c H^l H^m H_i^\dagger \epsilon_{jl}\epsilon_{km}, &  O_6 &= L^i L^j \bar{Q}_k \bar{u}^c H^l H^k H_i^\dagger \epsilon_{jl},& \nonumber\\
 O_7 &= L^i Q^j \bar{e}^c \bar{Q}_k H^k H^l H^m \epsilon_{il}\epsilon_{jm}, & O_{61} &= L^i L^j H^k H^l L^r e^c H_r^\dagger \epsilon_{ik}\epsilon_{jl},& \nonumber \\ 
 O_{66} &= L^i L^j H^k H^l Q^r d^c H_r^\dagger \epsilon_{ik}\epsilon_{jl}, & O_{71}  &= L^i L^j H^k H^l Q^r u^c H^s  \epsilon_{ik} \epsilon_{jl} \epsilon_{rs}, & 
\label{eq:dim9-4f3H}
\end{align}
Note that the operators $O_{61,66,71}$ are the products of $O_1$ and the three SM Yukawa operators. Another $12$ are six-fermion operators:
\begin{align}
 O_9 &= L^i L^j L^k e^c L^l e^c \epsilon_{ij}\epsilon_{kl}, & O_{10} & = L^i L^j L^k e^c Q^l d^c \epsilon_{ij}\epsilon_{kl},& \nonumber \\
 O_{11a}& = L^i L^j Q^k d^c Q^l d^c \epsilon_{ij}\epsilon_{kl}, & O_{11b} &= L^i L^j Q^k d^c Q^l d^c \epsilon_{ik}\epsilon_{jl}, & \nonumber \\
 O_{12a} &= L^i L^j \bar{Q}_i \bar{u}^c \bar{Q}_j \bar{u}^c, & O_{12b} &= L^i L^j \bar{Q}_k \bar{u}^c \bar{Q}_l \epsilon_{ij} \epsilon^{kl}, & \nonumber \\
 O_{13} &= L^i L^j \bar{Q}_i \bar{u}^c L^k e^c \epsilon_{jk}, &  O_{14a}& = L^i L^j \bar{Q}_k \bar{u}^c Q^k d^c \epsilon_{ij}, & O_{14b} &= L^i L^j \bar{Q}_i \bar{u}^c Q^k d^c \epsilon_{jk},& \nonumber \\
 O_{15} &= L^i L^j L^k d^c \bar{L}_i \bar{u}^c \epsilon_{jk},  & O_{16} &= L^i L^j \bar{e}^c d^c \bar{e}^c u^c \epsilon_{ij}, & \nonumber \\
 O_{17} &= L^i L^j d^c d^c \bar{d}^c \bar{u}^c \epsilon_{ij}, & O_{18} &= L^i L^j d^c u^c \bar{u}^c \bar{u}^c \epsilon_{ij}, & \nonumber \\
O_{19} &= L^i Q^j d^c d^c \bar{e}^c \bar{u}^c \epsilon_{ij}, & O_{20} &= L^i d^c \bar{Q}_i \bar{u}^c \bar{e}^c \bar{u}^c.  & 
\label{eq:dim9-6f}
\end{align}
Although absent from the BL list another such operator is $u^c u^c \bar{d}^c \bar{d}^c e^c e^c$, which generates the correct neutrino mass scale only for a very low lepton-number violation scale. In case it consists entirely of the first generation SM fermions it is strongly constrained by $0\nu\beta\beta$ (generated at tree level by this operator). The large number of dimension-11 operators can be found listed in Refs.~\cite{Babu:2001ex,deGouvea:2007qla}.

References~\cite{deGouvea:2007qla,Angel:2012ug} performed general analyses of diagram topologies for opening up these operators at tree-level using massive exotic scalars and either vector-like or Majorana fermion exotics, and consequently producing neutrino mass at various loop levels. The operators
\begin{equation}
O_2,\ \ O_{3b},\ \ O_{4a},\ \ O_5,\ \ O_6,\ \ O_{61},\ \ O_{66},\ \ O_{71}
\label{eq:Op-1loop}
\end{equation}
can give rise to $1$-loop neutrino mass models, while
\begin{equation}
O_2,\ \ O_{3a},\ \ O_{3b},\ \ O_{4a},\ \ O_{4b},\ \ O_{5-10},\ \ O_{11b},\ \ O_{12a},\ \ O_{13},\ \ O_{14b},\ \ O_{61},\ \ O_{66},\ \ O_{71}
\label{eq:Op-2loop}
\end{equation}
can produce $2$-loop models.  The set
\begin{equation}
O_{11a},\ \ O_{12b},\ \ O_{14a},\ \ O_{15-20}
\label{eq:Op-3loop}
\end{equation}
can form the basis for neutrino mass to be generated at three or more loops.

In each of these cases, one may derive an indicative upper bound on the scale of new physics from the requirement that at least one neutrino mass be at least $0.05$ eV in magnitude. For example, for operators involving first generation\footnote{The bound on the scale of new physics is generally higher for operators involving heavier quarks.} quarks this bound can be estimated as follows: Operator $O_{19}$, which can be opened up to give a $3$-loop neutrino mass contribution, has the lowest upper bound on the new physics scale of about $1$ TeV (apart from $u^c u^c \bar{d}^c \bar{d}^c e^c e^c$).  The highest is about $4 \times 10^9$ TeV for the $1$-loop case of $O_{4a}$.  These estimates come from an examination of the loop contribution to neutrino mass only, and do not take into account other phenomenological constraints that will exist for each complete model.  As part of that, any unknown coupling constants, such as Yukawas that involve the exotic fermions and/or scalars were set to unity.  In a realistic theory, many of these constants would be expected to be less than one, which would bring the scale of new physics to lower values.  In any case, one can see that the required new physics, even for $1$-loop models, is typically more testable than the type-I, II and III seesaw models.  Some high loop models, as the $O_{19}$ case demonstrates, have very low scales of new physics and some may even be ruled out already.  At the dimension-11 operator level, so not explicitly discussed here, there are even examples which can at best produce a $5$-loop neutrino mass contribution.  Those models are definitely already excluded.  Examples of full models that are associated with specific operators will be presented in later sections.

\vspace{5mm}

\noindent
{\em B.  Number of loops.}\ \ A complementary perspective on the spectrum of possible radiative neutrino mass models is provided by adopting the number of loops as the primary consideration rather than the type of $\Delta L = 2$ effective operator that dominates the new physics.  Equations~\ref{eq:Op-1loop}-\ref{eq:Op-3loop} already form the basis for such a classification for type~\ref{en:3} and type~\ref{en:6} scenarios, but a more general analysis will also capture the type~\ref{en:2} and type~\ref{en:5} possibilities.

At $j$-loop order, neutrino masses are typically given by
\begin{equation}
m_\nu \sim C \left(\frac{1}{16\pi^2}\right)^{j} \frac{v^2}{\Lambda} 
\end{equation}
for the $O_1$ associated options~\ref{en:2} and \ref{en:3}, and
\begin{equation}
m_\nu \sim C \left(\frac{1}{16\pi^2}\right)^{j} \frac{v^4}{\Lambda^3} 
\end{equation}
for the $O'_1$ cases of options~\ref{en:5} and \ref{en:6}, where  $v \equiv
\sqrt{2} \langle H^0 \rangle \simeq 100$ GeV, and $\Lambda$ is the new-physics
scale where lepton number is violated by two units. All coupling constants, and
for some models also certain mass-scale ratios, are absorbed in the
dimensionless coefficient $C$. In order to explain the atmospheric mass
splitting lower bound of $0.05$ eV, we obtain an upper limit on the new physics
scale $\Lambda$ of $10^5\, C$ TeV for 3-loop models and $10\, C$ TeV for 5-loop
models corresponding to the $O_1$ cases, and $10\, C^{1/3}$ TeV for the $O'_1$
case at 3-loop order. Constraints from flavor physics severely constrain the
scale of new physics and the couplings entering in $C$. In addition, in models
which feature explicit $\Delta L = 2$ lepton-number violation through trilinear
scalar interactions, the latter cannot be arbitrarily large because otherwise
they have issues with naturalness (see Ref.~\cite{Herrero-Garcia:2017xdu} for
the case of the Zee model) and charge/color breaking minima (see
Refs.~\cite{Frere:1983ag,AlvarezGaume:1983gj,Casas:1996de} for studies in the
context of supersymmetry and ref.~\cite{Herrero-Garcia:2014hfa} for the case of
the Zee-Babu model). Thus, apart from a few 4-loop
models~\cite{Nomura:2016seu,Nomura:2016fzs,CarcamoHernandez:2016pdu} which compensate the loop
suppression by a high multiplicity of particles in the loop, the vast majority
of radiative neutrino mass models generate neutrino mass at 1-, 2-, or 3-loop
level.  We therefore focus on these cases.

\vspace{5mm}

\begin{figure}
\centering
\begin{subfigure}[t]{0.45\linewidth}
  \centering
  \begin{tikzpicture}[node distance=1cm and 1cm]
   \coordinate[label=left:$L$] (nu1);
   \coordinate[vertex, right=of nu1] (v1);
   \draw[scalarnoarrow] (v1) arc[start angle=180, delta angle=-180, radius=1.1cm] node[pos=0.33,vertex] (v03) {} node[pos=0.667,vertex] (v07) {} node[vertex] (v2) {};
    \coordinate[right=of v2, label=right:$L$] (nu2);
    \coordinate[above=of nu1,xshift=0.4cm, yshift=0.8cm, label=left:$H$] (h1);
    \coordinate[above=of nu2,xshift=-0.4cm, yshift=0.8cm, label=right:$H$] (h2);
     
    \draw[fermion] (nu1)--(v1);
    \draw[fermion] (nu2)--(v2);
    \draw[fermionnoarrow] (v1)--(v2);
    \draw[scalar] (h1)--(v03); 
    \draw[scalar] (h2)--(v07); 
  \end{tikzpicture}
  \caption{T1-i}
\end{subfigure}
%
\begin{subfigure}[t]{0.45\linewidth}
  \centering
   \begin{tikzpicture}[node distance=1cm and 1cm]
     \coordinate[label=left:$L$] (nu1);
     \coordinate[vertex, right=of nu1] (v1);
     \coordinate[vertex, right=of v1] (lfv);
     \coordinate[vertex, above=of lfv] (v3);
     \coordinate[above=of v3,  label=above:$H$] (h1);
     \coordinate[below=of lfv, label=below:$H$] (h2);
     \coordinate[vertex, right=of lfv] (v2);
     \coordinate[right=of v2, label=right:$L$] (nu2);
     \draw[fermion] (nu1)--(v1);
     \draw[fermionnoarrow] (v1)--(lfv);
     \draw[fermionnoarrow] (lfv)--(v2);
     \draw[fermion] (nu2)--(v2);
     \draw[scalar] (h1) -- (v3);
     \draw[scalar] (h2) -- (lfv);
     \draw[scalarnoarrow] (v3) to[out=180,in=90] (v1);
     \draw[scalarnoarrow] (v3) to[out=0,in=90] (v2);
   \end{tikzpicture}
  \caption{T1-ii}
\end{subfigure}
\\

\begin{subfigure}[t]{0.45\linewidth}
 \centering
  \begin{tikzpicture}[node distance=1cm and 1cm]
   \coordinate[label=left:$L$] (nu1);
   \coordinate[vertex, right=of nu1] (v1);
   \draw[fermionnoarrow] (v1) arc[start angle=180, delta angle=-180, radius=1.1cm] node[pos=0.33,vertex] (v03) {} node[pos=0.667,vertex] (v07) {} node[vertex] (v2) {};
    \coordinate[right=of v2, label=right:$L$] (nu2);
    \coordinate[above=of nu1,xshift=0.4cm, yshift=0.8cm, label=left:$H$] (h1);
    \coordinate[above=of nu2,xshift=-0.4cm, yshift=0.8cm, label=right:$H$] (h2);
     
    \draw[fermion] (nu1)--(v1);
    \draw[fermion] (nu2)--(v2);
    \draw[scalarnoarrow] (v1)--(v2);
    \draw[scalar] (h1)--(v03); 
    \draw[scalar] (h2)--(v07); 
  \end{tikzpicture}
  \caption{T1-iii}
\end{subfigure}
\begin{subfigure}[t]{0.45\linewidth}
 \centering
  \begin{tikzpicture}[node distance=1cm and 1cm]
   \coordinate[label=left:$L$] (nu1);
   \coordinate[vertex, right=of nu1] (v1);
   \draw[scalarnoarrow] (v1) arc[start angle=180, delta angle=-180, radius=1.1cm] node[pos=0.5,vertex] (v05) {} node[vertex] (v2) {};
    \coordinate[right=of v2, label=right:$L$] (nu2);
    \coordinate[above=of nu1,xshift=0.5cm, yshift=1.0cm, label=left:$H$] (h1);
    \coordinate[above=of nu2,xshift=-0.5cm, yshift=1.0cm, label=right:$H$] (h2);

    \draw[scalar] (h1)--(v05); 
    \draw[scalar] (h2)--(v05); 
    \draw[fermion] (nu1)--(v1);
    \draw[fermion] (nu2)--(v2);
    \draw[fermionnoarrow] (v1)--(v2);
  \end{tikzpicture} 
  \caption{T3}
\end{subfigure}
\\
\begin{subfigure}[t]{0.45\linewidth}
 \centering
  \begin{tikzpicture}[node distance=1cm and 1cm]
   \coordinate[label=left:$L$] (nu1);
   \coordinate[vertex, right=of nu1] (v1);
   \draw[scalarnoarrow] (v1) arc[start angle=180, delta angle=-180, radius=1.1cm] node[pos=0.5,vertex] (v05) {} node[vertex] (v2) {};
    \coordinate[right=of v2, label=right:$L$] (nu2);
    \coordinate[vertex, above=of v05] (v3);
    \coordinate[left=2cm of v3, label=left:$H$] (h1);
    \coordinate[right=2cm of v3, label=right:$H$] (h2);

    \draw[scalar] (h1)--(v3); 
    \draw[scalar] (h2)--(v3); 
    \draw[scalarnoarrow] (v05)--(v3); 
    \draw[fermion] (nu1)--(v1);
    \draw[fermion] (nu2)--(v2);
    \draw[fermionnoarrow] (v1)--(v2);
  \end{tikzpicture}
  \caption{T4-2-i}
\end{subfigure}
\begin{subfigure}[t]{0.45\linewidth}
 \centering
  \begin{tikzpicture}[node distance=1cm and 1cm]
   \coordinate[label=left:$L$] (nu1);
   \coordinate[vertex, right=of nu1] (v3);
   \coordinate[above=of v3, label=above:$H$] (h1);
   \coordinate[vertex, right=of v3] (v1);
   \draw[scalarnoarrow] (v1) arc[start angle=180, delta angle=-180, radius=1.1cm] node[pos=0.5,vertex] (v05) {} node[vertex] (v2) {};
    \coordinate[above=of v05, label=above:$H$] (h2);
    \coordinate[right=of v2, label=right:$L$] (nu2);

    \draw[fermion] (nu1)--(v3);
    \draw[fermion] (nu2)--(v2);
    \draw[fermionnoarrow] (v3)--(v1)--(v2);
    \draw[scalar] (h2)--(v05);
    \draw[scalar] (h1)--(v3);
  \end{tikzpicture}
  \caption{T4-3-i}
\end{subfigure}
\caption[1-loop topologies.]{Feynman diagram topologies for 1-loop radiative neutrino mass generation with the Weinberg operator $O_1 = LLHH$. Dashed lines could be scalars or gauge bosons if allowed.}
\label{fig:WeinbergOneLoop}
\end{figure}
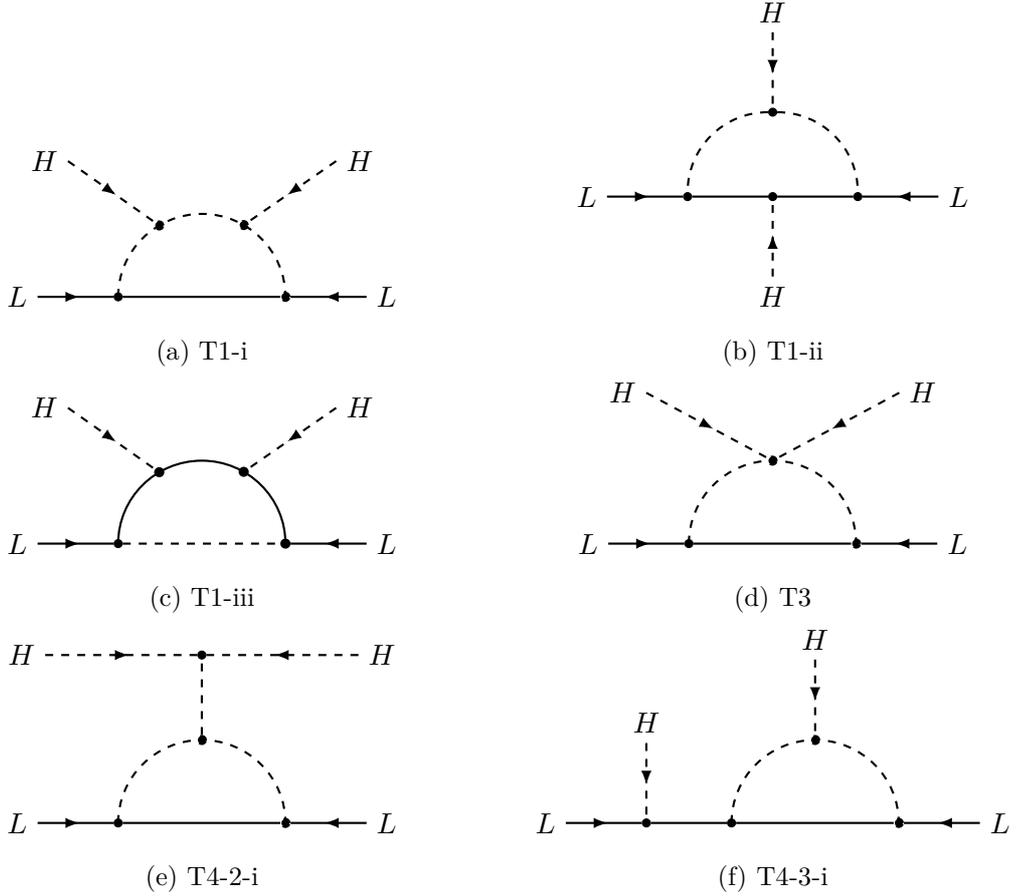

\noindent
{\em 1-loop topologies for $O_1 = LLHH$.}\ \ The opening up of the Weinberg operator at 1-loop level has been systematically studied in Refs.~\cite{Ma:1998dn, Bonnet:2012kz}. The authors of Ref.~\cite{Bonnet:2012kz} identified 12 topologies which contribute to neutrino mass.  Among all the topologies and possible Lorentz structures, topology T2 cannot be realized in a renormalizable theory.  For the other topologies, the expression for neutrino mass and the possible particle content for electroweak singlet, doublet, and triplet representations is listed in the appendix of Ref.~\cite{Bonnet:2012kz}.  The divergent ones, T4-1-i, T4-2-ii, T4-3-ii, T5 and T6, need counter-terms to absorb the divergences, which are indeed tree-level realisations of the Weinberg operators. Furthermore for T4-1-ii, there is no mechanism to forbid or suppress the tree-level  contribution from Weinberg operator, such as extra discrete symmetry or $U(1)$.  Therefore, there are in total six topologies which generate neutrino mass via a genuine\footnote{In a genuine n-loop neutrino mass model, only diagrams starting from n-loop order contribute to neutrino mass. There are no tree level or lower order loop contributions.} 1-loop diagram: T1-i, T1-ii, T1-iii, T3, T4-2-i, T4-3-i, which  are depicted in Fig.~\ref{fig:WeinbergOneLoop}. Depending on the particle content, the topologies do not rely on any additional symmetry. However, the topologies T4-x-i require a discrete $Z_2$ symmetry in addition to demanding Majorana fermions in the loop with lepton-number conserving couplings. This is difficult to achieve in a field theory, as lepton-number is necessarily broken by neutrino mass. For example, in topology T4-2-i the scalar connected to the two Higgs doublets $H$ is necessarily an electroweak triplet and thus its direct coupling to two lepton doublets $L$ is unavoidable. This coupling induces a type-II seesaw tree-level contribution to neutrino mass. Similar arguments hold for the other topologies T4-x-i.

\vspace{5mm}

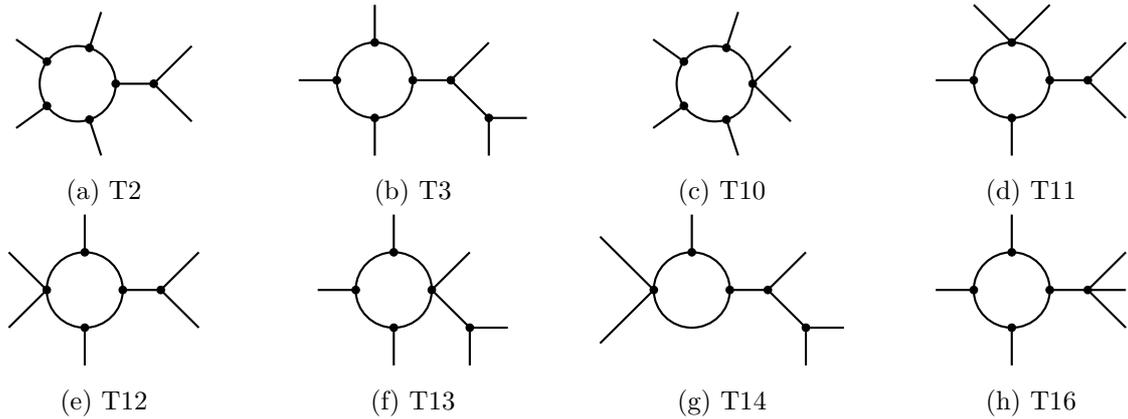
\begin{figure}
\centering

\begin{subfigure}[t]{0.24\linewidth}
\centering
\begin{tikzpicture}[scale=0.5]
\begin{feynman}[small, every edge={thick}]
\draw [thick] (0,0) circle [radius=1];

\vertex [dot] at (72:1) (l1) {};
\vertex [dot] at (144:1) (l2) {};
\vertex [dot] at (216:1) (l3) {};
\vertex [dot] at (288:1) (l4) {};
\vertex [dot] at (360:1) (l5) {};

\vertex at (72:2) (e1);
\vertex at (144:2) (e2);
\vertex at (216:2) (e3);
\vertex at (288:2) (e4);
\vertex [dot] at (360:2) (e5) {};

\vertex [above right=0.5 and 0.5 of e5] (f1);
\vertex [below right=0.5 and 0.5 of e5] (f2);

\diagram* {
  (l1) -- (e1),
  (l2) -- (e2),
  (l3) -- (e3),
  (l4) -- (e4),
  (l5) -- (e5),
  (e5) -- (f1),
  (e5) -- (f2),
};
\end{feynman}
\end{tikzpicture}
\caption{T2}
\label{subfig:T2}
\end{subfigure}
\begin{subfigure}[t]{0.24\linewidth}
\centering
\begin{tikzpicture}[scale=0.5]
\begin{feynman}[small, every edge={thick}]

\draw [thick] (0,0) circle [radius=1];

\vertex [dot] at (0:1) (l1) {};
\vertex [dot] at (90:1) (l2) {};
\vertex [dot] at (180:1) (l3) {};
\vertex [dot] at (270:1) (l4) {};

\vertex [dot] at (0:2) (e1) {};
\vertex at (90:2) (e2);
\vertex at (180:2) (e3);
\vertex at (270:2) (e4);

\vertex [above right=0.5 and 0.5 of e1] (f1);
\vertex [dot] [below right=0.5 and 0.5 of e1] (f2) {};
\vertex [right=0.5 of f2] (g1);
\vertex [below=0.5 of f2] (g2);

\diagram* {
 (l1) -- (e1),
 (l2) -- (e2),
 (l3) -- (e3),
 (l4) -- (e4),
 (e1) -- (f1),
 (e1) -- (f2),
 (f2) -- (g1),
 (f2) -- (g2),
};

\end{feynman}
\end{tikzpicture}
\caption{T3}
\label{subfig:T3}
\end{subfigure}
%
\begin{subfigure}[t]{0.24\linewidth}
\centering
\begin{tikzpicture}[scale=0.5]
\begin{feynman}[small, every edge={thick}]

\draw [thick] (0,0) circle [radius=1];

\vertex [dot] at (72:1) (l1) {};
\vertex [dot] at (144:1) (l2) {};
\vertex [dot] at (216:1) (l3) {};
\vertex [dot] at (288:1) (l4) {};
\vertex [dot] at (360:1) (l5) {};

\vertex at (72:2) (e1);
\vertex at (144:2) (e2);
\vertex at (216:2) (e3);
\vertex at (288:2) (e4);

\vertex [above right=0.5 and 0.5 of l5] (f1);
\vertex [below right=0.5 and 0.5 of l5] (f2);

\diagram* {
(l1) -- (e1),
  (l2) -- (e2),
  (l3) -- (e3),
  (l4) -- (e4),
  (l5) -- (f1),
  (l5) -- (f2),
};

\end{feynman}
\end{tikzpicture}
\caption{T10}
\label{subfig:T10}
\end{subfigure}
%
\begin{subfigure}[t]{0.24\linewidth}
\centering
\begin{tikzpicture}[scale=0.5]
\begin{feynman}[small, every edge={thick}]

\draw [thick] (0,0) circle [radius=1];

\vertex [dot] at (0:1) (l1) {};
\vertex [dot] at (90:1) (l2) {};
\vertex [dot] at (180:1) (l3) {};
\vertex [dot] at (270:1) (l4) {};

\vertex [dot] at (0:2) (e1) {};
\vertex at (180:2) (e3);
\vertex at (270:2) (e4);

\vertex [above left=0.5 and 0.5 of l2] (f1);
\vertex [above right=0.5 and 0.5 of l2] (f2);

\vertex [above right=0.5 and 0.5 of e1] (f3);
\vertex [below right=0.5 and 0.5 of e1] (f4);

\diagram* {
 (l1) -- (e1),
 (l2) -- (f1),
 (l2) -- (f2),
 (l3) -- (e3),
 (l4) -- (e4),
 (e1) -- (f3),
 (e1) -- (f4),
};

\end{feynman}
\end{tikzpicture}
\caption{T11}
\label{subfig:T11}
\end{subfigure}
%
\begin{subfigure}[t]{0.24\linewidth}
\centering
\begin{tikzpicture}[scale=0.5]
\begin{feynman}[small, every edge={thick}]

\draw [thick] (0,0) circle [radius=1];

\vertex [dot] at (0:1) (l1) {};
\vertex [dot] at (90:1) (l2) {};
\vertex [dot] at (180:1) (l3) {};
\vertex [dot] at (270:1) (l4) {};

\vertex [dot] at (0:2) (e1) {};
\vertex at (90:2) (e2);
\vertex at (270:2) (e4);

\vertex [above left=0.5 and 0.5 of l3] (f1);
\vertex [below left=0.5 and 0.5 of l3] (f2);

\vertex [above right=0.5 and 0.5 of e1] (f3);
\vertex [below right=0.5 and 0.5 of e1] (f4);

\diagram* {
 (l1) -- (e1),
 (l2) -- (e2),
 (l3) -- (f1),
 (l3) -- (f2),
 (l4) -- (e4),
 (e1) -- (f3),
 (e1) -- (f4),
};

\end{feynman}
\end{tikzpicture}
\caption{T12}
\label{subfig:T12}
\end{subfigure}
%
\begin{subfigure}[t]{0.24\linewidth}
\centering
\begin{tikzpicture}[scale=0.5]
\begin{feynman}[small, every edge={thick}]

\draw [thick] (0,0) circle [radius=1];

\vertex [dot] at (0:1) (l1) {};
\vertex [dot] at (90:1) (l2) {};
\vertex [dot] at (180:1) (l3) {};
\vertex [dot] at (270:1) (l4) {};

\vertex at (90:2) (e2);
\vertex at (180:2) (e3);
\vertex at (270:2) (e4);

\vertex [above right=0.5 and 0.5 of l1] (f1);
\vertex [dot] [below right=0.5 and 0.5 of l1] (f2) {};
\vertex [right=0.5 of f2] (g1);
\vertex [below=0.5 of f2] (g2);

\diagram* {
 (l2) -- (e2),
 (l3) -- (e3),
 (l4) -- (e4),
 (l1) -- (f1),
 (l1) -- (f2),
 (f2) -- (g1),
 (f2) -- (g2),
};

\end{feynman}
\end{tikzpicture}
\caption{T13}
\label{subfig:T13}
\end{subfigure}
%
\begin{subfigure}[t]{0.24\linewidth}
\centering
\begin{tikzpicture}[scale=0.5]
\begin{feynman}[small, every edge={thick}]

\draw [thick] (0,0) circle [radius=1];

\vertex [dot] at (0:1) (l1) {};
\vertex [dot] at (90:1) (l2) {};
\vertex [dot] at (180:1) (l3) {};

\vertex [dot] at (0:2) (e1) {};
\vertex at (90:2) (e2);
\vertex [above left=of l3] (f3);
\vertex [below left=of l3] (f4);

\vertex [above right=0.5 and 0.5 of e1] (f1);
\vertex [dot] [below right=0.5 and 0.5 of e1] (f2) {};
\vertex [right=0.5 of f2] (g1);
\vertex [below=0.5 of f2] (g2);

\diagram* {
 (l1) -- (e1),
 (l2) -- (e2),
 (e1) -- (f1),
 (e1) -- (f2),
 (f2) -- (g1),
 (f2) -- (g2),
 (l3) -- (f3),
 (l3) -- (f4),
};

\end{feynman}
\end{tikzpicture}
\caption{T14}
\label{subfig:T14}
\end{subfigure}
%
\begin{subfigure}[t]{0.24\linewidth}
\centering
\begin{tikzpicture}[scale=0.5]
\begin{feynman}[small, every edge={thick}]

\draw [thick] (0,0) circle [radius=1];

\vertex [dot] at (0:1) (l1) {};
\vertex [dot] at (90:1) (l2) {};
\vertex [dot] at (180:1) (l3) {};
\vertex [dot] at (270:1) (l4) {};

\vertex [dot] at (0:2) (e1) {};
\vertex at (90:2) (e2);
\vertex at (180:2) (e3);
\vertex at (270:2) (e4);

\vertex [above right=0.5 and 0.5 of e1] (f1);
\vertex [below right=0.5 and 0.5 of e1] (f2);
\vertex at (0:3) (f3);

\diagram* {
 (l1) -- (e1),
 (l2) -- (e2),
 (l3) -- (e3),
 (l4) -- (e4),
 (e1) -- (f1),
 (e1) -- (f2),
 (e1) -- (f3),
};

\end{feynman}
\end{tikzpicture}
\caption{T16}
\label{subfig:T16}
\end{subfigure}

\caption[1-loop topologies of the dimension-7 operator $O_1^\prime$.]{Topologies that {\em can} give rise to genuine 1-loop openings of the dimension-7 Weinberg operator $O'_1 = L L H H (H^\dagger H)$. }
\label{fig:topologies-1loop-dim7Weinberg}

\end{figure}

\noindent
{\em 1-loop topologies for $O'_1 = LLHH (H^\dagger H)$.}\ \ A similar analysis has been performed for $1$-loop topologies that give rise to the dimension-7 generalized Weinberg operator~\cite{Cepedello:2017eqf}. Of the $48$ possible topologies, only the eight displayed in Fig.~\ref{fig:topologies-1loop-dim7Weinberg} are relevant for genuine $1$-loop models.  For specific cases, not all of these eight diagrams will be realized. The three-point vertices can be Yukawa, gauge or cubic scalar interactions, while the four-point vertices only contain scalar and gauge bosons.

\vspace{5mm}

\begin{figure}
\centering

\begin{subfigure}[t]{0.32\linewidth}
\begin{tikzpicture}
\begin{feynman}[small, every edge={thick}]
\vertex (a);
\vertex [dot, right=of a] (b) {};
\vertex [dot, right=of b] (c) {};
\vertex [dot, right= of c] (d) {};
\vertex [right=of d] (e);
\vertex [below=of a] (f);
\vertex [dot, below=of b] (g) {};
\vertex [dot, below=of c] (k) {};
\vertex [dot, below=of d] (i) {};
\vertex [below=of e] (j);
\diagram* {
(a) -- (b),
(b) -- (c),
(c) -- (d),
(d) -- (e),
(b) -- (g),
(c) -- (k),
(d) -- (i),
(f) -- (g),
(g) -- (k),
(k) -- (i),
(i) -- (j),
};
\end{feynman}
\end{tikzpicture}
\caption{$T2^B_1$}
\end{subfigure}
\begin{subfigure}[t]{0.32\linewidth}
\begin{tikzpicture}
\begin{feynman}[small, every edge={thick}]
\vertex (a);
\vertex [dot, right=of a] (b) {};
\vertex [dot, right=of b] (c) {};
\vertex [dot, right=of c] (d) {};
\vertex [right=of d] (e);
\vertex [below=of a] (g);
\vertex [dot, below=0.5 of b] (f) {};
\vertex [dot, below=0.5 of f] (h) {};
\vertex [dot, below=of c] (i) {};
\vertex [below=of e] (j);
\diagram* {
(a) -- (b),
(b) -- (c),
(c) -- (d),
(d) -- (e),
(b) -- (f),
(f) -- (h),
(c) -- (i),
(g) -- (h),
(h) -- (i),
(i) -- (j),
(f) -- (d),
};

\end{feynman}
\end{tikzpicture}
\caption{$T2^B_2$}
\end{subfigure}
\begin{subfigure}[t]{0.32\linewidth}
\begin{tikzpicture}
\begin{feynman}[small, every edge={thick}]
\vertex (a);
\vertex [dot, right=of a] (b) {};
\vertex [dot, right=2.0 of b] (c) {};
\vertex [right=of c] (d);
\vertex[below=of a] (e);
\vertex[dot, below=of b] (f) {};
\vertex[dot, right=of f] (g) {};
\vertex[dot, right=of g] (h) {};
\vertex[right=of h] (i);
\diagram* {
(a) -- (b),
(b) -- (c),
(c) -- (d),
(e) -- (f),
(f) -- (g),
(g) -- (h),
(h) -- (i),
(b) -- (f),
(b) -- (g),
(c) -- (h),
};
\end{feynman}
\end{tikzpicture}
\caption{$T2^B_3$}
\end{subfigure}
\begin{subfigure}[t]{0.32\linewidth}
\begin{tikzpicture}
\begin{feynman}[small, every edge={thick}]
\vertex (a);
\vertex [dot, right=of a] (b) {};
\vertex [dot, right=of b] (c) {};
\vertex[below=of a] (f);
\vertex[dot, right=of f] (g) {};
\vertex[dot, right=of g] (h) {};
\vertex[dot, below right=0.5 and 1.0 of c] (d) {};
\vertex[right=2.0 of c] (e);
\vertex[right=2.0 of h] (i);
\diagram* {
(a) -- (b),
(b) -- (c),
(c) -- (d),
(d) -- (e),
(f) -- (g),
(g) -- (h),
(h) -- (d),
(d) -- (i),
(b) -- (h),
(g) -- (c),
};
\end{feynman}
\end{tikzpicture}
\caption{$T2^T_1$}
\end{subfigure}
\begin{subfigure}[t]{0.32\linewidth}
\begin{tikzpicture}
\begin{feynman}[small, every edge={thick}]
\vertex (a);
\vertex [dot, right=of a] (b) {};
\vertex [dot, right=of b] (c) {};
\vertex [dot, right=of c] (d) {};
\vertex [right=of d] (e);
\vertex [below=of a] (f);
\vertex [dot, right=of f] (g) {};
\vertex [dot, right=2.0 of g] (h) {};
\vertex [right=of h] (i);
\diagram* {
(a) -- (b),
(b) -- (c),
(c) -- (d),
(d) -- (e),
(f) -- (g),
(g) -- (h),
(h) -- (i),
(b) -- (g),
(c) -- (h),
(g) -- (d),
};
\end{feynman}
\end{tikzpicture}
\caption{$T2^T_2$}
\end{subfigure}
\begin{subfigure}[t]{0.32\linewidth}
\begin{tikzpicture}
\begin{feynman}[small, every edge={thick}]
\vertex (a);
\vertex [dot, right=of a] (b) {};
\vertex [dot, right=of b] (c) {};
\vertex[below=of a] (f);
\vertex[dot, right=of f] (g) {};
\vertex[dot, right=of g] (h) {};
\vertex[dot, below right=0.5 and 1.0 of c] (d) {};
\vertex[right=2.0 of c] (e);
\vertex[right=2.0 of h] (i);
\diagram* {
(a) -- (b),
(b) -- (c),
(c) -- (d),
(d) -- (e),
(f) -- (g),
(g) -- (h),
(h) -- (d),
(d) -- (i),
(b) -- (g),
(h) -- (c),
};
\end{feynman}
\end{tikzpicture}
\caption{$T2^T_3$}
\end{subfigure}
\caption[Genuine 2-loop topologies of the Weinberg operator.]{Topologies for genuine $2$-loop completions of the Weinberg operator $O_1 = L L H H$. Solid dots denote interaction vertices. Crossed lines without a dot at the intersection denote a non-planar configuration.}
\label{fig:topologies-2loop-Weinberg}
\end{figure}
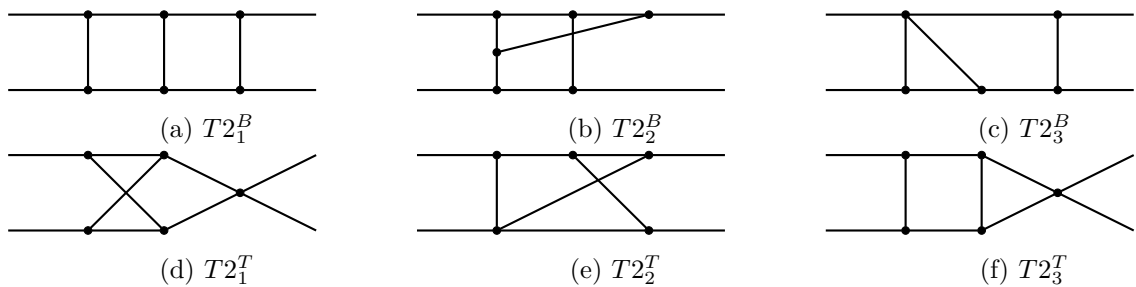

\noindent
{\em 2-loop topologies for $O_1 = LLHH$.}\ \ A systematic analysis of 2-loop openings of $O_1$ was performed in Ref.~\cite{Sierra:2014rxa}.  Figure~\ref{fig:topologies-2loop-Weinberg} displays the topologies identified in this study as able to contribute to genuine $2$-loop models. There are additional 2-loop diagrams -- that were termed ``class II'' -- that have the form of one of the 1-loop topologies of Fig.~\ref{fig:WeinbergOneLoop} with one the vertices expanded into a 1-loop subgraph.  They remark the class II topologies may be useful for justifying why a certain vertex has an unusually small magnitude.

\vspace{5mm}

\noindent
{\em C. Other considerations.}\ \ We now briefly survey other perspectives on classifying or discriminating between neutrino mass models. 

One suggested criterion is complexity~\cite{Law:2013dya}. While recognising that sometimes nature appears to favor minimal possibilities (in an Occam's razor approach), and at other times not (e.g.\ the old problem of why there are three families), it does make sense to rank neutrino mass models on some sensible measure of how complex they are. Reference~\cite{Law:2013dya} proposes a hierarchy based on (i) whether or not the model relies on the imposition of {\em ad hoc} symmetries, (ii) the number of exotic multiplets required, and (iii) the number of new parameters. Interestingly, they construct radiative models that are even simpler, on the basis of these criteria, than the $1$-loop Zee-Wolfenstein model~\cite{Zee:1980ai,Wolfenstein:1980sy}. However, like the Zee-Wolfenstein model, while these models generate nonzero neutrino masses, they fail phenomenologically. Thus, we must conclude that if nature utilises the radiative mechanism, it will be non-minimal.

Another consideration for Majorana mass models is the important phenomenological connection to $0\nu \beta \beta$ decay~\cite{Bonnet:2012kh,delAguila:2012nu,Helo:2016vsi}. Just as Majorana neutrino mass models may be systematically constructed through opening up $\Delta L = 2$ effective operators, models for $0\nu\beta\beta$ decay can be analysed by opening up the $\bar{u} \bar{u} d d \bar{e} \bar{e}$ family of operators.  The neutrino mass and $0\nu\beta\beta$ decay  considerations are of course connected, but the nature of the relationship is model-dependent. An interesting situation would emerge in a hypothetical future where $0\nu\beta\beta$ decay is observed, but the standard Majorana neutrino exchange contribution through $m_{\beta\beta}$ is contradicted by, for example, cosmological upper bounds on the absolute neutrino mass scale.  That would point to a non-minimal framework, which may be connected with radiative neutrino mass generation.

A further interesting aspect is the existence or otherwise of a deep theoretical reason for a given radiative model.  At first sight, each such model looks random.  However, some of them can be connected with, for example, grand unified theories (GUTs).  One simple point to make is that exotics, such as scalar leptoquarks, that often feature in radiative models can be components of higher-dimension multiplets of $SU(5)$ and $SO(10)$. Also, by contributing to renormalization group running, some of them can assist with gauge coupling constant unification~\cite{Hagedorn:2016dze}. If they are to be light enough to play these roles, while other exotics within the multiplets have, for example, GUT-scale masses, then we face a similar issue to the famous doublet-triplet splitting problem.  Nevertheless, this is a starting point for investigating the possible deeper origin of some of the required exotics.  Another interesting GUT-related matter was analysed in depth in Ref.~\cite{deGouvea:2014lva}.  A necessary condition for a $\Delta L = 2$ operator of a certain mass dimension to be consistent with a GUT origin is that it occurs as a term in an effective operator of the {\em same} mass dimension derived with grand unified gauge invariance imposed.  For example, the dimension-7 operator $O_{3a}$ from Eq.~\ref{eq:dim7-BLlist} does not appear as a component in any $SU(5)$ operator of the same dimension. On the other hand, other SM operators are embedded in the same GUT operator, with only one of them being able of giving the dominant contribution to neutrino masses. In addition to the question of the mere existence of SM-level operators in GUT decompositions, grand unification also imposes relations between SM-level operators, including some that violate baryon number and generate B$-$L violating nucleon decays and/or neutron-antineutron oscillations, leading to additional constraints. In the end, the authors of Ref.~\cite{deGouvea:2014lva} conclude that only a small subset of SM $\Delta L = 2$ operators are consistent with grand unification.

Another strategy for uncovering a deeper origin for a radiative model is by
asking if a given model has some close connection with the solution of
important particle physics problems beyond just the origin of neutrino mass.
One that has been explored at length in the literature is a possible connection
to dark matter.  Examples of such models will be given in more detail in later
sections. Here, we simply mention some systematic analyses of what new
symmetries can be imposed in radiative models to stabilize dark
matter~\cite{Farzan:2012ev,Restrepo:2013aga}. Reference~\cite{Farzan:2012ev}
classified the symmetries $G_\nu$ that can be imposed in order to ensure that
the first nonzero contribution to $O_1$ occurs at a given loop order, by
forbidding all potential lower-order contributions. All standard model
particles are singlets under $G_\nu$, implying that the lightest of the exotics
that do transform under this symmetry must be stable if the symmetry remains
exact, establishing a connection with dark matter.
Reference~\cite{Restrepo:2013aga} performed a systematic analysis of radiative
models in a certain class in order to find those that have viable dark matter
candidates.  The considered models are those that generate mass at $1$-loop
level using exotics that are at most triplets under weak isospin, and where the
stabilising symmetry is $Z_2$.  They found $35$ viable models. A similar
analysis, but requiring $2$-loop neutrino mass generation, can be found in
Ref.~\cite{Simoes:2017kqb}. 

Besides dark matter, radiative neutrino mass models may also be connected to other physics beyond the SM such as the anomalous magnetic moment of the muon, the strong CP problem, the baryon asymmetry of the Universe or B-physics anomalies, among others. Phenomenology related to radiative neutrino mass models is briefly discussed in Sec.~\ref{sec:pheno} in general and an example of a possible connection to the recent B-physics anomalies is presented in Sec.~\ref{sec:leptoquarks}.


\section{Radiative generation of neutrino masses} 
\label{sec:models} We adopt
the classification of radiative neutrino mass models according to their Feynman
diagram topology,\footnote{Note that diagrams with scalar or vector bosons are equivalent from a topological point of view.} but refer to the other classification schemes where
appropriate. In particular, we indicate the lowest-dimensional non-trivial
$\Delta L=2$ operator
which is generated beyond the Weinberg operator $LLHH$.
These $\Delta L=2$ operators capture light particles which are in the loop to
generate neutrino mass and are very useful to identify relevant low-energy phenomenology. 

In the subsections \ref{sec:1loop} - \ref{sec:3loop} we classify Majorana
neutrino mass models proposed in the literature according to their topology and
specifically discuss models with SM gauge bosons in the loop in
Sec.~\ref{sec:derivatives}. In Sec.~\ref{sec:Dirac} we review Dirac neutrino mass models and briefly comment on models based on the gauge group $SU(3)_{\rm c}\times
SU(3)_{\rm L}\times U(1)_{\rm X}$ in Sec.~\ref{sec:331Models}. 
\subsection{1-loop Majorana neutrino mass models}
\label{sec:1loop}
\begin{figure} \centering \begin{subfigure}[t]{0.3\linewidth} \centering
	\begin{tikzpicture}[node distance=1cm and 1cm]
		\coordinate[label=left:$H$] (h1); \coordinate[vertex,
			right=0.8cm of h1] (v1); \draw[fermionnoarrow] (v1)
				arc[start angle=-180, delta angle=180,
				radius=0.7cm] node[pos=0.5,vertex] (v05) {}
				node[vertex] (v2) {}; \coordinate[right=0.8cm
					of v2, label=right:$H$] (h2);
				\coordinate[vertex, below=0.7cm of v05] (v3);
				\coordinate[left=of v3,
					xshift=-0.5cm,label=left:$L$] (nu1);
				\coordinate[right=of v3, xshift=0.5cm,
					label=right:$L$] (nu2);

    \draw[scalar] (h1)--(v1); \draw[scalar] (h2)--(v2); \draw[scalarnoarrow]
    (v05)--(v3); \draw[fermion] (nu1)--(v3); \draw[fermion] (nu2)--(v3);
    \draw[fermionnoarrow] (v1)--(v2); \end{tikzpicture} \caption{T4-1-i}
    \label{fig:T41i} \end{subfigure}
    \hspace{0.03\linewidth}%
\begin{subfigure}[t]{0.3\linewidth} \centering \begin{tikzpicture}[node
	distance=1cm and 1cm] \coordinate[label=left:$H$] (h1);
	\coordinate[vertex, right=0.8cm of h1] (v1); \draw[scalarnoarrow] (v1)
		arc[start angle=-180, delta angle=180, radius=0.7cm]
		node[pos=0.5,vertex] (v05) {} node[vertex] (v2) {};
	\coordinate[right=0.8cm of v2, label=right:$H$] (h2);
	\coordinate[vertex, below=0.7cm of v05] (v3); \coordinate[left=of v3,
		xshift=-0.5cm,label=left:$L$] (nu1); \coordinate[right=of v3,
			xshift=0.5cm, label=right:$L$] (nu2);

    \draw[scalar] (h1)--(v1); \draw[scalar] (h2)--(v2); \draw[scalarnoarrow]
    (v05)--(v3); \draw[fermion] (nu1)--(v3); \draw[fermion] (nu2)--(v3);
    \draw[scalarnoarrow] (v1)--(v2); \end{tikzpicture} \caption{T4-1-ii}
    \label{fig:T41ii} \end{subfigure}
    \hspace{0.03\linewidth}%
\begin{subfigure}[t]{0.3\linewidth} \centering \begin{tikzpicture}[node
	distance=1cm and 1cm] \coordinate[label=left:$L$] (nu1);
	\coordinate[vertex, right=0.8cm of nu1] (v1); \draw[fermionnoarrow]
		(v1) arc[start angle=180, delta angle=-180, radius=0.7cm]
		node[pos=0.5,vertex] (v05) {} node[vertex] (v2) {};
	\coordinate[right=0.8cm of v2, label=right:$L$] (nu2);
	\coordinate[vertex, above=0.7cm of v05] (v3); \coordinate[left=of v3,
		xshift=-0.5cm,label=left:$H$] (h1); \coordinate[right=of v3,
			xshift=0.5cm, label=right:$H$] (h2);

    \draw[scalar] (h1)--(v3); \draw[scalar] (h2)--(v3); \draw[scalarnoarrow]
    (v05)--(v3); \draw[fermion] (nu1)--(v1); \draw[fermion] (nu2)--(v2);
    \draw[scalarnoarrow] (v1)--(v2); \end{tikzpicture} \caption{T4-2-ii}
    \label{fig:T42ii} \end{subfigure} \caption{Non-genuine topologies of the
    Weinberg operator.} \label{fig:WeinbergOneLoopNonGenuine} \end{figure}
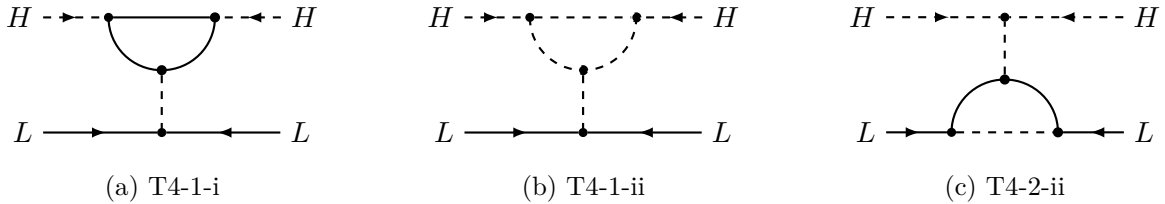

This section is divided into several parts: (i) 1-loop UV completions of the
Weinberg operator, (ii) 1-loop seesaws, (iii) UV completions with additional VEV insertions,
(iv) 1-loop UV completions of the higher dimensional operators and (v) other 1-loop models. Notice that the first 
part includes models with multi-Higgs doublets, while the second part discusses external
fields which transform under an extended symmetry. Besides the genuine
topologies discussed in Sec.~\ref{sec:schemes}, there are models based on the
non-genuine 1-loop topologies in Fig.~\ref{fig:WeinbergOneLoopNonGenuine}. 

\subsubsection{Weinberg operator $LLHH$} 
We follow the general classification
of UV completions of the Weinberg operator at 1-loop~\cite{Bonnet:2012kz}
discussed in Sec.~\ref{sec:RadiativeNeutrinoMassClassification}. The six
genuine topologies are shown in Fig.~\ref{fig:WeinbergOneLoop}. Analytic expressions for all 1-loop topologies are listed in the appendix of Ref.~\cite{Bonnet:2012kz}.

Here we list the theories falling into respective categories.  As the
topologies stay the same while incorporating multiple Higgs doublets, theories
with more than one Higgs doublet will also be listed here.  
Models in which the generation of neutrino mass relies on additional VEVs
connected to the neutrino mass loop diagram are discussed in
Sec.~\ref{subsubsec:1LoopAdditionalVEVs}. We first discuss the models based on topology T3, the only one with a quartic scalar interaction, before moving on to the other topologies.

\textbf{T3:} Topology T3 is one of the most well-studied. It was
first proposed in Ref.~\cite{Ma:1998dn} and its first realization, the
scotogenic model with a second electroweak scalar doublet and sterile fermion singlets
(at least two) both odd under a $Z_2$ symmetry, was later proposed in
Ref.~\cite{Ma:2006km}. See Sec.~\ref{subsec:scotogenic} for a detailed
discussion of the model. Its appeal lies in the simultaneous explanation of
dark matter, which is stabilized by a $Z_2$ symmetry. A crucial ingredient is the quartic scalar interaction $(H^\dagger \eta)^2$ (see Eq.~\ref{eq:scotogenicScalarPotential}) of the SM Higgs boson $H$ with the electroweak scalar doublet $\eta$ in the loop. This scalar interaction splits the masses of the neutral scalar and pseudoscalar components of $\eta$. Neutrino masses vanish in the limit of degenerate neutral $\eta$ scalar masses. Several variants of the
scotogenic model have been proposed in the literature: with triplet instead of
singlet fermions~\cite{Ma:2008cu,Chao:2012sz,vonderPahlen:2016cbw}, an extension with an
additional singlet scalar~\cite{Farzan:2009ji}, one fermionic singlet and two
additional electroweak scalar doublets~\cite{Hehn:2012kz}, scalar
triplets~\cite{Lu:2016dbc}, colored scalars and
fermions~\cite{FileviezPerez:2009ud,Liao:2009fm}, a vector-like fermionic
lepton doublet, a triplet scalar, and a
neutral~\cite{Farzan:2010mr,Okada:2015vwh} or charged~\cite{Brdar:2013iea}
singlet scalar, vector-like doublet and singlet fermions and doublet scalar,
which contains a doubly charged scalar~\cite{Aoki:2011yk}, higher SU(2)
representations~\cite{Cai:2011qr,Chen:2011bc,Law:2013saa,Chowdhury:2015sla}, an
extended discrete symmetry with $Z_2\times Z_2$~\cite{Chen:2009gd,Patra:2014sua}
or $Z_2\times \mathrm{CP}$~\cite{Ferreira:2016sbb}, a discrete flavor symmetry
based on $S_3$~\cite{Fortes:2017ndr},
$A_4$~\cite{Ma:2008ym,Bhattacharya:2013mpa,Hernandez:2013dta,Campos:2014lla},
$\Delta(27)$~\cite{Ma:2013xqa,Ma:2014eka}, which is either softly-broken or via
electroweak doublets, and its embedding in (grand) unified
theories~\cite{Gu:2008zf,Adulpravitchai:2009re,Parida:2011wh, Ma:2013nga,Campos:2014lla}. 
Finally, the authors of Ref.~\cite{Megrelidze:2016fcs} proposed the generation of neutrino mass via lepton-number-violating soft supersymmetry-breaking terms. In particular the generation of the dimension-4 term $(\tilde L H_u)^2$ with left-handed sleptons $\tilde L$ leads to models based on the topology T-3 with supersymmetric particles in the loop. Another variant involves a global continuous dark symmetry~\cite{Ma:2013yga,generalized}, termed the generalized scotogenic model.

\textbf{T1-i:} Reference~\cite{Ma:2006uv} discusses a supersymmetrized version
of the scotogenic model, which is based on topology T3 and we discuss in detail
in Sec.~\ref{subsec:scotogenic}. The topology necessarily differs from T3 because the
term $(H^\dagger \eta)^2$ is
not introduced by D-terms. An embedding of this model in SU(5) is given in
Ref.~\cite{Ma:2007kt}. In a non-supersymmetric context, the same topology is
discussed in Ref.~\cite{Farzan:2009ji}, which introduces one real singlet scalar, in
the context of a (dark) left-right symmetric
model~\cite{Khalil:2009nb,Ma:2009tc}, and in
Refs.~\cite{Budhi:2014gxa,Kashiwase:2015yla, Budhi:2015sha}, which introduce
multiple singlet scalars to connect the two external Higgs fields. The
term $(H^\dagger \eta)^2$, which is essential to generate topology T3, is neglected in
Refs.~\cite{Budhi:2014gxa,Kashiwase:2015yla,Budhi:2015sha} and thus neutrino
mass is generated via topology T1-i. One of the singlet scalars in the neutrino
mass model can be the inflaton via a non-minimal coupling with the
Ricci-scalar. The term $(H^\dagger \eta)^2$ can be explicitly forbidden by imposing a
$U(1)$ symmetry, which is softly broken by the CP-violating mass term $\chi^2$ of a complex scalar field $\chi$~\cite{Arhrib:2015dez}. Finally the
authors of Ref.~\cite{Lu:2016ucn} proposed a model with electroweak singlet and triplet
scalars as well as fermions and study the dark matter phenomenology and
leptogenesis.

\textbf{T1-ii:} Among the models based on the topology T1-ii, there are four
possible operators which models are based on. Besides models with only heavy
new particles, there are models with
SM charged leptons, down-type quarks, or up-type quarks in the loop, which are
based on the operators $O_2$ and $O_3$, respectively.  We first discuss the
models based on operator $O_2$.  The first radiative Majorana neutrino mass
model, the Zee model~\cite{Zee:1980ai}, is based on this operator.
See Sec.~\ref{sec:ZeeModel} for a detailed discussion of its phenomenology.
Several variants of the Zee model exist in the literature. The minimal
Zee-Wolfenstein model~\cite{Wolfenstein:1980sy} with a $Z_2$ symmetry to forbid
tree-level FCNCs has been excluded by neutrino oscillation
data~\cite{Koide:2001xy,He:2003ih}, while the general version with both Higgs
doublets coupling to the leptons is
allowed~\cite{He:2011hs,Herrero-Garcia:2017xdu}. Imposing a $Z_4$
symmetry~\cite{Babu:2013pma} allows to explain neutrino data and forbid
tree-level FCNCs in the quark sector. 
Previously in Ref.~\cite{Aranda:2011rt} a flavor-dependent $Z_4$ symmetry was used 
to obtain specific flavor structures in the quark and lepton sector.
A supersymmetric version of the Zee
model has been proposed in
Refs.~\cite{Leontaris:1987me,Haba:1999iw,Cheung:1999az,Kanemura:2015maa}. Its
embedding into a grand unified theory has been discussed in
Refs.~\cite{Zee:1980ai,Tamvakis:1985if,Perez:2016qbo}, and in models with extra
dimensions in Refs.~\cite{Chang:2003gv,Chang:2010ic}.

		Other flavor symmetries beyond $Z_4$ have been studied in
		Refs.~\cite{Babu:1989wn,Babu:1990wv,Koide:2000jm,Kitabayashi:2001ex,Adhikary:2006wi,Fukuyama:2010ff,Aranda:2010im,Aranda:2011dx},
		and Ref.~\cite{Okamoto:1998st} studied the Zee model when the
		third generation transforms under a separate $SU(2)\times U(1)$
		group. References~\cite{Babu:1989wn,Babu:1990wv} studied large
		transition magnetic moments of the electron neutrino, which was
		an early, now excluded, explanation for the solar neutrino
		anomaly.  General group theoretic considerations about the
		possible particle content in the loop are discussed in
		Ref.~\cite{Ma:1998dn}. 
	
		Models with multiple leptoquarks, which mix among each other,
		also generate neutrino mass via topology T1-ii. We discuss this
		possibility in more detail in Sec.~\ref{sec:leptoquarks-1}.
		They induce the operator $O_3$ if the leptoquark couples
		down-type quarks to neutrinos. Well-studied examples of
		leptoquarks are	down-type squarks in R-parity violating SUSY
		models, which generate neutrino masses, as was first
		demonstrated in Ref.~\cite{Hall:1983id}.  Specific examples
		with multiple leptoquarks which mix with each other were
		discussed in
		Refs.~\cite{Nieves:1981tv,Chua:1999si,Mahanta:1999xd,AristizabalSierra:2007nf,Helo:2015fba,Pas:2015hca,Cheung:2016fjo,Dorsner:2017wwn}.
		There are several supersymmetric
		models~\cite{Chua:1999si,Kong:2000dx,Davidson:2000uc,Koide:2003mh,Koide:2003ad,Koide:2004xf}
		which generate neutrino mass via different down-type quarks or
		charged leptons in the loop and consequently induce the
		operators $O_{3}$ and $O_2$, respectively.  Finally, there are
		models with only heavy particles in the loop such as the inert
		Zee model~\cite{Longas:2015sxk} or supersymmetric models with
		R-parity conservation~\cite{Ma:1999ni,Suematsu:2009ia}.

\textbf{T1-iii:} This topology was first proposed in Ref.~\cite{Ma:1998dn} and
it naturally appears in the supersymmetrized version of the scotogenic
model~\cite{Ma:2006uv,Ma:2007kt,Ma:2008ba,Ma:2008zh,Fukuoka:2009cu,Suematsu:2010nd,Fukuoka:2010kx,Bhattacharya:2013nya,Kanemura:2013uva,Figueiredo:2014gpa,Kanemura:2014cka,Hundi:2015kea}
together with topology T1-i. The topology can be used to implement the radiative inverse
seesaw~\cite{Fraser:2014yha,Restrepo:2015ura,Chiang:2017tai}, which resembles
the structure of the inverse seesaw~\cite{Mohapatra:1986aw,Mohapatra:1986bd}.
This model has been extended by a softly-broken non-Abelian flavor symmetry
group~\cite{Ma:2015pma,Natale:2016xob,Ma:2016nkf} in order to explain the
flavor structure in the lepton sector.  The SUSY model in Ref.~\cite{Ma:2011zm}
generates neutrino mass via sneutrinos and neutralinos in the loop. This
mechanism was first pointed out in Ref.~\cite{Hirsch:1997vz}. In the realization of
Ref.~\cite{Ma:2011zm}, the masses of the real and imaginary parts of the
sneutrinos are split by the VEV of a scalar triplet, which only couples to the
sneutrinos via a soft-breaking term and thus does not induce the ordinary
type-II seesaw.  Similarly it has been used in a model with vector-like
down-type quarks~\cite{Cai:2014kra,Popov:2016fzr}, which requires mixing of the SM quarks
with the new vector-like quarks. This model leads to the operator $O_3$.

\subsubsection{1-loop seesaws and soft-breaking terms}

	For completeness we also include the two possible 1-loop seesaw topologies T4-2-i and T4-3-i which have been identified in Ref.~\cite{Bonnet:2012kz}. Topology T4-2-i always involves a electroweak scalar
      triplet like in the type-II seesaw mechanism and topology T4-3-i contains an electroweak singlet or triplet fermion like in the type-I or type-III seesaw mechanism, respectively. Based on our knowledge,
      there are currently no models based on topologies T4-2-i and T4-3-i in
      the literature.
      
      Finally, although the topology T4-2-ii shown in
      Fig.~\ref{fig:T42ii} has been discarded in Ref.~\cite{Bonnet:2012kz},
      because it is generally accompanied by the tree-level type-II seesaw
      mechanism, there are three models based on this
      topology~\cite{Ma:2015xla,Fraser:2015mhb,Guo:2016dzl}. They break lepton
      number softly by a dimension-2 term and thus there is no tree-level
      contribution by forbidding the ``hard-breaking'' dimension-4 terms 
      which are required for the type-II seesaw mechanism. Similar constructions may be possible for other topologies and lead to new interesting models.

\subsubsection{Additional VEV insertions} \label{subsubsec:1LoopAdditionalVEVs}
The above discussed classification technically does not cover models with
additional scalar fields, which contribute to neutrino mass via their vacuum
expectation value in contrast to being a propagating degree of freedom in the
loop. Inspired by the above classification, we similarly classify these new
models according to the topologies in Fig.~\ref{fig:WeinbergOneLoop} by
disregarding the additional VEV insertions.

	\textbf{T1-i:} There are several radiative neutrino mass models which
	are based on a $U(1)$ symmetry, which is commonly broken to a remnant
	$Z_2$ symmetry: there are models based on a global Peccei-Quinn
	$U(1)_{\rm PQ}$ symmetry~\cite{Dasgupta:2013cwa,Ma:2014yka}, which
	connects neutrino mass to the strong CP problem, a local $U(1)_{{\rm B}-{\rm L}}$
	symmetry~\cite{Ho:2016aye,Ma:2016nnn,Nomura:2017vzp} and local dark
	$U(1)$ symmetry~\cite{Chang:2011kv, Lindner:2013awa,Kownacki:2016hpm}.
	The authors of Ref.~\cite{Ho:2016aye} systematically study radiative
	neutrino mass generation at 1-loop (but also 2-, and 3-loop) level based on a
	gauged $U(1)_{{\rm B}-{\rm L}}$ symmetry, which is broken to a $Z_N$ symmetry.
	The models in Refs.~\cite{Chang:2011kv,Dasgupta:2013cwa,Lindner:2013awa,Adhikari:2015woo,Kownacki:2016hpm} also have a contribution to neutrino mass at 2-loop order based on a	Cheng-Li-Babu-Zee (CLBZ) topology.
	
	\textbf{T1-ii:} All of the models with additional VEV insertions rely
	on the breaking of a symmetry: left-right
	symmetry~\cite{Babu:1988qv,FileviezPerez:2017zwm,FileviezPerez:2016erl},
	a more general $SU(2)_1\times SU(2)_2$ symmetry~\cite{Hue:2016nya}, a flavor
	symmetry~\cite{Ma:1997nq,Ma:1998dp,Dicus:2001ph}, $U(1)_{{\rm B}-{\rm L}}$~\cite{Sahu:2008aw}, and dilation symmetry~\cite{Foot:2007ay}. All
	these models lead to the operator $O_2$. Reference~\cite{Foot:2007ay}
	discusses in particular the following two 1-loop models: the scale-invariant Zee model and a
	scale-invariant model with leptoquarks which induces $O_3$. Finally, there is the inert Zee model with a flavor
	symmetry~\cite{Chen:2009ata,Arbelaez:2016mhg}.

\textbf{T1-iii:} The model in Ref.~\cite{Nomura:2016jnl} relies on the
VEVs of a scalar triplet and a septuplet which are subject to strong constraints from electroweak precision tests in particular from the $T$ (or $\rho$) parameter. The minimization of the potential is
not discussed, but the VEVs can in principle be introduced via the linear term
in the scalar potential, which leads to the operator
$O_1^{\prime\prime\prime\prime}$ at 2-loop level, because the linear term for
the septuplet is only induced at the 1-loop level. The topology can also be generated by new heavy lepton-like doublets and sterile fermions, which are charged under a new gauged dark $U(1)$ in addition to a $Z_2$ symmetry~\cite{Nomura:2017tzj}. 

\textbf{T3:} There are several variants of the scotogenic model with additional
VEV insertions. Most of them are based on an extended symmetry sector, such as
a discrete $Z_3$ instead of a $Z_2$ symmetry \cite{Haba:2011nb,Okada:2014zea},
dilation
symmetry~\cite{Foot:2007ay,Okada:2014nea,Ahriche:2016ixu,Ahriche:2016cio}, a
gauged $U(1)_{{\rm B}-{\rm L}}$~\cite{Kanemura:2011vm,Kajiyama:2012xg,Okada:2012np,Dasgupta:2016odo,Seto:2016pks},
global $U(1)_{{\rm B}-{\rm L}}$~\cite{Machado:2017ksd}, a general gauged
$U(1)$~\cite{Kashiwase:2015joo,Yu:2016lof,Ko:2017quv}, continuous $U(1)$ flavor
symmetry~\cite{Baek:2015mna,Baek:2015fea}, a discrete flavor symmetry based on
$D_6$~\cite{Kajiyama:2006ww},
$A_4$~\cite{Ahn:2010cc,Ahn:2010ui,Ahn:2012cga,Ma:2012ez,Holthausen:2012wz} or
$S_4$~\cite{Mukherjee:2017pzq}, and different LR symmetric models without a
bidoublet~\cite{Borah:2016lrl}. Apart from additional symmetries, the mixing of
the fermionic singlet with a fermionic triplet in the loop requires the VEV of
an electroweak triplet with vanishing
hypercharge~\cite{Hirsch:2013ola,Rocha-Moran:2016enp,Merle:2016scw}. Finally,
the two models discussed in Refs.~\cite{Okada:2013iba,Okada:2014nsa} rely on a
similar topology as the scotogenic model, but with triplet VEVs instead of electroweak
doublet VEVs.

\textbf{T4-2-i:} Based on our knowledge, there are currently no models based on
topology T4-2-i in the literature.

 \textbf{T4-3-i:} Ref.~\cite{Wang:2015saa} proposed a model
 which reduces to topology T4-3-i after breaking of the $U(1)_{{\rm B}-{\rm L}}$
 symmetry. As the Majorana mass term for the fermionic pure singlet is not introduced, there is no inverse seesaw contribution to
 neutrino mass after the breaking of the $U(1)_{{\rm B}-{\rm L}}$ symmetry and neutrino masses are generated at 1-loop level.
 
 \textbf{T4-1-i/ii:} These types of models contain a triplet scalar which
 couples to the lepton doublet as per the tree-level type-II seesaw. However,
 the neutral component of the triplet scalar gets an induced VEV at 1-loop and
 thus generates neutrino masses effectively at 1-loop. The model in
 Ref.~\cite{Nomura:2017emk} is based on topology T4-1-i shown in
 Fig.~\ref{fig:T41i}, which is finite  due to additional VEV insertions on the
 fermion line. The model in Ref.~\cite{Kanemura:2012rj} is based on topology
 T4-1-ii shown in Fig.~\ref{fig:T41ii}. The tree-level contribution is
 forbidden by a discrete symmetry and renormalizability of the theory. However
 at loop-level neutrino mass is generated by a dimension-7 operator $LLHHs_1^2$
 with two additional SM singlet fields $s_1$.  Note in both cases an extra
 symmetry such as $U(1)_{{\rm B}-{\rm L}}$ or a discrete symmetry and lepton number is
 needed to forbid the contribution from the tree-level type-II seesaw. Topology
 T4-1-ii is also induced in the SUSY model in
 Refs.~\cite{Franceschini:2013aha,Figueiredo:2014gpa} after the breaking of SUSY
 and the discrete $Z_4$ symmetry.

\subsubsection{Higher-dimensional Weinberg-like operators}

Apart from UV completions of the Weinberg operator, there are a few models
which induce one of the higher dimensional operators with additional Higgs
doublets at 1-loop level.

\begin{figure}[bt] \centering 
	\begin{minipage}{0.4\linewidth}\centering
		\begin{tikzpicture}[scale=0.7] \begin{feynman}[small,
	every edge={thick}] \draw [thick] (0,0) circle [radius=1]; \vertex
		[dot] at (0:1) (l1) {}; \vertex [dot] at (90:1) (l2) {};
	\vertex [dot] at (180:1) (l3) {}; \vertex [dot] at (270:1) (l4) {};
\vertex at (0:2) (e1); \vertex at (270:2) (e4); \vertex [above left=0.5 and 0.5
of l3] (f1); \vertex [below left=0.5 and 0.5 of l3] (f2); \vertex [above
left=0.5 and 0.5 of l2] (f3); \vertex [above right=0.5 and 0.5 of l2] (f4);
\diagram* { (l1) -- (e1), (l3) -- (f1),
(l3) -- (f2), (l4) -- (e4), (l2) -- (f3), (l2) -- (f4), }; \end{feynman}
\end{tikzpicture}	\captionof{figure}{Non-genuine 1-loop topology T31 for operator
$O_1^\prime$.} \label{fig:T31} \end{minipage}
\hfill
\begin{minipage}{0.5\linewidth}\centering
			\begin{tikzpicture}[node distance=1cm and 1cm]
		\coordinate[label=left:$L$] (nu1);
		\coordinate[vertex,right=of nu1] (v1);
		\coordinate[vertex,right=of v1] (v1a);
		\coordinate[cross,right=of v1a] (m);
		\coordinate[vertex,right=of m] (v2a);
		\coordinate[vertex,right=of v2a] (v2);
		\coordinate[label=right:$L$,right=of v2] (nu2);
		\coordinate[above=of v1,label=above:$H$] (h1); 
		\coordinate[above=of v2,label=above:$H$] (h2); 
		\coordinate[cross,above=of m] (mus); 

    \draw[scalar] (h1)--(v1); 
    \draw[scalar] (h2)--(v2); 
    \draw[fermion] (nu1) -- (v1);
    \draw[fermion] (nu2) -- (v2);
    \draw[fermionnoarrow] (v1) -- (v1a) node[midway,cross] {};
    \draw[fermionnoarrow] (v2) -- (v2a) node[midway,cross] {};
    \draw[fermionnoarrow] (v1a) -- (m);
    \draw[fermionnoarrow] (v2a) -- (m);
    \draw[scalarnoarrow] (v1a) to[out=90,in=180] (mus);
    \draw[scalarnoarrow] (v2a) to[out=90,in=0] (mus);
    \end{tikzpicture} 
    \captionof{figure}{Radiative inverse seesaw.}
    \label{fig:RadInvSeesawAhriche}
    \end{minipage}
	\end{figure}

\textbf{Dimension-7
($O_1^\prime$):} The first model which induced the dimension-7 operator
$O_1^\prime$ at 1-loop level in a two Higgs doublet model was proposed in
Ref.~\cite{Kanemura:2010bq}. It was realized using at most adjoint
representations and an additional softly-broken $Z_5$ symmetry and an exact
$Z_2$ symmetry and thus allows to use the topologies T12
(Fig.~\ref{subfig:T12}) and T31 (Fig.~\ref{fig:T31}), which would otherwise be
accompanied by the dimension-5 operator $O_1$.  If the Zee model is extended by
a triplet Majoron~\cite{Chang:1988aa,Santamaria:1988fh} the operator
$O_2^\prime= LLLe^c H (H^\dagger H)$ is induced at tree-level. After closing
the loop of charged leptons via topology T3 (Fig.~\ref{subfig:T3}), the
dimension-7 operator $O_1^\prime$ is obtained.
Reference~\cite{Cepedello:2017eqf} systematically studies the possible 1-loop
topologies of $O_1^\prime$ and explicitly shows several models: the only
genuine model without representations beyond the adjoint of SU(2) is based on
topology T11, while the other models use quadruplets or even larger
representations to realize the other genuine topologies.

\textbf{Dimension-9 ($O_1^{\prime\prime}$):} In Refs.~\cite{Law:2012mj,Baldes:2013eva} neutrino masses are generated via a radiative inverse seesaw. The mass of the
additional SM singlets is induced at tree-level and then first transmitted to
the neutral components of new electroweak doublets via a 1-loop diagram, before
it induces neutrino mass via the seesaw. It leads to the dimension-9 operator
$O_1^{\prime\prime}$ via the four VEV insertions on the scalar line of the
1-loop diagram. There is also a 2-loop contribution, which may dominate
neutrino mass depending on the masses of the new particles. 

\textbf{Dimension-11 ($O_1^{\prime\prime\prime}$):} The model proposed in
Ref.~\cite{Aranda:2015xoa} relies on the VEV of a 7-plet $\chi$, which is
induced via a non-renormalizable coupling, linear in $\chi$, to six electroweak
Higgs doublets. 

As can be seen from the discussion above, in order to generate Weinberg-like
effective operators at dimension larger than five, typically extra symmetries
(in some cases large discrete symmetries), new large representations, a large
number of fields or a combination of all the previous need to be invoked. This
makes the model-building of such scenarios much more involved than for the case
of the Weinberg operator.

\subsubsection{Other 1-loop models}
	Apart from the models in the general
	classification~\cite{Bonnet:2012kz}, it is possible to generate
	neutrino mass via a radiative inverse seesaw mechanism shown in
	Fig.~\ref{fig:RadInvSeesawAhriche} at 1-loop order, which has been
	proposed in Ref.~\cite{Ahriche:2016acx}. Tree-level contributions are
	forbidden by a softly-broken $Z_4$ symmetry. The soft-breaking is
	indicated by the cross on the scalar line. Note the cross on the fermion
	line in the loop denotes a Majorana mass term, while the other two
	denote Dirac mass terms. 
	
	Finally we would like to comment on one further possibility to generate
	neutrino mass at 1-loop order. If the neutrino masses vanish at tree-level in type-I seesaw model, then 1-loop electroweak corrections give the
	leading contribution~\cite{Pilaftsis:1991ug}:\footnote{The finite
		1-loop
	corrections to the active neutrino mass matrix in the seesaw model were
first discussed in Ref.~\cite{Grimus:1989pu} with an arbitrary number of
right-handed neutrinos, left-handed lepton doublets, and Higgs doublets. The
finite 1-loop corrections are particularly important in case of delicate
cancellations in the tree-level neutrino mass terms, which have been studied in
Ref.~\cite{AristizabalSierra:2011mn} using the result of
Ref.~\cite{Grimus:2002nk}.}
	non-zero neutrino masses are induced
	by finite 1-loop diagrams with either a $Z$-boson or a Higgs boson. The
	UV divergent part of the 1-loop corrections to the Weinberg operator
	cancel due to the absence of a tree-level contribution.
	This has been explicitly shown in Ref.~\cite{Pilaftsis:1991ug} with a
	calculation in the mass basis. In terms of the classification of 1-loop
	topologies, these diagrams
	correspond to the topologies T3 and T1-iii for the Higgs and $Z$-boson in the loop, respectively. The vanishing
	of the tree-level contribution can be achieved using a specific texture
	in the seesaw model with SM singlet fermions
	$S$~\cite{Dev:2012sg} in addition to the right-handed neutrinos $N$
	\begin{equation}
		\begin{pmatrix}
			0 & m_D & 0 \\
			. & \mu_R & M_N^T\\
			. & . & \mu_S
		\end{pmatrix}
	\end{equation}
	in the basis $(\nu, N, S)$. In the limit $\mu_S\to 0$ the tree-level
	contribution to the active neutrinos exactly vanishes and neutrino masses are
	generated at 1-loop order. This construction has been denoted
	\emph{minimal radiative inverse seesaw}~\cite{Dev:2012sg}. 
	
	This texture can be obtained by imposing a $U(1)$ symmetry under which $S$ is charged. After it
	is spontaneously broken by the VEV of a SM singlet scalar $\eta$, the
	Yukawa interaction $SN\eta$ generates the term $M_N$ without generating
	a Majorana mass term $\mu_S$ for the fermionic singlets $S$ or a coupling of $S$ to the SM lepton
	doublets $L$ at the renormalizable level.

\subsection{2-loop Majorana neutrino mass models}
\label{sec:2loop}

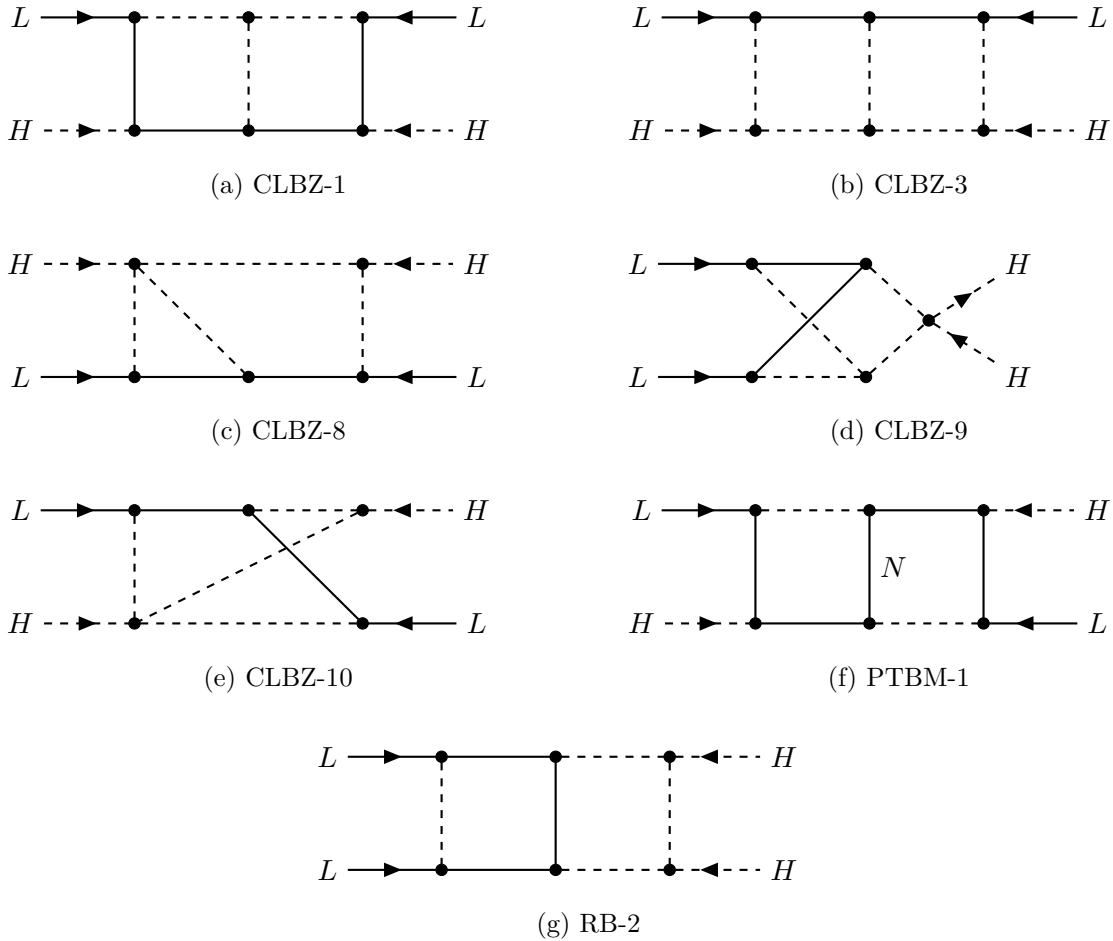
\begin{figure}[tb]\centering

\begin{subfigure}[t]{0.45\linewidth} \begin{tikzpicture} \begin{feynman}[every
	edge={thick}] \vertex (a) {$L$}; \vertex [dot, right=of a] (b) {};
	\vertex [dot, right=of b] (c) {}; \vertex [dot, right= of c] (d) {};
	\vertex [right=of d] (e) {$L$}; \vertex [below=of a] (f) {$H$}; \vertex
	[dot, below=of b] (g) {}; \vertex [dot, below=of c] (k) {}; \vertex
	[dot, below=of d] (i) {}; \vertex [below=of e] (j) {$H$}; \diagram* {
		(a) -- [fermion] (b), (b) --  [scalar](c), (c) -- [scalar] (d),
(d) --  [anti fermion] (e), (b) --  (g), (c) -- [scalar] (k), (d) -- (i), (f)
-- [charged scalar] (g), (g) -- (k), (k) -- (i), (i) -- [anti charged scalar]
(j), }; \end{feynman} \end{tikzpicture} \caption{CLBZ-1} \label{fig:CLBZ-1}
\end{subfigure}
    \hspace{0.04\linewidth}%
\begin{subfigure}[t]{0.45\linewidth} \begin{tikzpicture} \begin{feynman}[every
	edge={thick}] \vertex (a) {$L$}; \vertex [dot, right=of a] (b) {};
	\vertex [dot, right=of b] (c) {}; \vertex [dot, right= of c] (d) {};
	\vertex [right=of d] (e) {$L$}; \vertex [below=of a] (f) {$H$}; \vertex
	[dot, below=of b] (g) {}; \vertex [dot, below=of c] (k) {}; \vertex
	[dot, below=of d] (i) {}; \vertex [below=of e] (j) {$H$}; \diagram* {
		(a) -- [fermion] (b), (b) -- (c), (c) -- (d), (d) --  [anti
		fermion] (e), (b) --  [scalar] (g), (c) -- [scalar]  (k), (d)
		-- [scalar] (i), (f) -- [charged scalar] (g), (g) -- [scalar]
		(k), (k) -- [scalar] (i), (i) -- [anti charged scalar](j), };
	\end{feynman} \end{tikzpicture} \caption{CLBZ-3} \label{fig:CLBZ-3}
\end{subfigure}

\vspace{3ex}
\begin{subfigure}[t]{0.45\linewidth} \begin{tikzpicture} \begin{feynman}[every
	edge={thick}] \vertex (a) {$H$}; \vertex [dot, right=of a] (b) {};
	\vertex [dot, right=3 of b] (c) {}; \vertex [right=of c] (d) {$H$};
	\vertex[below=of a] (e) {$L$}; \vertex[dot, below=of b] (f) {};
	\vertex[dot, right=of f] (g) {}; \vertex[dot, below=of c] (h) {};
	\vertex[right=of h] (i) {$L$}; \diagram* { (a) -- [charged scalar] (b),
		(b) -- [scalar] (c), (c) --  [anti charged scalar](d), (e) --
[fermion] (f), (f) -- (g), (g) -- (h), (h) -- [anti fermion] (i), (b) --
[scalar] (f), (b) -- [scalar] (g), (c) -- [scalar] (h), }; \end{feynman}
\end{tikzpicture} \caption{CLBZ-8} \label{fig:CLBZ-8} \end{subfigure}
    \hspace{0.04\linewidth}%
\begin{subfigure}[t]{0.45\linewidth} \begin{tikzpicture} \begin{feynman}[every
	edge={thick}] \vertex (a) {$L$}; \vertex [dot, right=of a] (b) {};
	\vertex [dot, right=of b] (c) {}; \vertex[below=of a] (f) {$L$};
	\vertex[dot, right=of f] (g) {}; \vertex[dot, right=of g] (h) {};
	\vertex[right=2.0 of c] (e) {$H$}; \vertex[right=2.0 of h] (i) {$H$};
	\vertex[dot, right=of $(c)!0.5!(g)$] (d) {}; \diagram* { (a) --
		[fermion] (b), (b) -- (c), (c) -- [scalar] (d), (d) -- [charged
scalar] (e), (f) -- [fermion] (g), (g) -- [scalar] (h), (h) -- [scalar] (d),
(d) -- [anti charged scalar] (i), (b) -- [scalar] (h), (g) -- (c), };
\end{feynman} \end{tikzpicture} \caption{CLBZ-9} \label{fig:CLBZ-9}
\end{subfigure}

\vspace{3ex}
\begin{subfigure}[t]{0.45\linewidth}
	\begin{tikzpicture} 
	\begin{feynman}[every edge={thick}] 
		\vertex (a) {$L$}; 
		\vertex [dot, right=of a] (b) {};
		\vertex [dot, right=of b] (c) {}; 
		\vertex [dot, right= of c] (d) {};
		\vertex [right=of d] (e) {$H$}; 
		\vertex [below=of a] (f) {$H$}; 
		\vertex [dot, below=of b] (g) {}; 
		\vertex [below=of c] (k) {}; 
		\vertex	[dot, below=of d] (i) {}; 
		\vertex [below=of e] (j) {$L$}; 
		\diagram* {
		(a) -- [fermion] (b),
		(b) --  (c),
		(c) -- [scalar] (d),
		(d) --  [anti charged scalar] (e),
		(b) -- [scalar] (g),
		(c) --  (i), 
		(d) -- [scalar] (g),
		(f) -- [charged scalar] (g),
		(g) -- [scalar] (i),
		(i) -- [anti fermion] (j),
		}; 
	\end{feynman} 
\end{tikzpicture} \caption{CLBZ-10} \label{fig:CLBZ-10}
\end{subfigure}
    \hspace{0.04\linewidth}%
\begin{subfigure}[t]{0.45\linewidth} \begin{tikzpicture} \begin{feynman}[every
	edge={thick}] \vertex (a) {$L$}; \vertex [dot, right=of a] (b) {};
	\vertex [dot, right=of b] (c) {}; \vertex [dot, right= of c] (d) {};
	\vertex [right=of d] (e) {$H$}; \vertex [below=of a] (f) {$H$}; \vertex
	[dot, below=of b] (g) {}; \vertex [dot, below=of c] (k) {}; \vertex
	[dot, below=of d] (i) {}; \vertex [below=of e] (j) {$L$}; \diagram* {
		(a) -- [fermion] (b), (b) -- [scalar] (c), (c) --  (d), (d) --
		[anti charged scalar] (e), (b) --  (g), (c) -- [edge
		label=\(N\)] (k), (d) --  (i), (f) -- [charged scalar] (g), (g)
		--  (k), (k) -- [scalar] (i), (i) -- [anti fermion] (j), };
	\end{feynman} \end{tikzpicture} \caption{PTBM-1} \label{fig:PTBM-1}%
\end{subfigure}

\vspace{3ex}
\begin{subfigure}[t]{0.45\linewidth} \begin{tikzpicture} \begin{feynman}[every
	edge={thick}] \vertex (a) {$L$}; \vertex [dot, right=of a] (b) {};
	\vertex [dot, right=of b] (c) {}; \vertex [dot, right= of c] (d) {};
	\vertex [right=of d] (e) {$H$}; \vertex [below=of a] (f) {$L$}; \vertex
	[dot, below=of b] (g) {}; \vertex [dot, below=of c] (k) {}; \vertex
	[dot, below=of d] (i) {}; \vertex [below=of e] (j) {$H$}; \diagram* {
		(a) -- [fermion] (b), (b) -- (c), (c) --  [scalar] (d), (d) --
	[anti charged scalar] (e), (b) -- [scalar] (g), (c) -- (k), (d) --
[scalar] (i), (f) -- [fermion] (g), (g) --  (k), (k) -- [scalar] (i), (i) --
[anti charged scalar] (j), }; \end{feynman} \end{tikzpicture} \caption{RB-2}
\label{fig:RB-2} \end{subfigure} \caption{Relevant genuine 2-loop topologies.}
\label{fig:genuine-2-loop} \end{figure}

The possible 2-loop topologies of the Weinberg operator have been discussed in
Ref.~\cite{Sierra:2014rxa}. We will closely follow this classification. All
possible genuine 2-loop topologies are shown in
Fig.~\ref{fig:topologies-2loop-Weinberg}. Analytic expressions for the 2-loop diagrams are summarized in the appendix of Ref.~\cite{Sierra:2014rxa} and are based on the results in Refs.~\cite{McDonald:2003zj,Angel:2013hla}. Most topologies can be considered as
variations of a few 2-loop models discussed in the literature: (i) variations
of the Cheng-Li-Babu-Zee (CLBZ)
topology~\cite{Cheng:1980qt,Zee:1985id,Babu:1988ki}, (ii) the
Petcov-Toshev-Babu-Ma (PTBM)
topology~\cite{Petcov:1984nz,Babu:1988ig,Branco:1988ex}, and the so-called
rainbow (RB) topology~\cite{Sierra:2014rxa}. In the following we will further
distinguish between fermion and scalar lines and show in Figs.~\ref{fig:genuine-2-loop} and \ref{fig:ISC} the relevant diagrams of
genuine topologies and the internal-scalar-correction (ISC)-type topology which are used in the following
discussion. The first two
subsections discuss models based on genuine topologies, the third one models
based on non-genuine topologies, and the last one models based on multiple
topologies.

%

\subsubsection{Genuine 2-loop topologies} The relevant diagrams for the
genuine topologies are shown in Fig.~\ref{fig:genuine-2-loop}.

\textbf{CLBZ-1:} The topology CLBZ-1 is displayed in Fig.~\ref{fig:CLBZ-1}.
The first model was independently
proposed and studied by Zee~\cite{Zee:1985id} and Babu~\cite{Babu:1988ki}, and
is commonly called Zee-Babu model (See a more detailed discussion in
Sec.~\ref{sec:ZeeBabuModel}). It also leads to the operator $O_9$. A scale-invariant version of the model has been proposed in Ref.~\cite{Foot:2007ay}. It has been
extended to include a softly-broken continuous $L_e-L_\mu-L_\tau$ flavor
symmetry~\cite{Lavoura:2000kg,Kitabayashi:2000nf} or discrete flavor
symmetry~\cite{Araki:2010kq}, and has been embedded in a SUSY
model~\cite{Aoki:2010ib,Haba:2011cja}.  The same topology has also been used
for models with quarks instead of charged leptons inside the loop. They rely on
the introduction of a leptoquark and a diquark~\cite{Kohda:2012sr,Nomura:2016ask,Chang:2016zll} and lead to operator
$O_{11}$. Similarly, there is a version without light fields in the
loop~\cite{Ma:2007gq,Ding:2016wbd,Nomura:2016rjf,Ho:2016aye}. The models in Ref.~\cite{Ho:2016aye} are part of a systematic study of models based on a gauged $U(1)_{{\rm B}-{\rm L}}$ which is broken to a $Z_N$ symmetry.

\textbf{CLBZ-3:} Topology CLBZ-3 is depicted in Fig.~\ref{fig:CLBZ-3} and only
differs from topology CLBZ-1 in the way how the Higgs VEVs are attached to the
loop diagram: Topology CLBZ-3 has the Higgs VEVs attached to two of the scalar
lines, while they are attached to the internal fermion lines for CLBZ-1.
Reference~\cite{Cheng:1980qt} listed several possible neutrino mass models,
including the first 2-loop model which was based on topology CLBZ-3 with an effective scalar coupling. A possible UV completion was presented with an electroweak quintuplet scalar. This UV completion leads to the operator $O_{33}=\bar e^c \bar e^c L^i L^j e^c e^c H^k H^l \epsilon_{ik} \epsilon_{jl}$ (with an additional VEV insertion from an electroweak quintuplet scalar). All models~\cite{Okada:2014qsa,Aoki:2014cja,Ho:2016aye,Ho:2017fte,Baek:2017qos} based on
topology CLBZ-3 only contain heavy fields.

\textbf{CLBZ-8:} The topology is shown in Fig.~\ref{fig:CLBZ-8}.  Variants of
the Zee-Babu model have also been embedded in grand unified
theories~\cite{Wu:1980hz}. In case of SU(5), there is a 5-plet of matter
particles in the loop which leads to the effective operators $O_9$ and
$O_{11}$.

\textbf{CLBZ-9:} Topology CLBZ-9 which is displayed in Fig.~\ref{fig:CLBZ-9} has been utilized in a model with two diquarks~\cite{Popov:2016fzr}.

\textbf{CLBZ-10:} The same paper also introduces another model with two diquarks which is based on topology CLBZ-10, shown in Fig.~\ref{fig:CLBZ-10}.

\textbf{PTBM-1:} The first model to utilize the topology \ref{fig:PTBM-1},
although in presence of a tree-level contribution, was presented in
Refs.~\cite{Petcov:1984nz,Babu:1988ig,Branco:1988ex,Babu:1988wk}. Neutrino mass receives a
2-loop correction via the exchange of two $W$-bosons as shown in
Fig.~\ref{fig:PTBM-1}. This idea has been recently revived and experimentally
excluded in the context of extra chiral
generations~\cite{Aparici:2011nu}, but the mechanism can still work in the case of vector-like leptons.
 Lepton number is violated by the SM singlet Majorana
fermion $N$ in the center of the diagram and thus there is a tree-level contribution in addition to the 2-loop contribution to neutrino mass. Lepton number can equally well be broken by the type-III seesaw, when the
fermionic singlet is replaced by a fermionic triplet~\cite{Liao:2009nq}. The
model in Ref.~\cite{Babu:2010vp} has one of the $W$-bosons replaced by scalar
leptoquarks and it is consequently not accompanied by a tree-level
contribution. The 1-loop contribution induced by the mixing of the leptoquarks
vanishes, because the left-chiral coupling of one of the leptoquarks is
switched off~\cite{Cai:2017wry}. All models with $W$ bosons will lead to
operators with derivatives in the classification according to $\Delta L=2$
operators. Finally, Ref.~\cite{Angel:2013hla} proposed a model with a scalar
leptoquark and colored octet fermion. 

\subsubsection{Genuine topologies with additional VEV insertions}

Similar to the 1-loop models, we also categorize the models with additional VEV
insertions following the classification of Ref.~\cite{Sierra:2014rxa}.

\textbf{CLBZ-1:} There are several models based on topology CLBZ-1 (shown in
Fig.~\ref{fig:CLBZ-1}), which all induce the operator $O_9$.
Reference~\cite{Bamba:2008jq} discusses a possible connection of neutrino mass
with dark energy. Reference~\cite{Porto:2008hb} proposed a model with one
electroweak Higgs doublet field per lepton generation, an extension of the so-called private Higgs scenario. Finally, Ref.~\cite{Lindner:2011it} discusses an
extension of the Zee-Babu model by a global $U(1)_{{\rm B}-{\rm L}}$ symmetry, which is
spontaneously broken to a $Z_2$ subgroup. This implies the existence of a
Majoron and a DM candidate.

\textbf{CLBZ-3:} Reference~\cite{Chang:1988dq} proposed a variant of the Zee-Babu
model with an additional triplet Majoron, which is based on topology CLBZ-3
which is displayed in Fig.~\ref{fig:CLBZ-3}.

\textbf{CLBZ-9:} The topology CLBZ-9 is depicted in Fig.~\ref{fig:CLBZ-9}. The
model in Ref.~\cite{Guo:2012ne} is based on a dark $U(1)$ symmetry with only
heavy fields in the loop. 

\textbf{RB-2:} The model proposed in Ref.~\cite{Kajiyama:2013zla} is based on
$U(1)_{{\rm B}-{\rm L}}$, which is broken to $Z_2$. Apart from the VEV breaking
$U(1)_{{\rm B}-{\rm L}}$, neutrino mass is generated by a diagram with topology RB-2
which is shown in Fig.~\ref{fig:RB-2}.

\subsubsection{Non-genuine topologies} 
\begin{figure}[bt]\centering \begin{subfigure}[t]{0.45\linewidth}
	\begin{tikzpicture} \begin{feynman}[small, every edge={thick}] \vertex
		(nu1) at (-2.5,0)  {$L$}; \vertex (nu2) at (2.5,0)  {$L$};
		\vertex [dot,right=of nu1] (v1) {}; \vertex [dot,left=of nu2]
		(v2) {}; \vertex [dot,right=of v1] (v1a) {}; \vertex
		[dot,left=of v2] (v2a) {}; \vertex (c) at ($(v1)!0.5!(v2)$);
		\vertex [dot] (ca) at ($(c)+(0,0.5)$) {}; \vertex [dot] (cb) at
		($(c)+(0,2)$) {};

\vertex [above=of ca]  (ha) {$H$}; \vertex [above=of cb]  (hb) {$H$};
	
\diagram* { (nu1) -- [fermion] (v1), (nu2) --  [fermion] (v2),
	(v1) -- (v1a), (v2) -- (v2a), (v1a) -- (v2a), (ha) -- [charged scalar]
	(ca), (hb) -- [charged scalar] (cb), (v1) -- [scalar,out=90,in=180]
	(cb), (v2) -- [scalar,out=90,in=0] (cb), (v1a) --
	[scalar,out=90,in=180] (ca), (v2a) -- [scalar,out=90,in=0] (ca), };
\end{feynman} \end{tikzpicture}	\caption{NG-RB-1} \label{fig:NG-RB-1}
\end{subfigure}
\begin{subfigure}[t]{0.45\linewidth} \begin{tikzpicture} \begin{feynman}[small,
	every edge={thick}] \draw [thick,dashed] (0,0) circle [radius=1];
	\vertex [dot] at (0:1) (l1) {}; \vertex[cross,black]  at (90:1) (l2)
	{}; \vertex [dot] at (180:1) (l3) {}; \vertex  at (270:1) (l4) {};
\vertex (nu1) at (-3,-1.5)  {$L$}; \vertex (nu2) at (3,-1.5)  {$L$}; \vertex
[above right=of l1] (h1) {$H$}; \vertex [above left=of l3] (h3) {$H$}; \vertex
[dot,right=of nu1] (v1) {}; \vertex [dot,left=of nu2] (v2) {}; \diagram* {
	(nu1) -- [fermion] (v1), (nu2) --  [fermion] (v2),
	(v1) -- (v2), (v2) -- [scalar,bend right] (l1), (v1) --  [scalar,bend
	left] (l3), (l1) -- [anti charged scalar] (h1), (l3) -- [anti charged
	scalar] (h3), }; \end{feynman} \end{tikzpicture}
\caption{ISC-type} \label{fig:ISC} \end{subfigure}
\caption{Non-genuine 2-loop topologies.} \label{fig:NG-2loop} \end{figure}
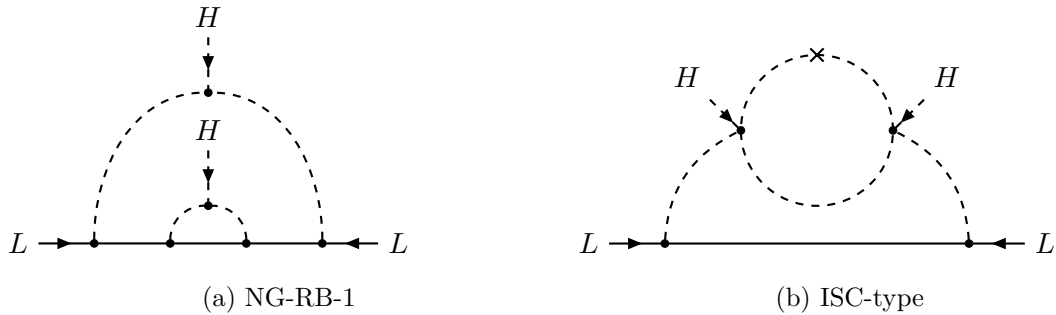

The relevant non-genuine 2-loop
topologies are shown in Fig.~\ref{fig:NG-2loop}.

\textbf{NG-RB-1:} The non-genuine topology NG-RB-1
(Fig.~\ref{fig:NG-RB-1}) is generated in Ref.~\cite{Nomura:2016pgg}. There are
no lower-order contributions due to the $U(1)$ symmetry, which is broken to
$Z_2$ as in the above-mentioned models.

\textbf{Other non-genuine topologies:} There are several models which generate
vertices or masses of particles at loop level. The models in
Refs.~\cite{Aoki:2013gzs,Aoki:2014lha} realize an ISC-type topology which is shown in Fig.~\ref{fig:ISC} by softly breaking lepton
number with a dimension-2 scalar mass insertion in the internal scalar loop.
Similarly, Ref.~\cite{Ma:2007yx} discusses a supersymmetric model where the
scalar-quartic coupling is induced after supersymmetry is softly broken and
thus an ISC-type topology is induced for neutrino mass.  The models in
Refs.~\cite{Kajiyama:2013rla,Baek:2013fsa} have only heavy particles in the
loop and can be considered as a 1-loop scotogenic model, where the Majorana
mass term for the SM singlet fermions is generated at 1-loop order. Thus neutrino
mass is effectively generated at 2-loop order. It can be considered as an RB-type
topology. In contrast to the topology RB-2, the SM Higgs fields are attached to the
outer scalar line (the one on the left in Fig.~\ref{fig:RB-2}). Both models
break $U(1)_{{\rm B}-{\rm L}}$ to a discrete $Z_N$ subgroup. Reference~\cite{Ghosh:2017vhe}
proposes another model based on an RB-type topology, where the Higgs fields
couple to the fermions in the outer loop. The model features a stable dark
matter candidate due to an imposed $Z_2$ symmetry. Moreover, neutrino mass
relies on the spontaneous breaking of an extended lepton number symmetry to a
discrete $Z_2$ subgroup. 
The models in
Refs.~\cite{Ma:1989tz,Nasri:2001ax,Wei:2010ww,Ma:2012xj,Nomura:2017ezy} realize the
type-I seesaw by generating the Dirac mass terms at 1-loop order, and the model
in Ref.~\cite{Okada:2015nca} generates a radiative type-II seesaw contribution
by generating the triplet VEV at 1-loop level, and thus the Weinberg operator
at 2-loop level. 
Finally, Refs.~\cite{Witten:1979nr,Rodriguez:2013rma} firstly generate the
right-handed neutrino mass at 2-loop level in the context of a GUT, which
induces the active neutrino mass via the usual seesaw mechanism.
Similarly Refs.~\cite{Ma:2009gu,Baldes:2013eva} realize a radiative inverse seesaw. The mass of additional singlets is generated at 2-loop order. The model is based on an additional gauged $U(1)$ symmetry (which is spontaneously broken to its $Z_2$ subgroup) to forbid the generation of neutrino mass at tree-level via the seesaw mechanism. The model can explain the matter-antimatter asymmetry of the universe, but not account for the dark matter abundance~\cite{Baldes:2013eva}.

\subsubsection{Models based on several topologies} Several models in the
literature~\cite{Babu:1989pz,Babu:2011vb,Kanemura:2011mw,Ma:2013mga,Baek:2014awa,Kanemura:2014rpa,Kashiwase:2015pra,Ho:2016aye,Nomura:2016run,Nomura:2016dnf,Chen:2006vn,Chen:2007dc,delAguila:2011gr,Chen:2012vm,delAguila:2013zba,King:2014uha,Geng:2015sza,Megrelidze:2016fcs} are based on multiple 2-loop topologies. 
We highlight three examples. Reference~\cite{Megrelidze:2016fcs} proposed
	to generate neutrino mass via lepton-number-violating soft
	supersymmetry-breaking terms using the
	so-called type-II-B soft seesaw with electroweak triplet superfields.
	Integrating out the scalar components of the electroweak triplets leads
	to the dimension-5 lepton-number-violating term $(\tilde L \tilde
	H_u)^2$. Neutrino mass is generated at 2-loop via a diagram based on
	topology CLBZ-1 and diagrams which generate the couplings of the
	scalar component of the electroweak triplet superfield to two lepton doublets $L$ on the one hand and the two electroweak Higgs doublets $H_u$ on the other hand at the 1-loop level.
Another interesting class of models are based on internal electroweak
gauge bosons, which are based on CLBZ-type topologies and discussed in
Refs.~\cite{Chen:2006vn,Chen:2007dc,delAguila:2011gr,Chen:2012vm,delAguila:2013zba,King:2014uha,Geng:2015sza}.
All of them introduce a doubly-charged scalar and a coupling of the
doubly-charged scalar to two $W$-bosons, which can be achieved via a mixing of
the doubly-charged scalar with the doubly-charged scalar in an electroweak
triplet scalar. Neutrino mass is typically generated via topologies CLBZ-1 and
CLBZ-3 and induces the operator \begin{equation} \mathcal{O}^{\rm RR} = \bar
	e_{\rm R} e_{\rm R}^c (H^\dagger D^\mu \tilde H) (H^\dagger D_\mu \tilde H)\;. 
\label{eq:RR}
\end{equation} This possibility is further discussed in
Sec.~\ref{sec:derivatives}.
Gauge bosons similarly can play an important role in the
	generation of neutrino mass in extended technicolor (ETC) models as discussed in
	Refs.~\cite{Appelquist:2002me,Appelquist:2003uu,Appelquist:2003hn}.
	These models contain many SM singlet fermions and only a few elements
	of the neutral fermion mass matrix are directly generated by
	condensates, while many elements are generated at 1-loop (or higher loop) level via
	loop diagrams with ETC gauge bosons. In particular the relevant
	Dirac mass terms relevant for the active neutrino masses are generated
	at 1-loop level and thus neutrino mass is effectively generated at 2-loop (or even higher loop) level.


\subsection{3-loop Majorana neutrino mass models}
\label{sec:3loop}

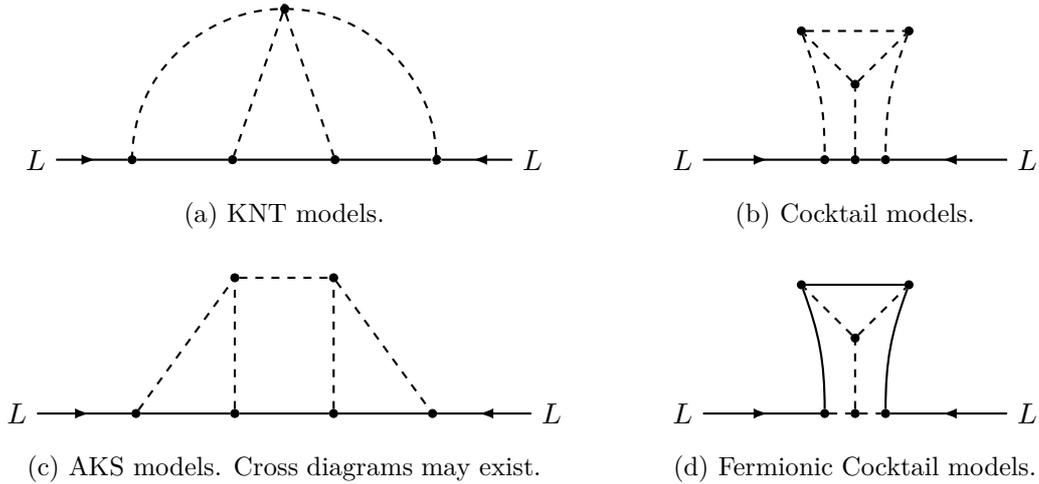
\begin{figure}[tb] \centering
\begin{subfigure}[t]{0.5\linewidth}\centering \begin{tikzpicture}[node distance=1cm and
	1cm] \coordinate[label=left:$L$] (nu1); \coordinate[vertex, right
		of=nu1] (v1); \draw[scalarnoarrow] (v1) arc[start angle=180,
			delta angle=-180, radius=2.0cm] node[pos=0.5,vertex]
			(v05) {} node[vertex] (v2) {}; \coordinate[right of=v2,
				label=right:$L$] (nu2); \coordinate[vertex]
					(v03) at ($(v1)!0.33!(v2)$);
				\coordinate[vertex] (v06) at
					($(v1)!0.667!(v2)$);

    \draw[fermionnoarrow] (v1)--(v03)--(v06)--(v2); \draw[scalarnoarrow]
    (v05)--(v03); \draw[scalarnoarrow] (v05)--(v06); \draw[fermion]
    (nu1)--(v1); \draw[fermion] (nu2)--(v2); \end{tikzpicture} \caption{KNT
    models.} \label{fig:KNT} \end{subfigure}
\begin{subfigure}[t]{0.4\linewidth}\centering \begin{tikzpicture}[node
	distance=1cm and 1cm] \coordinate[label=left:$L$] (nu1);
	\coordinate[vertex, right of=nu1,xshift=0.6cm] (v1a);
	\coordinate[vertex, right of=v1a,xshift=-0.6cm] (v1);
	\coordinate[vertex,right of=v1,xshift=-0.6cm] (v1b); \coordinate[right
		of=v1b, label=right:$L$,xshift=0.6cm] (nu2);
	\coordinate[vertex, above of=v1] (v2); \coordinate[vertex, above left
		of=v2] (v3a); \coordinate[vertex, above right of=v2] (v3b);
		\draw[fermionnoarrow] (v1a)--(v1)--(v1b);
		\draw[scalarnoarrow] (v1)--(v2); \draw[scalarnoarrow]
			(v2)--(v3a); \draw[scalarnoarrow] (v2)--(v3b);
			\draw[scalarnoarrow] (v3a)--(v3b);
			\draw[scalarnoarrow] (v3a) to [bend left=10] (v1a);
			\draw[scalarnoarrow] (v3b) to [bend right=10] (v1b);
    \draw[fermion] (nu1)--(v1a); \draw[fermion] (nu2)--(v1b); \end{tikzpicture}
    \caption{Cocktail models.} \label{fig:cocktail} \end{subfigure}\\
    \vspace{3ex}
 \begin{subfigure}[t]{0.5\linewidth}\centering \begin{tikzpicture}[node
	distance=1.3cm and 1.3cm] \coordinate[label=left:$L$] (nu1);
	\coordinate[vertex, right of=nu1] (v1); \coordinate[vertex, right
		of=v1] (v3); \coordinate[vertex, right of=v3] (v4);
		\coordinate[vertex, right of=v4] (v2); \coordinate[vertex,
			above of=v3, yshift=0.5cm] (v5); \coordinate[vertex,
				above of=v4, yshift=0.5cm] (v6);
			\coordinate[right of=v2, label=right:$L$] (nu2);

    \draw[fermion] (nu1)--(v1); \draw[fermion] (nu2)--(v2);
    \draw[scalarnoarrow] (v1)--(v5); \draw[scalarnoarrow] (v5)--(v6);
    \draw[scalarnoarrow] (v6)--(v2); \draw[scalarnoarrow] (v3)--(v5);
    \draw[scalarnoarrow] (v4)--(v6); \draw[fermionnoarrow]
	    (v1)--(v3)--(v4)--(v2); \end{tikzpicture} \caption{AKS models. Cross diagrams may exist.} \label{fig:trapezoid} \end{subfigure} 
\begin{subfigure}[t]{0.4\linewidth}\centering \begin{tikzpicture}[node
	distance=1cm and 1cm] \coordinate[label=left:$L$] (nu1);
	\coordinate[vertex, right of=nu1,xshift=0.6cm] (v1a);
	\coordinate[vertex, right of=v1a,xshift=-0.6cm] (v1);
	\coordinate[vertex,right of=v1,xshift=-0.6cm] (v1b); \coordinate[right
		of=v1b, label=right:$L$,xshift=0.6cm] (nu2);
	\coordinate[vertex, above of=v1] (v2); \coordinate[vertex, above left
		of=v2] (v3a); \coordinate[vertex, above right of=v2] (v3b);
		\draw[scalarnoarrow] (v1a)--(v1)--(v1b);
		\draw[scalarnoarrow] (v1)--(v2); \draw[scalarnoarrow]
			(v2)--(v3a); \draw[scalarnoarrow] (v2)--(v3b);
			\draw[fermionnoarrow] (v3a)--(v3b);
		\draw[fermionnoarrow] (v3a) to [bend left=10] (v1a);
	\draw[fermionnoarrow] (v3b) to [bend right=10] (v1b); \draw[fermion]
	(nu1)--(v1a); \draw[fermion] (nu2)--(v1b); \end{tikzpicture}
	\caption{Fermionic Cocktail models.} \label{fig:fermioniccocktail}
\end{subfigure} 
\caption{3-loop model topologies. Note that we do not specify the Higgs insertions.} \label{fig:3looptop}
\end{figure}

Unlike 1-loop and 2-loop topologies, there is no systematic classification of all 3-loop topologies. Thus we restrict ourselves to the existing 3-loop models in the literature and do not consider other topologies or different fermion flow for the given topologies. Most of the existing 3-loop models can be
categorized in four basic types of diagrams shown in Fig.~\ref{fig:3looptop} where we do not specify the Higgs insertions.  
The remaining models are either based on a combination of the listed topologies
or the combination of a loop-induced vertex at 1- or 2-loop inside a loop
diagram.

\textbf{The KNT models:} The first 3-loop radiative neutrino mass model was
proposed in Ref.~\cite{Krauss:2002px} with the topology shown in
Fig.~\ref{fig:KNT} by Krauss, Nasri and Trodden (KNT) and it leads to the
operator $O_9$.  We refer to radiative neutrino mass models sharing the same
topology as KNT models and discuss them in more detail in Sec.~\ref{sec:KNT}. A
systematic study with several different variants can be found in
Ref.~\cite{Chen:2014ska}.  The models of
Refs.~\cite{Cheung:2004xm,Ahriche:2013zwa,Ahriche:2014xra,Chen:2014ska,Ahriche:2014cda,Ahriche:2014oda,Ahriche:2015wha,Ahriche:2015loa,Ahriche:2015taa}
also generate the operator $O_9$, the models of
Refs.~\cite{Chen:2014ska,Nomura:2016ezz,Cheung:2016frv} the operator $O_{11}$ with down-type
quarks, while the models in
Refs.~\cite{Okada:2014oda,Chen:2014ska,Okada:2016rav,Cheung:2016ypw} only have new heavy
states in the loop. 

\textbf{AKS-type models:} Neutrino mass can also arise at 3-loop order from the diagram shown in Fig.~\ref{fig:trapezoid}.  The first model of such
topology was proposed by Aoki, Kanemura, and Seto (AKS) in Ref.~\cite{Aoki:2008av} and is based on the operator $\bar e^c \bar e^c H_1^i H_2^j H_1^k H_2^l \epsilon_{ij} \epsilon_{kl}$ with two Higgs doublets $H_i$. We will refer to it as the AKS model and more generally to models based
on this topology as AKS-type models. It contains a second Higgs doublet and
several $SU(2)_{\rm L}$ singlets.  The exotic particles can also be all electroweak
singlets~\cite{Gu:2016xno,Ho:2016aye}. The model in Ref.~\cite{Gu:2016xno}
leads to the operator $O_9$.   Other variants include colored exotic particles
such as leptoquarks~\cite{Cheung:2017efc,  Culjak:2015qja, Chen:2014ska}, which
generate the operators $O_{11,12}$, or electroweak
multiplets~\cite{Chen:2014ska,Okada:2015hia,Ko:2016sxg} generating the
operators $O_{1,9}$. Note cross diagrams may be allowed in specific models.  

\textbf{Cocktail models:} The third class of models are based on the two
cocktail diagrams shown in Figs.~\ref{fig:cocktail} and
\ref{fig:fermioniccocktail}.  The name for the diagram has been coined by
Ref.~\cite{Gustafsson:2012vj}, which proposed a 3-loop model with two
$W$-bosons based on topology \ref{fig:cocktail} and consequently generated the
operator $\mathcal{O}^{\rm RR}$, which are discussed in more detail in
Sec.~\ref{sec:derivatives}. The same model has also been studied in
Ref.~\cite{Geng:2014gua}. The models in
Refs.~\cite{Hatanaka:2014tba,Alcaide:2017xoe} are based on the
same topology, but with $W$ bosons replaced by scalars. While
Ref.~\cite{Alcaide:2017xoe} induces operator $\mathcal{O}^{\rm RR}$,
the model of Ref.~\cite{Hatanaka:2014tba} leads to operator $O_9$. Finally,
the fermionic cocktail topology \ref{fig:fermioniccocktail} is used in the models of
Refs.~\cite{Nishiwaki:2015iqa,Kanemura:2015bli}, both of which generate operator
$O_9$. 

Apart from the three classes of models, there are a few models which do not
uniquely fit in any of the three classes. The model in Ref.~\cite{Jin:2015cla}
is based on topologies~\ref{fig:KNT} and \ref{fig:trapezoid} with two
$W$-bosons and thus generates the operator $\mathcal{O}^{\rm RR}$.
Reference~\cite{Nomura:2016vxr} generates the mass of new exotic fermions at
2-loop level via a CLBZ-type diagram, which in turn generate neutrino mass at
1-loop. Reference~\cite{Geng:2016auy} studies a 2-loop model based on the operator
$O_8$, which itself is generated at 1-loop order. 

Most of the 3-loop models need to impose extra discrete symmetries such as
$Z_2$ or a continuous $U(1)$ symmetry to forbid lower-loop or tree-level contributions, unless
accidental symmetries exist and thus partly require other VEV insertions.  One
example is to employ higher dimensional representation of
$SU(2)_{\rm L}$~\cite{Ahriche:2015wha}, e.g. septuplet, in the spirit of minimal dark
matter \cite{Cirelli:2005uq,Cirelli:2009uv} such that  undesirable couplings
are forbidden by the SM gauge group alone.  Due to the existence of the extra
imposed or accidental symmetries, 3-loop  models serve as a natural playground
for DM physics.  

\subsection{Models with gauge bosons} \label{sec:derivatives}

The first model~\cite{Petcov:1984nz,Babu:1988ig,Branco:1988ex} with gauge bosons in the loop uses the topology PTBM-1 and leads to operators built from two lepton doublets including covariant derivatives.
However it also has a tree-level contribution, while models
based on operators with right-handed charged leptons are genuine radiative
neutrino mass models.

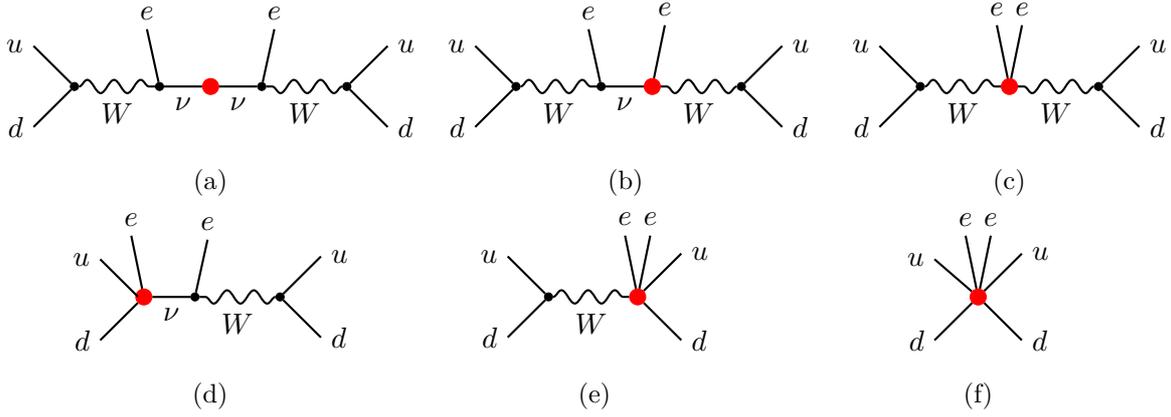
\begin{figure}[bt]\centering
		\begin{subfigure}[t]{0.35\linewidth} \centering
		\begin{tikzpicture}[node distance=1cm and 1cm,scale=0.9]
     \coordinate[label=left:$d$] (nu1);
     \coordinate[vertex, above right=0.7 of nu1] (v1);
     \coordinate[ above left=0.7 of v1, label=left:$u$] (h1);
     
     \coordinate[vertex, right=of v1] (lfv);
     \coordinate[above=0.7 of lfv,xshift=-1ex,label=above:$e$] (e1) ;
     
     \coordinate[blob, right=0.5 of lfv] (c);

     \coordinate[vertex, right=0.5 of c] (lfv2);
     \coordinate[ above=0.7 of lfv2,xshift=1ex,label=above:$e$] (e2) ;

     \coordinate[vertex, right=of lfv2] (v2);
     \coordinate[below right=0.7 of v2, label=right:$d$] (nu2);
     \coordinate[above right=0.7 of v2, label=right:$u$] (h2);

     \draw[fermionnoarrow] (nu1)--(v1);
     \draw[fermionnoarrow] (nu2)--(v2);
     \draw[fermionnoarrow] (h1) -- (v1);
     \draw[fermionnoarrow] (h2) -- (v2);
	
     \draw[photon] (v1) -- node[below,black,yshift=-0.5ex]{$W$}  (lfv);
     \draw[photon] (v2) -- node[below,black,yshift=-0.5ex]{$W$}  (lfv2);

     \draw[fermionnoarrow] (lfv) -- (e1);
     \draw[fermionnoarrow] (lfv2) -- (e2);
	
     \draw[fermionnoarrow] (lfv) -- node[midway,below] {$\nu$} (c);
     \draw[fermionnoarrow] (lfv2) -- node[midway,below] {$\nu$} (c);

   \end{tikzpicture}
\caption{}
\label{fig:conts_0nubb-a}
\end{subfigure}
	\begin{subfigure}[t]{0.3\linewidth} \centering
		\begin{tikzpicture}[node distance=1cm and 1cm,scale=0.9]
     \coordinate[label=left:$d$] (nu1);
     \coordinate[vertex, above right=0.7 of nu1] (v1);
     \coordinate[ above left=0.7 of v1, label=left:$u$] (h1);
     
     \coordinate[vertex, right=of v1] (lfv);
     \coordinate[above=0.7 of lfv,xshift=-1ex,label=above:$e$] (e1) ;
     
     \coordinate[blob, right=0.5 of lfv] (lfv2);

     \coordinate[ above=0.7 of lfv2,xshift=1ex,label=above:$e$] (e2) ;

     \coordinate[vertex, right=of lfv2] (v2);
     \coordinate[below right=0.7 of v2, label=right:$d$] (nu2);
     \coordinate[above right=0.7 of v2, label=right:$u$] (h2);

     \draw[fermionnoarrow] (nu1)--(v1);
     \draw[fermionnoarrow] (nu2)--(v2);
     \draw[fermionnoarrow] (h1) -- (v1);
     \draw[fermionnoarrow] (h2) -- (v2);
	
     \draw[photon] (v1) -- node[below,black,yshift=-0.5ex]{$W$}  (lfv);
     \draw[photon] (v2) -- node[below,black,yshift=-0.5ex]{$W$}  (lfv2);

     \draw[fermionnoarrow] (lfv) -- (e1);
     \draw[fermionnoarrow] (lfv2) -- (e2);
	
     \draw[fermionnoarrow] (lfv) -- node[midway,below] {$\nu$} (c);

   \end{tikzpicture}
\caption{}
\label{fig:conts_0nubb-b}
\end{subfigure}
	\begin{subfigure}[t]{0.3\linewidth} \centering
		\begin{tikzpicture}[node distance=1cm and 1cm,scale=0.9]
     \coordinate[label=left:$d$] (nu1);
     \coordinate[vertex, above right=0.7 of nu1] (v1);
     \coordinate[ above left=0.7 of v1, label=left:$u$] (h1);
     
     \coordinate[blob, right=of v1] (lfv);
     \coordinate[above=0.7 of lfv,xshift=-1ex,label=above:$e$] (e1) ;
     
     \coordinate[ above=0.7 of lfv,xshift=1ex,label=above:$e$] (e2) ;

     \coordinate[vertex, right=of lfv] (v2);
     \coordinate[below right=0.7 of v2, label=right:$d$] (nu2);
     \coordinate[above right=0.7 of v2, label=right:$u$] (h2);

     \draw[fermionnoarrow] (nu1)--(v1);
     \draw[fermionnoarrow] (nu2)--(v2);
     \draw[fermionnoarrow] (h1) -- (v1);
     \draw[fermionnoarrow] (h2) -- (v2);
	
     \draw[photon] (v1) -- node[below,black,yshift=-0.5ex]{$W$}  (lfv);
     \draw[photon] (v2) -- node[below,black,yshift=-0.5ex]{$W$}  (lfv);

     \draw[fermionnoarrow] (lfv) -- (e1);
     \draw[fermionnoarrow] (lfv) -- (e2);
	
   \end{tikzpicture}
\caption{}
\label{fig:conts_0nubb-c}
\end{subfigure}

		\begin{subfigure}[t]{0.3\linewidth} \centering
		\begin{tikzpicture}[node distance=1cm and 1cm,scale=0.9]
     \coordinate[label=left:$d$] (nu1);
     \coordinate[blob, above right=0.7 of nu1] (lfv);
     \coordinate[ above left=0.7 of v1, label=left:$u$] (h1);
     
     \coordinate[above=0.7 of lfv,xshift=-1ex,label=above:$e$] (e1) ;
     
     \coordinate[vertex, right=0.5 of lfv] (lfv2);
     \coordinate[ above=0.7 of lfv2,xshift=1ex,label=above:$e$] (e2) ;

     \coordinate[vertex, right=of lfv2] (v2);
     \coordinate[below right=0.7 of v2, label=right:$d$] (nu2);
     \coordinate[above right=0.7 of v2, label=right:$u$] (h2);

     \draw[fermionnoarrow] (nu1)--(v1);
     \draw[fermionnoarrow] (nu2)--(v2);
     \draw[fermionnoarrow] (h1) -- (v1);
     \draw[fermionnoarrow] (h2) -- (v2);
	
     \draw[photon] (v2) -- node[below,black,yshift=-0.5ex]{$W$}  (lfv2);

     \draw[fermionnoarrow] (lfv) -- (e1);
     \draw[fermionnoarrow] (lfv2) -- (e2);
	
     \draw[fermionnoarrow] (lfv2) -- node[midway,below] {$\nu$} (lfv);

   \end{tikzpicture}
\caption{}
\label{fig:conts_0nubb-d}
\end{subfigure}
		\begin{subfigure}[t]{0.3\linewidth} \centering
		\begin{tikzpicture}[node distance=1cm and 1cm,scale=0.9]
     \coordinate[label=left:$d$] (nu1);
     \coordinate[vertex, above right=0.7 of nu1] (v1);
     \coordinate[ above left=0.7 of v1, label=left:$u$] (h1);
     
     \coordinate[blob, right=of v1] (lfv);
     \coordinate[above=0.7 of lfv,xshift=-1ex,label=above:$e$] (e1) ;
     
     \coordinate[ above=0.7 of lfv,xshift=1ex,label=above:$e$] (e2) ;

     \coordinate[below right=0.7 of lfv, label=right:$d$] (nu2);
     \coordinate[above right=0.7 of lfv, label=right:$u$] (h2);

     \draw[fermionnoarrow] (nu1)--(v1);
     \draw[fermionnoarrow] (nu2)--(lfv);
     \draw[fermionnoarrow] (h1) -- (v1);
     \draw[fermionnoarrow] (h2) -- (lfv);
	
     \draw[photon] (v1) -- node[below,black,yshift=-0.5ex]{$W$}  (lfv);

     \draw[fermionnoarrow] (lfv) -- (e1);
     \draw[fermionnoarrow] (lfv) -- (e2);

   \end{tikzpicture}
\caption{}
\label{fig:conts_0nubb-e}
\end{subfigure}
		\begin{subfigure}[t]{0.3\linewidth} \centering
		\begin{tikzpicture}[node distance=1cm and 1cm,scale=0.9]
     \coordinate[label=left:$d$] (nu1);
     \coordinate[blob, above right=0.7 of nu1] (lfv);
     \coordinate[ above left=0.7 of v1, label=left:$u$] (h1);
     
     \coordinate[above=0.7 of lfv,xshift=-1ex,label=above:$e$] (e1) ;
     
     \coordinate[ above=0.7 of lfv,xshift=1ex,label=above:$e$] (e2) ;

     \coordinate[below right=0.7 of lfv, label=right:$d$] (nu2);
     \coordinate[above right=0.7 of lfv, label=right:$u$] (h2);

     \draw[fermionnoarrow] (nu1)--(lfv);
     \draw[fermionnoarrow] (nu2)--(lfv);
     \draw[fermionnoarrow] (h1) -- (lfv);
     \draw[fermionnoarrow] (h2) -- (lfv);
	
     \draw[fermionnoarrow] (lfv) -- (e1);
     \draw[fermionnoarrow] (lfv) -- (e2);

   \end{tikzpicture}
\caption{}
\label{fig:conts_0nubb-f}
\end{subfigure}
\caption[Possible contributions to $0\nu\beta\beta$.]{Possible contributions to $0\nu\beta\beta$. The red dot indicates the $\Delta L=2$ effective vertex. Figure reproduced from Ref.~\cite{delAguila:2013zba}.} \label{fig:conts_0nubb} \end{figure}

In Ref.~\cite{delAguila:2012nu} two LNV effective operators with gauge bosons, i.e.~present in covariant derivatives, were considered, which allowed to have
neutrinoless double beta decay rates generated at tree level thanks to new
couplings to the SM leptons.\footnote{In general, $0\nu\beta\beta$ is
generated in these models at a lower order than neutrino masses.}
Interestingly, depending on the chirality of the outgoing leptons in
$0\nu\beta\beta$, there are two new operators (beyond the standard contribution
from the Weinberg operator which involves left-handed electrons). For left-right (LR)
chiralities of the outgoing electrons, there is a dimension-7 operator:\footnote{There are other operators which, however, are simultaneously
generated with the Weinberg operator, which dominates as it is dimension
5~\cite{delAguila:2012nu}.}
\begin{equation}
	\mathcal{O}^{\rm LR} = (H^\dagger D^\mu \tilde H) (H^\dagger \overline{e_{\rm R}} \gamma_\mu \tilde L) \,.
\label{eq:LR} 
\end{equation}
For right-right (RR) chiralities, there is a dimension-9 operator $\mathcal{O}^{\rm RR}$ as define in 
Eq.~\ref{eq:RR}.
After electroweak symmetry breaking, these operators generate the relevant
vertices for $0\nu\beta\beta$ at tree level: $W_\mu^- \overline{e_{\rm R}}
\gamma^\mu \nu^c_{\rm L}$ and $W^-_\mu W^{-\mu} \overline{e_{\rm R}} e^c_{\rm
R}$, respectively. The contributions of $\mathcal{O}^{\rm LR}$ and
$\mathcal{O}^{\rm RR}$ to $0\nu\beta\beta$ are depicted in Figs.~\ref{fig:conts_0nubb-b} and \ref{fig:conts_0nubb-c} respectively, where the red point denotes the
effective operator insertion.

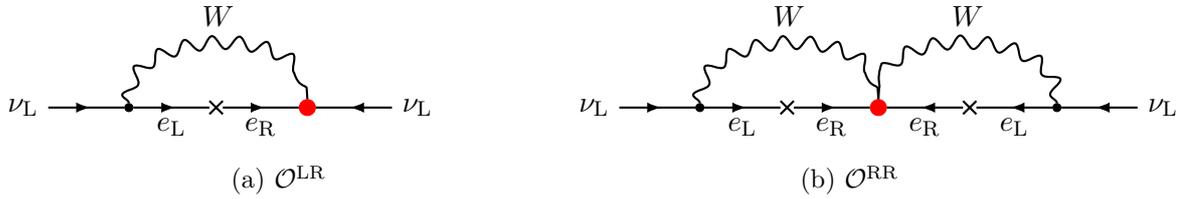
\begin{figure}[bt]\centering
	\begin{subfigure}[t]{0.45\linewidth}
		\begin{tikzpicture}[node distance=1cm and 1cm]
			\coordinate[label=left:$\nu_{\rm L}$] (nu1);
     \coordinate[vertex, right=of nu1] (v1);
     \coordinate[cross, right=of v1] (lfv);
     \coordinate[blob, right=of lfv] (v2);
     \coordinate[right=of v2, label=right:$\nu_{\rm L}$] (nu2);

     \draw[fermion] (nu1)--(v1);
     \draw[fermion] (v1) -- node[below]{$e_{\rm L}$} (lfv);
     \draw[fermion] (lfv) -- node[below]{$e_{\rm R}$}  (v2);
     \draw[fermion] (nu2)--(v2);
     \draw[photon] (v1) to[in=90,out=90] node[midway,above,yshift=1ex] {$W$} (v2);
   \end{tikzpicture}
   \caption{$\mathcal{O}^{\rm LR}$}
   \label{fig:O-LR}
   \end{subfigure}	
	\begin{subfigure}[t]{0.45\linewidth}
		\begin{tikzpicture}[node distance=1cm and 1cm]
			\coordinate[label=left:$\nu_{\rm L}$] (nu1);
     \coordinate[vertex, right=of nu1] (v1);
     \coordinate[cross, right=of v1] (lfv);
     \coordinate[blob, right=of lfv] (v2);
     \coordinate[cross, right=of v2] (lfv2);
     \coordinate[vertex, right=of lfv2] (v3);

     \coordinate[right=of v3, label=right:$\nu_{\rm L}$] (nu2);

     \draw[fermion] (nu1)--(v1);
     \draw[fermion] (nu2)--(v3);
     \draw[fermion] (v1) -- node[below]{$e_{\rm L}$} (lfv);
     \draw[fermion] (lfv) -- node[below]{$e_{\rm R}$}  (v2);
     \draw[fermion] (v3) -- node[below]{$e_{\rm L}$} (lfv2);
     \draw[fermion] (lfv2) -- node[below]{$e_{\rm R}$}  (v2);
 
     \draw[photon] (v1) to[in=90,out=90] node[midway,above,yshift=1ex] {$W$} (v2);
     \draw[photon] (v3) to[in=90,out=90] node[midway,above,yshift=1ex] {$W$} (v2);
   \end{tikzpicture}
   \caption{$\mathcal{O}^{\rm RR}$}
   \label{fig:O-RR}
   
   \end{subfigure}	
   \caption[Topologies with $W$-bosons.]{Lowest order contributions of $\mathcal{O}^{\rm LR}$ (left, at 1-loop order) and $\mathcal{O}^{\rm RR}$ (right, at 2-loop order) to neutrino masses. The red dot indicates the $\Delta L=2$ effective vertex. Figure reproduced following Ref.~\cite{delAguila:2012nu}.}
\label{fig:Der_ops}
\end{figure}

The lowest order contributions from these operators to neutrino masses occur
at 1- and 2-loop orders, respectively, via the diagrams of Fig.~\ref{fig:Der_ops}. The dominant contributions come from matching (see
also Refs.~\cite{Babu:2001ex, deGouvea:2007qla, Angel:2012ug, deGouvea:2014lva}
for estimates of the matching contributions to neutrino masses of LNV
operators), which using dimensional analysis can be estimated to be given
by~\cite{delAguila:2012nu}:
\begin{equation}
(m_\nu)^{\rm LR}_{ab} \simeq \frac{v}{16 \pi^2 \Lambda} \left(m_a C^{\rm LR}_{ab} + m_b C^{\rm LR}_{ba}\right) 
\end{equation}
for $\mathcal{O}^{\rm LR}$ and by 
\begin{equation}
(m_\nu)^{\rm RR}_{ab} \simeq \frac{1}{(16 \pi^2)^2 \Lambda} m_a C^{\rm RR}_{ab} m_b \,
\end{equation}
for $\mathcal{O}^{\rm RR}$.
Notice that the appearance of the chirality-flipping charged lepton masses is
expected in order to violate lepton number in the LH neutrinos, which naturally
generates textures in the neutrino mass matrix.

Possible tree-level UV completions which have new contributions to
$0\nu\beta\beta$ at tree level were outlined in Ref.~\cite{delAguila:2012nu}.
See also Ref.~\cite{delAguila:2013zba}, which provides a summary of two
examples of models generating $\mathcal{O}^{\rm LR}$ and $\mathcal{O}^{\rm
RR}$, respectively. The UV model of $\mathcal{O}^{\rm
RR}$~\cite{delAguila:2011gr} generates $0\nu\beta\beta$ at tree level, while
neutrino masses are generated as expected at 2-loop order. It includes a
doubly-charged singlet, a $Y=1$ triplet scalar and a real singlet.  In order to
prevent tree-level neutrino masses as in type-II seesaw via the latter field, a
discrete $Z_2$ symmetry, which was spontaneously broken by the VEV of the
singlet, was added. Recently a variation has been studied, in which the $Z_2$
symmetry is exact, such that there is a good dark matter candidate, which is a
mixture of singlet and triplet~\cite{Alcaide:2017xoe}. In this case, the
contributions to $0\nu\beta\beta$ and to neutrino masses are further shifted by
one extra loop, i.e., they are generated at 1- and 3-loop orders, respectively.
References~\cite{Gustafsson:2012vj,Gustafsson:2014vpa} studied also a specific model
with a dark matter candidate, named \emph{the cocktail model}, which generated
$\mathcal{O}^{\rm RR}$ at 1-loop order, i.e.~$0\nu\beta\beta$ at 1-loop order and
therefore neutrino masses at 3-loop order. It includes a singly-charged singlet,
a doubly-charged singlet and a $Y=1$ scalar doublet, together with a discrete
symmetry $Z_2$ under which all the new fields except the doubly-charged are
odd. Other models generating $\mathcal{O}^{\rm RR}$ were presented in Refs.~\cite{Chen:2006vn,Chen:2007dc,Chen:2012vm,King:2014uha,Geng:2015sza,Liu:2016mpf}. 

\subsection{Radiative Dirac neutrino mass models} \label{sec:Dirac}
Although Majorana neutrinos are the main focus of research, Dirac neutrinos are
a viable possibility to explain neutrino mass. It is noteworthy that the first
radiative neutrino mass model~\cite{Cheng:1977ir} was based on Dirac neutrinos.
In recent years, there has been an increased interest in Dirac neutrinos and,
in particular, there are a few systematic studies on the generation of Dirac
neutrino mass beyond the simple Yukawa interaction, which include both
tree-level and loop-level realizations, besides several newly-proposed
radiative Dirac neutrino mass models, which we will outline below.

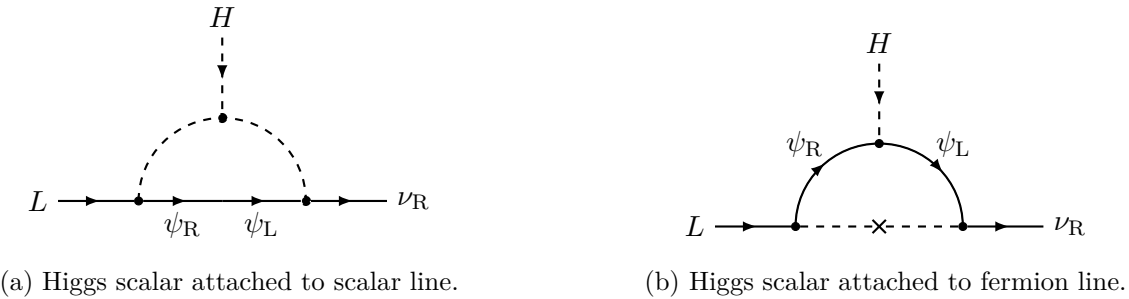
\begin{figure}[tb]\centering \begin{subfigure}[t]{0.45\linewidth} \centering
	\begin{tikzpicture}[node distance=1cm and 1cm]
		\coordinate[label=left:$L$] (nu1); \coordinate[vertex, right=of
			nu1] (v1); \draw[scalarnoarrow] (v1) arc[start
				angle=180, delta angle=-180, radius=1.1cm]
				node[pos=0.5,vertex] (v05) {} node[vertex] (v2)
				{}; \coordinate[right=of v2,
					label=right:$\nu_{\rm R}$] (nu2);
				\coordinate[above=of v05,label=above:$H$] (h);
				\coordinate (v3) at ($(v1)!0.5!(v2)$);
				\draw[scalar] (h) -- (v05); \draw[fermion]
					(nu1)--(v1); \draw[fermion]
						(v2)--(nu2); \draw[fermion]
							(v1)--(v3)
							node[midway,below]
							{$\psi_{\rm R}$};
					\draw[fermion] (v3)--(v2)
						node[midway,below] {$\psi_{\rm L}$};
			\end{tikzpicture} \caption{Higgs scalar attached to
		scalar line.} \label{fig:DiracOneLoopFermionicSoftMass}
	\end{subfigure} \hspace{1cm} \begin{subfigure}[t]{0.45\linewidth}
		\centering \begin{tikzpicture}[node distance=1cm and 1cm]
			\coordinate[label=left:$L$] (nu1); \coordinate[vertex,
				right=of nu1] (v1); \draw (v1) arc[start
					angle=180, delta angle=-180,
					radius=1.1cm] node[pos=0.5,vertex]
					(v05) {}  node[vertex] (v2) {};
				\draw[fermion] (v1) arc[start angle=180, delta
					angle=-90, radius=1.1cm] (v05)
					node[midway,above, xshift=-0.2cm] {$\psi_{\rm R}$};
				\draw[fermion] (v05) arc[start angle=90, delta
					angle=-90, radius=1.1cm] (v2)
					node[midway,above, xshift=0.2cm] {$\psi_{\rm L}$};
				\coordinate[right=of v2, label=right:$\nu_{\rm R}$]
					(nu2); \coordinate[above=of
						v05,label=above:$H$] (h);
					\coordinate[cross,black] (v3) at
						($(v1)!0.5!(v2)$);
					\draw[scalar] (h) -- (v05);
					\draw[fermion] (nu1)--(v1);
				\draw[fermion] (v2)--(nu2);
			\draw[scalarnoarrow] (v1)--(v3) node[midway,below] {};
		\draw[scalarnoarrow] (v3)--(v2) node[midway,below] {};
	\end{tikzpicture} \caption{Higgs scalar attached to fermion line.}
	\label{fig:DiracOneLoopScalarSoftMass} \end{subfigure}
\caption{Generation of the Dirac neutrino Yukawa coupling at 1-loop order.}
\label{fig:DiracOneLoop} \end{figure}

References~\cite{Ma:2016mwh,Wang:2016lve} performed a study of Dirac neutrino
mass according to topology at tree-level and 1-loop level. There are only two
possible one-particle-irreducible topologies for the Dirac Yukawa coupling at
1-loop, which are shown in Fig.~\ref{fig:DiracOneLoop}. The simplest radiative
Dirac neutrino mass models are based on a softly-broken $Z_2$ symmetry, which
is required to forbid the tree-level contribution, and generate the topologies in Fig.~\ref{fig:DiracOneLoop}. Reference~\cite{Wang:2017mcy} studied scotogenic-type models with a $U(1)_{{\rm
B}-{\rm L}}$ symmetry at 1- and 2-loop order. Finally, Ref.~\cite{Kanemura:2016ixx}
takes a model-independent approach and discusses the possible flavor
structures of the induced Dirac mass term under a number of constraints: The
fermion line only contains leptons and each lepton type can appear at most
once.

\textbf{1-loop Models:} Many of the proposed 1-loop models are realized in a
left-right symmetric context without SU(2) triplet
scalars~\cite{Cheng:1977ir,Mohapatra:1987hh,Mohapatra:1987nx,Balakrishna:1988bn,Gu:2007ug,Ma:2016mwh,Borah:2016hqn,Borah:2017leo}.
Reference~\cite{Rajpoot:1990gy} attempted the generation of Dirac neutrino
masses in the context of a model where hypercharge emerges as diagonal subgroup
of $U(1)_{\rm L}\times U(1)_{\rm R}$. To our knowledge the generation of Dirac neutrino
mass at 1-loop level with a softly broken $Z_2$ was first suggested in
Ref.~\cite{Kanemura:2011jj} based on topology \ref{fig:DiracOneLoopScalarSoftMass}. Reference~\cite{Farzan:2012sa} implements the
first scotogenic model of Dirac neutrino mass by using a dark $Z_2$ and
softly-broken $Z_2$. Both of these possibilities have been studied in more
detail in the systematic studies outlined above. Another way to explain the
smallness of Dirac neutrino mass is via a small loop-induced
VEV~\cite{Kanemura:2013qva}.  Finally, Ref.~\cite{Borah:2016zbd} discusses a
left-right symmetric model with pseudo-Dirac neutrinos. The tree-level Majorana
mass terms are not allowed, because the bidoublet is absent and the coupling of
the left-handed triplet to leptons is forbidden by a discrete symmetry.

\textbf{2-loop Models:} Two explicit models of Dirac neutrino mass have been
discussed in Refs.~\cite{Bonilla:2016diq,Kanemura:2017haa} apart from the
general classification~\cite{Wang:2017mcy}. They are both based on a $U(1)$
symmetry, a dark $U(1)$ and lepton number, respectively. The $U(1)$ symmetry is
broken to a discrete subgroup and thus both models feature a stable dark matter
candidate.

\textbf{3-loop Model:} Finally, a Dirac neutrino mass term can also be
induced via a global chiral anomaly term~\cite{Pilaftsis:2012hq}. The
five-dimensional anomaly term $a F_{\mu\nu} \tilde F^{\mu\nu}$  with the
pseudo-scalar $a$ and the (dual) field strength tensor $F_{\mu\nu}$ ($\tilde
F_{\mu\nu}$) is induced at 1-loop level and leads to a Dirac mass term at
2-loop order, being effectively a 3-loop contribution.


\subsection{331 models} \label{sec:331Models}
\begin{figure}[tb]\centering 
	\begin{tikzpicture}[node distance=1cm and 1cm]
		\coordinate[label=left:$\nu_{\rm L}$] (nu1); 
	\coordinate[vertex, right=of nu1] (v1); 
	\draw[photon] (v1) arc[start angle=180, delta angle=-180, radius=1.1cm] node[pos=0.5,vertex] (v05) {} node[vertex] (v2) {}; 
	\coordinate[right=of v2, label=right:$\nu_{\rm L}$] (nu2); 
	\coordinate[above left=of v05,label=above:$\braket{H_2}$] (h1);
	\coordinate[above right=of v05,label=above:$\braket{\chi}$] (h2);
	\coordinate[vertex] (v3) at ($(v1)!0.5!(v2)$);
	\coordinate[below=of v3,label=below:$\braket{H_1}$] (v03);
	\draw[scalar] (h1) -- (v05);
	\draw[scalar] (h2) -- (v05);
	\draw[scalar] (v03) -- (v3);
	\draw[fermion]	(nu1)--(v1);
	\draw[fermion] (nu2)--(v2);
	\draw[fermion] (v1)--(v3) node[midway,below] {$\nu_{\rm L}$};
	\draw[fermion] (v3)--(v2) node[midway,below] {$N^c$};
	\end{tikzpicture} 
	\caption{Neutrino mass generation from gauged lepton number violation.}
\label{fig:331gauge} \end{figure}
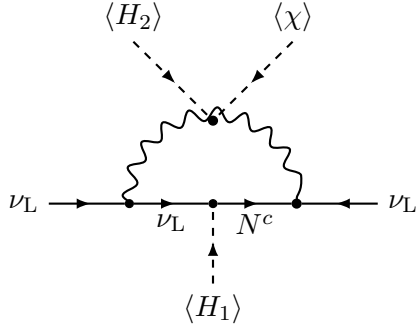

Another interesting class of models is based on the extended gauge group
$SU(3)_{\rm c}\times SU(3)_{\rm L}\times U(1)_{\rm X}$. The SM gauge group can be embedded in
several different ways and is determined by how the hypercharge generator is
related to the generator $T_8$ of $SU(3)_{\rm L}$ and the generator $X$ of $U(1)_{\rm X}$,
\begin{equation} Y= \beta\, T_8 + X\;, \end{equation} where $\beta$ is a
	continuous parameter. In addition to one radiative Dirac neutrino mass
	model~\cite{Gutierrez:2005rq}, several radiative Majorana neutrino mass
	models have been proposed at 1-loop
	level~\cite{Okamoto:1999cf,Kitabayashi:2000wh,Kitabayashi:2000nq,Kitabayashi:2001dg,Kitabayashi:2001jp,Kitabayashi:2001hx,Kitabayashi:2002jd,Chang:2003sx,Chang:2003rp,Ponce:2006au,Salazar:2007ym,Dong:2008ya,Boucenna:2014ela,Boucenna:2014dia,Deppisch:2016jzl,Okada:2015bxa,CarcamoHernandez:2017kra},
	2-loop
	order~\cite{Kitabayashi:2001id,Aizawa:2004qf,Sanchez:2004at,Chang:2006aa,Benavides:2010zw,CarcamoHernandez:2017cwi}, 3-loop order~\cite{Hernandez:2015hrt},
	and even at 4-loop order~\cite{CarcamoHernandez:2016pdu}. As lepton
	number violation (LNV) in 331 models and in particular neutrino mass
	generation has been discussed in a recent
	review~\cite{Fonseca:2016xsy}, we refer the interested reader to it for
	a detailed discussion. However, we highlight one model based on gauged
	lepton number
	violation~\cite{Boucenna:2014ela,Boucenna:2014dia,Deppisch:2016jzl},
	which generates neutrino mass via lepton number violation in the 1-loop
	diagram shown in Fig.~\ref{fig:331gauge} with the $SU(3)_{\rm L}\times
	U(1)_{\rm X}$ gauge bosons, where $H_i$ denotes the SM Higgs doublets,
	$\braket{\chi}$ the VEV in the third component of $SU(3)_{\rm L}$ and $N^c$
	the third partner of $\nu_{\rm L}$ in the triplet of $SU(3)_{\rm L}$. Note that
	lepton number is broken by the mixing of the gauge bosons in the vertex
	at the top of the diagram.


\section{Phenomenology}
\label{sec:pheno}

In this section we revisit the most relevant phenomenological implications of radiative neutrino mass models. The possible signals are very model dependent, as each radiative model has its own particularities that should be studied on a case-by-case basis. However, in the following we will try to discuss generic predictions of these models, making use of simplified scenarios and/or of effective operators, and referring to particular examples when necessary.


\subsection{Universality violations and non-standard interactions} \label{sec:univ_NSI}
In the SM, leptonic decays mediated by gauge interactions are universal. Several scenarios of physics beyond the SM have universality violations, that is, decays into different families (up to phase space-factors) are no longer identical\footnote{Higher order effects break universality in a tiny amount due to Higgs interactions, i.e, by the charged lepton Yukawa couplings.}. These may or may not be related to neutrino masses, as lepton number is not violated in these interactions. Indeed, for instance a two Higgs doublet model with general Yukawa interactions breaks universality, irrespective of neutrino masses. In tree-level neutrino mass models, there are also violations of universality, mediated by the (singly) charged scalar boson in the type-II seesaw model, or due to the non-unitarity of the leptonic mixing matrix in type-I and type-III seesaw models when the extra neutral fermions are heavy~\cite{Antusch:2006vwa,Abada:2007ux}.

In some of the radiative models there can be violations of universality. One illustrative example of this case is due to the presence of a singly-charged singlet $h$ with mass $m_{h^+}$ (as in the Zee and Zee-Babu models, see Sec.~\ref{sec:modelleptons}). The relevant interaction is $\overline{\tilde{L}}f\,L h^{+}$, where $f$ is an antisymmetric matrix in flavor space
and $\tilde L \equiv i\tau_2 C L^c = i\tau_2 C \bar L^T$. Integrating out the singlet, one obtains the following dimension 6 effective operator~\cite{Bilenky:1993bt}
\begin{equation}\label{eq:singlet_effective}
\mathcal{L}_{\rm eff} \subset\, \dfrac{1}{m_{h^+}^2} (\overline{e_{\rm L}}\,f^\dagger\,  \nu^c_{\rm L})(\overline{\nu_{\rm L}^c}\,f\,  e_{\rm L})\,.
\end{equation}
One can see that this operator involves left-handed leptons, like charged currents in the SM.\footnote{In models with an extra Higgs doublet coupled to the leptons, other operators can be formed by integrating out the second Higgs doublet. In those cases, the electrons involved are right-handed and therefore there is no interference with the $W$ boson. An example is the Zee model, see Ref.~\cite{Herrero-Garcia:2017xdu}.} This implies that it interferes constructively with the $W$ boson, modifying among others the muon decay rate~\cite{Nebot:2007bc}. Therefore, the Fermi constant which is extracted from muon decay in the SM, $G_\mu^{\rm SM}$, and that in a model with a singly-charged singlet, $G_{\mu}^{\rm h}$, are different, i.e., $G_{\mu}^{\rm SM} \neq G_{\mu}^{\rm h}$. Their ratio obeys to leading order in $f$:
\begin{equation}
\left(\dfrac{G^{\rm h}_\mu}{G_\mu^{\rm SM}}\right)^2 \simeq 1 + \dfrac{\sqrt{2}}{G_F m_{h^+}^2} |f^{e\mu}|^2\,.
\end{equation}
The \emph{new} Fermi constant $G_{\mu}^{\rm h}$ is subject to different constraints. For example, from measurements of the unitarity of the CKM matrix, as the Fermi constant extracted from hadronic decays should be equivalent to that from leptonic decays, we can bound $f_{e\mu}$:
\begin{equation}
|V^{\rm exp}_{ud}|^2 + |V^{\rm exp}_{us}|^2 + |V^{\rm exp}_{ub}|^2 = \left(\dfrac{G_\mu^{\rm SM}}{G^{\rm h}_{\mu}}\right)^2 = 1 - \dfrac{\sqrt{2}}{G_F m_{h^+}^2}  |f^{e\mu}|^2\,.
\end{equation}
Also leptonic decays which in the SM are mediated by charged-current interactions are not universal anymore. The ratio of leptonic decays among the different generations can be tested via the effective couplings given by
\begin{align}
	\left(\dfrac{G^{\rm h}_{\tau \to e}}{G^{\rm h}_{\mu \to e}} \right)^2 & \approx 1 + \dfrac{\sqrt{2}}{G_F m_{h^+}^2} (|f^{e\tau}|^2-|f^{e \mu}|^2)\,, \label{eq:gtm}\\
	\left(\dfrac{G^{\rm h}_{\tau \to \mu}}{G^{\rm h}_{\mu \to e}}\right)^2  &\approx 1 + \dfrac{\sqrt{2}}{G_F m_{h^+}^2} (|f^{\mu \tau}|^2-|f^{e \mu}|^2)\,,\label{eq:gte}\\
	\left(\dfrac{G^{\rm h}_{\tau \to \mu}}{G^{\rm h}_{\tau \to e}}\right)^2 &\approx 1 + \dfrac{\sqrt{2}}{G_F m_{h^+}^2} \left(|f^{\mu \tau}|^2 - |f^{e \tau}|^2\right)\,. \label{eq:gme}
\end{align}
All these lead to strong limits on the $f$ couplings depending on the mass on the singlet~\cite{Herrero-Garcia:2014hfa}.

Furthermore, the new singly-charged scalar via the effective operator in Eq.~\ref{eq:singlet_effective} induces neutrino interactions that cannot be described by $W$-boson exchange and are termed non-standard neutrino interactions (NSIs). Equation~\ref{eq:singlet_effective} is usually rewritten after a Fierz identity as
\begin{equation}
	\mathcal{L}_{d=6}^{\rm NSI} = 2\sqrt{2} G_F\, \varepsilon^{\rho \sigma}_{\alpha \beta} \,(\overline{\nu_\alpha} \gamma^\mu {\rm P_L} \nu_\beta)\, (\overline{e_\rho} \gamma_\mu {\rm P_L} e_\sigma)\,, 
\end{equation}
where $\varepsilon^{\rho \sigma}_{\alpha \beta}$ are the NSI parameters given by
\begin{equation}
\varepsilon^{\rho \sigma}_{\alpha \beta} = \dfrac{f_{\sigma \beta} (f_{\rho \alpha})^*}{\sqrt{2} G_F m_{h^+}^2} \,.
\label{eq:NSIdef}
\end{equation}
These could be in principle probed at neutrino oscillation experiments.
However, typically whenever NSIs are induced, lepton flavor violating (LFV)
processes are also generated, which are subject to stronger constraints. This
is particularly the case for the four-lepton dimension 6 operators, due to
gauge invariance. Models with large NSI are difficult to construct, and
typically involve light mediators~\cite{Farzan:2015doa,Farzan:2015hkd}. We
refer the reader to Refs.~\cite{Davidson:2003ha,Ibarra:2004pe,Gavela:2008ra,Biggio:2009nt,Biggio:2009kv,Antusch:2008tz,Ohlsson:2012kf}
for studies of NSIs and their theoretical constraints.


\subsection{Lepton flavor violation}
\label{subsec:LFV}

One of the common predictions shared by most neutrino mass models
(radiative or not) is the existence of LFV processes
involving charged leptons with observable rates in some cases. Indeed, neutrino
oscillations imply that lepton flavors are violated in neutrino
interactions, and as in the SM neutrinos come in $SU(2)$ doublets
together with the charged leptons, also violations of lepton flavors
involving the latter are expected.  Which is the most constraining
LFV observable is, however, a model-dependent question. It is thus
convenient to use a parametrization that allows for a
model-independent description of these processes. For each of the
models one can then compute the relevant coefficients and apply the
following formalism.
We follow the notation and conventions of Ref.~\cite{Porod:2014xia}.\footnote{See Refs.~\cite{Lee:1977qz,Lee:1977tib,Marciano:1977wx} for pioneering work on LFV processes.}

The general LFV Lagrangian can be written as
\begin{equation} \label{eq:L-LFV}
\mathcal{L}_{\text{LFV}} = \mathcal{L}_{\ell \ell \gamma} + \mathcal{L}_{\ell \ell Z} 
+ \mathcal{L}_{\ell \ell h} + \mathcal{L}_{4 \ell} + \mathcal{L}_{2 \ell 2q} \, .
\end{equation}
The first term contains the $\ell - \ell - \gamma$ interaction
Lagrangian, given by
\begin{equation} \label{eq:L-llg}
	\mathcal{L}_{\ell \ell \gamma} = e \, \bar \ell_\beta \left[ \gamma^\mu
	\left(K_1^L {\rm P_L} + K_1^R {\rm P_R} \right) + i m_{\ell_\alpha}
\sigma^{\mu \nu} q_\nu \left(K_2^L {\rm P_L} + K_2^R {\rm P_R} \right) \right] \ell_\alpha
A_\mu + {\rm H.c.} \, ,
\end{equation}
where $e$ is the electric charge, $q$ is the photon momentum, ${\rm P_{L,R}}
= \frac{1}{2} (1 \mp \gamma_5)$ are the standard chirality projectors
and the indices $\{\alpha,\beta\}$ denote the lepton flavors. The
first term in Eq.~\ref{eq:L-llg} corresponds to the monopole
interaction between a photon and a pair of leptons whereas the second
is a dipole interaction term. In this parametrization the form factors $K_1^{L,R}$ vanish when the photon is on-shell, i.e.~in the limit of $q^2\to0$. Similarly, the interaction Lagrangians
with the $Z$ and Higgs bosons are given by\footnote{Note the different choice of Lorentz structures in Eqs.~\ref{eq:L-llg} and \ref{eq:L-llZ}. The two forms can be related via the Gordon-identity.}
\begin{equation} \label{eq:L-llZ}
	\mathcal{L}_{\ell \ell Z} = \bar \ell_\beta \left[ \gamma^\mu \left(R_1^L {\rm P_L} + R_1^R {\rm P_R} \right) + p^\mu \left(R_2^L {\rm P_L} + R_2^R {\rm P_R} \right) \right] \ell_\alpha Z_\mu \, ,
\end{equation}
where $p$ is the $\ell_\beta$ 4-momentum, and
\begin{equation} \label{eq:L-llh}
	\mathcal{L}_{\ell \ell h} = \bar \ell_\beta \left(S_L {\rm P_L} + S_R {\rm P_R} \right) \ell_\alpha \, h 
\end{equation}
with the SM Higgs $h$. The general 4-lepton interaction Lagrangian can be written
as
\begin{equation}
	\mathcal{L}_{4 \ell} = \sum_{\substack{I={\rm S,V,T}\\X,Y={\rm L,R}}} A_{XY}^I \bar \ell_\beta \Gamma_I {\rm P}_X \ell_\alpha \bar \ell_\delta \Gamma_I {\rm P}_Y \ell_\gamma + {\rm H.c.} \, , \label{eq:L-4L}
\end{equation}
where in this case the indices $\{\alpha,\beta,\gamma,\delta\}$
denote the lepton flavors and we have defined $\Gamma_{\rm S} = 1$,
$\Gamma_{\rm V} = \gamma_\mu$ and $\Gamma_{\rm T} = \sigma_{\mu \nu}$. It is clear
that the Lagrangian in Eq.~\ref{eq:L-4L} contains all possible terms
allowed by Lorentz invariance. Finally, the general $2 \ell 2 q$
4-fermion interaction Lagrangian (at the quark level) can be split in
two pieces
\begin{equation}
\mathcal{L}_{2 \ell 2q} = \mathcal{L}_{2 \ell 2d} + \mathcal{L}_{2 \ell 2u} \, , \label{eq:L-2L2Q}
\end{equation}
where
\begin{eqnarray}
\mathcal{L}_{2 \ell 2d} = 
&& \sum_{\substack{I={\rm S,V,T}\\X,Y={\rm L,R}}} B_{XY}^I \bar \ell_\beta \Gamma_I {\rm P}_X \ell_\alpha \bar d_\gamma \Gamma_I {\rm P}_Y d_\gamma + {\rm H.c.}  \, , \label{eq:L-2L2D} \\
\mathcal{L}_{2 \ell 2u} = && \left. \mathcal{L}_{2 \ell 2d} \right|_{d \to u, \, B \to C} \label{eq:L-2L2U} \, .
\end{eqnarray}
Here $\gamma$ denotes the $d$-quark flavor and we are neglecting the
possibility of quark flavor violation, which is beyond the scope of
this review.~\footnote{Reference~\cite{Carpentier:2010ue} provides a comprehensive collection of constraints on quark flavor violating operators.}

The parametrization used implies that the operators appearing in Eqs.~\ref{eq:L-4L}, \ref{eq:L-2L2D} and
\ref{eq:L-2L2U} have canonical dimension six. Therefore, the Wilson
coefficients $A_{XY}^I$, $B_{XY}^I$ and $C_{XY}^I$ scale as
$1/\Lambda^2$, where $\Lambda$ is the new physics energy scale at
which they are generated. 
Note this scale is unrelated to the scale at which lepton number is violated. 
The same comment applies to the dipole
coefficients $K_2^{L,R}$ in Eq.~\ref{eq:L-llg}. In contrast, the
rest of the coefficients discussed in this section, $K_1^{L,R}$,
$R_{1,2}^{L,R}$ and $S_{L,R}$, are dimensionless (although their leading
new physics contribution appears at order $v^2/\Lambda^2$). If we
restrict the discussion to flavor violating coefficients, they all
vanish in the SM. Therefore, they encode the effects induced by the
new degrees of freedom present in specific models.

It should be noted that all operators in the general LFV Lagrangian in
Eqs.~\ref{eq:L-llg}-\ref{eq:L-2L2U} break gauge invariance. For
instance, they contain new charged lepton interactions, but not the
analogous new interactions for the neutrinos, their $SU(2)_{\rm L}$ doublet
partners which are partly discussed in the previous subsection. This type of parametrization of LFV effects is correct at
energies below the electroweak symmetry breaking scale, but it may
miss relevant correlations between operators that are connected by
gauge invariance in the underlying new physics theory. See for instance
Ref.~\cite{Pruna:2014asa} for a discussion of LFV in terms of
gauge-invariant operators.

We now proceed to discuss the LFV processes with the most promising
experimental perspectives in the near future. We will provide simple
analytical expressions in terms of the coefficients of the general LFV
Lagrangian and highlight some radiative neutrino mass models with
specific features leading to non-standard expectations for these
processes. By no means this will cover all the models constrained by
these processes, but will serve as a review of the novel LFV scenarios
in radiative neutrino mass models.

Note, however, that there are other processes, which may yield
stringent constraints in particular models: for instance in models
with leptoquarks, the latter can mediate semi-leptonic $\tau$-decays
and leptonic meson decays at tree level. The LFV decays $Z \to
\ell_\alpha \bar \ell_\beta$ have also been investigated in several
radiative models, although they typically have very low rates, see for
instance \cite{Ghosal:2001ep,Li:2016dou}.

\subsubsection{$\ell_\alpha \rightarrow \ell_\beta \gamma$}

The most popular LFV process is $\ell_\alpha \rightarrow \ell_\beta
\gamma$. There are basically two reasons for this: (1) for many years,
the experiments looking for the radiative process $\mu \to e \gamma$
have been leading the experimental developments, with the publication
of increasingly tighter bounds, and (2) in many models of interest these
are the processes where one expects the highest rates. In fact, many
phenomenological studies have completely focused on these decays,
neglecting other LFV processes that may also be relevant.

\begin{table}[tb!]
\centering
\begin{tabular}{ccc}
\toprule
LFV Process BR & Present Bound & Future Sensitivity  \\
\midrule
    $\mu \to  e \gamma$ & $4.2\times 10^{-13}$~\cite{TheMEG:2016wtm}  & $6\times 10^{-14}$~\cite{Baldini:2013ke} \\
    $\tau \to e \gamma$ & $3.3 \times 10^{-8}$~\cite{Aubert:2009ag}& $ \sim3\times10^{-9}$~\cite{Aushev:2010bq}\\
    $\tau \to \mu \gamma$ & $4.4 \times 10^{-8}$~\cite{Aubert:2009ag}& $ \sim3\times10^{-9}$~\cite{Aushev:2010bq} \\
\bottomrule
\end{tabular}
\caption{Current experimental bounds and future sensitivities for $\ell_\alpha \rightarrow \ell_\beta \gamma$ branching ratios.}
\label{tab:llgamma}
\end{table}

The experimental situation in radiative LFV decays is summarized in
Tab.~\ref{tab:llgamma}. As one can easily see in this table, muon
observables have the best experimental limits. This is due to the
existing high-intensity muon beams. The current limit for the $\mu \to
e \gamma$ branching ratio has been obtained by the MEG experiment,
$\text{BR}(\mu \to e \gamma) < 4.2 \cdot 10^{-13}$
\cite{TheMEG:2016wtm}, slightly improving the previous bound also
obtained by the same collaboration. This bound is expected to be
improved by about one order of magnitude in the MEG-II
upgrade~\cite{Baldini:2013ke}. The bounds in $\tau$ decays are weaker,
 with the branching ratios bounded to be below $\sim
10^{-8}$, and some improvements are expected as well in future
B-factories~\cite{Aushev:2010bq}.

The decay width for $\ell_\alpha \rightarrow \ell_\beta \gamma$ is
given by~\cite{Hisano:1995cp}
\begin{equation}
\Gamma \left( \ell_\alpha \to \ell_\beta \gamma \right) = \frac{\alpha m_{\ell_\alpha}^5}{4} \left( |K_2^L|^2 + |K_2^R|^2 \right) \, ,
\end{equation}
where $\alpha$ is the fine structure constant. Only the dipole
coefficients $K_2^{L,R}$, defined in Eq.~\ref{eq:L-llg}, contribute to
this process. General expressions for these coefficients can be found
in Ref. \cite{Lavoura:2003xp}.

The $\mu \to e \gamma$ limit is typically the most constraining one in
most radiative neutrino mass models. One can usually evade it by
adopting specific Yukawa textures that reduce the $\mu-e$
flavor-violating entries (see for example Ref.~\cite{Schmidt:2012yg})
or simply by globally reducing the Yukawa couplings by increasing the
new physics scale. However, in some cases this is not possible. A
simple example of such situation is the scotogenic
model~\cite{Ma:2006km} with a fermionic dark matter candidate. The
singlet fermions in the scotogenic model only couple to the SM
particles via the Yukawa couplings. Therefore, these Yukawa couplings
must be sizable in order to thermally produce singlet fermions in the
early universe in sufficient amounts so as to reproduce the observed
DM relic density. This leads to some tension between the DM relic
density requirement and the current bounds on LFV processes, although
viable regions of the parameter space still exist
\cite{Schmidt:2012yg,Vicente:2014wga}. 
In contrast, in other radiative models the tight
connection between neutrino masses and LFV implies suppressed
$\ell_\alpha \to \ell_\beta \gamma$ rates. This is the case of
bilinear R-parity violating models
\cite{Cheung:2001sb,Abada:2001zh,Carvalho:2002bq}, see
Sec.~\ref{sec:RPV} for a detailed discussion of this type of
supersymmetric neutrino mass models.

\subsubsection{$\ell_\alpha \rightarrow \ell_\beta \ell_\delta \ell_\delta$}

We now consider the $\ell_\alpha \rightarrow \ell_\beta \ell_\delta
\ell_\delta$ 3-body decays. One can distinguish three categories:
$\ell_\alpha \rightarrow \ell_\beta \overline{\ell_\beta} \ell_\beta$,
$\ell_\alpha \rightarrow \ell_\beta \overline{\ell_\delta}
\ell_\delta$ (with $\beta \ne \delta$) and $\ell_\alpha \rightarrow
\overline{\ell_\beta} \ell_\delta \ell_\delta$ (also with $\beta \ne
\delta$). These processes have received less attention even though the
experimental limits on their branching ratios are of the same order as
for the analogous $\ell_\alpha \rightarrow \ell_\beta \gamma$
decays. We summarize the current experimental bounds and future
sensitivities for the $\ell_\alpha \rightarrow \ell_\beta
\ell_\delta \ell_\delta$ 3-body decays in Tab.~\ref{tab:l3l}. We note that an impressive improvement of four orders
of magnitude is expected in the $\mu \rightarrow e e e$ branching
ratio sensitivity thanks to the Mu3e experiment at
PSI~\cite{Blondel:2013ia}.

\begin{table}[tb!]
\centering
\begin{tabular}{ccc}
\toprule
LFV Process BR & Present Bound & Future Sensitivity  \\
\midrule
    $\mu \rightarrow e e e$ &  $1.0 \times 10^{-12}$~\cite{Bellgardt:1987du} &  $\sim10^{-16}$~\cite{Blondel:2013ia} \\
    $\tau \rightarrow e e e$ & $2.7\times10^{-8}$~\cite{Hayasaka:2010np} &  $\sim 10^{-9}$~\cite{Aushev:2010bq} \\
    $\tau \rightarrow \mu \mu \mu$ & $2.1\times10^{-8}$~\cite{Hayasaka:2010np} & $\sim 10^{-9}$~\cite{Aushev:2010bq} \\
    $\tau^- \rightarrow e^- \mu^+ \mu^-$ &  $2.7\times10^{-8}$~\cite{Hayasaka:2010np} & $\sim 10^{-9}$~\cite{Aushev:2010bq} \\
    $\tau^- \rightarrow \mu^- e^+ e^-$ &  $1.8\times10^{-8}$~\cite{Hayasaka:2010np} & $\sim 10^{-9}$~\cite{Aushev:2010bq} \\    
$\tau^- \rightarrow e^+ \mu^- \mu^-$ &  $1.7\times10^{-8}$~\cite{Hayasaka:2010np} & $\sim 10^{-9}$~\cite{Aushev:2010bq} \\
$\tau^- \rightarrow \mu^+ e^- e^-$ &  $1.5\times10^{-8}$~\cite{Hayasaka:2010np} & $\sim 10^{-9}$~\cite{Aushev:2010bq} \\
\bottomrule
\end{tabular}
\caption{Current experimental bounds and future sensitivities for
  $\ell_\alpha \rightarrow \ell_\beta \ell_\delta \ell_\delta$
  branching ratios.}
\label{tab:l3l}
\end{table}

The $\ell_\alpha \rightarrow \ell_\beta \ell_\delta \ell_\delta$ decay
width receives contributions from several operators of the general LFV
Lagrangian. In the case of the first category, $\ell_\alpha
\rightarrow \ell_\beta \overline{\ell_\beta} \ell_\beta$, the decay
width is given by~\cite{Porod:2014xia}
\begin{eqnarray}
\Gamma \left( \ell_\alpha \to \ell_\beta \overline{\ell_\beta} \ell_\beta \right) &=& \frac{m_{\ell_\alpha}^5}{512 \pi^3} \left[ e^4 \, \left( \left| K_2^L \right|^2 + \left| K_2^R \right|^2 \right) \left( \frac{16}{3} \ln{\frac{m_{\ell_\alpha}}{m_{\ell_\beta}}} - \frac{22}{3} \right) \right. \label{eq:L3Lwidth} \\
&+& \frac{1}{24} \left( \left| A_{LL}^S \right|^2 + \left| A_{RR}^S \right|^2 \right) + \frac{1}{12} \left( \left| A_{LR}^S \right|^2 + \left| A_{RL}^S \right|^2 \right) \nonumber \\
&+& \frac{2}{3} \left( \left| \hat A_{LL}^V \right|^2 + \left| \hat A_{RR}^V \right|^2 \right) + \frac{1}{3} \left( \left| \hat A_{LR}^V \right|^2 + \left| \hat A_{RL}^V \right|^2 \right) + 6 \left( \left| A_{LL}^T \right|^2 + \left| A_{RT}^T \right|^2 \right) \nonumber \\
&+& \frac{e^2}{3} \left( K_2^L A_{RL}^{S \ast} + K_2^R A_{LR}^{S \ast} + \cc \right) - \frac{2 e^2}{3} \left( K_2^L \hat A_{RL}^{V \ast} + K_2^R \hat A_{LR}^{V \ast} + \cc \right) \nonumber \\
&-& \frac{4 e^2}{3} \left( K_2^L \hat A_{RR}^{V \ast} + K_2^R \hat A_{LL}^{V \ast} + \cc \right) \nonumber \\
&-& \left. \frac{1}{2} \left( A_{LL}^S A_{LL}^{T \ast} + A_{RR}^S A_{RR}^{T \ast} + \cc \right) - \frac{1}{6} \left( A_{LR}^S \hat A_{LR}^{V \ast} + A_{RL}^S \hat A_{RL}^{V \ast} + \cc \right) \right]  \nonumber \, ,
\end{eqnarray}
in case of the second category, $\ell_\alpha \rightarrow \ell_\beta
\overline{\ell_\delta} \ell_\delta$ (with $\beta \ne \delta$), the
expression is given by~\cite{Abada:2014kba}
\begin{eqnarray}
\Gamma \left( \ell_\alpha \rightarrow \ell_\beta \overline{\ell_\delta}
\ell_\delta \right) &=& \frac{m_{\ell_\alpha}^5}{512 \pi^3} \left[ e^4 \, \left( \left| K_2^L \right|^2 + \left| K_2^R \right|^2 \right) \left( \frac{16}{3} \ln{\frac{m_{\ell_\alpha}}{m_{\ell_\gamma}}} - 8 \right) \right. \label{eq:L3Lwidth2} \\
&+& \frac{1}{12} \left( \left| A_{LL}^S \right|^2 + \left| A_{RR}^S \right|^2 \right) + \frac{1}{12} \left( \left| A_{LR}^S \right|^2 + \left| A_{RL}^S \right|^2 \right) \nonumber \\
&+& \frac{1}{3} \left( \left| \hat A_{LL}^V \right|^2 + \left| \hat A_{RR}^V \right|^2 \right) + \frac{1}{3} \left( \left| \hat A_{LR}^V \right|^2 + \left| \hat A_{RL}^V \right|^2 \right) + 4 \left( \left| A_{LL}^T \right|^2 + \left| A_{RR}^T \right|^2 \right) \nonumber \\
&-& \left. \frac{2 e^2}{3} \left( K_2^L \hat A_{RL}^{V \ast} + K_2^R \hat A_{LR}^{V \ast} + K_2^L \hat A_{RR}^{V \ast} + K_2^R \hat A_{LL}^{V \ast} + \cc \right) \right]  \nonumber \, ,
\end{eqnarray}
whereas for the third category, $\ell_\alpha \rightarrow
\overline{\ell_\beta} \ell_\delta \ell_\delta$ (with $\beta \ne
\delta$), the decay width is given by~\cite{Abada:2014kba}
\begin{eqnarray}
\Gamma \left( \ell_\alpha \rightarrow \overline{\ell_\beta} \ell_\delta
\ell_\delta \right) &=& \frac{m_{\ell_\alpha}^5}{512 \pi^3} \left[ \frac{1}{24} \left( \left| A_{LL}^S \right|^2 + \left| A_{RR}^S \right|^2 \right) + \frac{1}{12} \left( \left| A_{LR}^S \right|^2 + \left| A_{RL}^S \right|^2 \right)  \right. \label{eq:L3Lwidth3} \\
&+& \frac{2}{3} \left( \left| \hat A_{LL}^V \right|^2 + \left| \hat A_{RR}^V \right|^2 \right) + \frac{1}{3} \left( \left| \hat A_{LR}^V \right|^2 + \left| \hat A_{RL}^V \right|^2 \right) + 6 \left( \left| A_{LL}^T \right|^2 + \left| A_{RR}^T \right|^2 \right) \nonumber \\
&-& \left. \frac{1}{2} \left( A_{LL}^S A_{LL}^{T \ast} + A_{RR}^S A_{RR}^{T \ast} + \cc \right) - \frac{1}{6} \left( A_{LR}^S \hat A_{LR}^{V \ast} + A_{RL}^S \hat A_{RL}^{V \ast} + \cc \right) \right]  \nonumber \, .
\end{eqnarray}
Here we have defined
\begin{equation}
\hat A_{XY}^V = A_{XY}^V + e^2 K_1^X \qquad \left( X,Y = L,R \right) \, .
\end{equation}
The masses of the leptons in the final state have been neglected in
Eqs.~\ref{eq:L3Lwidth}, \ref{eq:L3Lwidth2} and
\ref{eq:L3Lwidth3}, with the exception of the contributions given by
the dipole coefficients $K_2^{L,R}$, where infrared divergences would
otherwise occur.

The dipole coefficients $K_2^{L,R}$, which contribute to $\ell_\alpha
\rightarrow \ell_\beta \gamma$, also contribute $\ell_\alpha
\rightarrow \ell_\beta \ell_\delta \ell_\delta$. It is easy to see
how: the Feynman diagram contributing to $\ell_\alpha \rightarrow
\ell_\beta \gamma$ can always be supplemented with a flavor-conserving
$\ell_\delta-\ell_\delta-\gamma$ additional vertex resulting in a
diagram contributing to $\ell_\alpha \rightarrow \ell_\beta
\ell_\delta \ell_\delta$.\footnote{We clarify that this is only true
  for the processes $\ell_\alpha \rightarrow \ell_\beta
  \overline{\ell_\beta} \ell_\beta$ and $\ell_\alpha \rightarrow
  \ell_\beta \overline{\ell_\delta} \ell_\delta$ (with $\beta \ne
  \delta$). The process $\ell_\alpha \rightarrow \overline{\ell_\beta}
  \ell_\delta \ell_\delta$ (with $\beta \ne \delta$) does not receive
  contributions from penguin diagrams, but only from boxes.} In fact,
such diagrams have been shown to be dominant in many models, the most
popular example being the Minimal Supersymmetric Standard Model (MSSM). In
this case, known as \textit{dipole dominance scenario}, a simple
proportionality between the decays widths of both LFV decays can be
established. For example, in the $\beta = \delta$ case this
proportionality leads to
\begin{equation}
\text{BR}(\ell_\alpha \rightarrow \ell_\beta \overline{\ell_\beta} \ell_\beta)
\simeq \frac{\alpha}{3 \pi}
\left(\ln\left(\frac{m^2_{\alpha}}{m^2_{\beta}}\right) - \frac{11}{4} \right)
\text{BR}(\ell_\alpha \rightarrow \ell_\beta \gamma) \, ,
\end{equation}
which implies $\text{BR}\left(\ell_\alpha \rightarrow \ell_\beta
\overline{\ell_\beta} \ell_\beta \right) \ll \text{BR}
\left(\ell_\alpha \rightarrow \ell_\beta \gamma\right)$, making the
radiative decay the most constraining process.

The dipole dominance assumption is present in many works discussing
LFV phenomenology. However, it can be easily broken in many radiative
neutrino mass models. This can happen in two ways:\footnote{In some
  models, cancellations due to certain Yukawa textures can affect some
  decays (like $\mu \to e \gamma$), but it is virtually impossible to
  cancel all radiative decays simultaneously.}
\begin{itemize}

\item {\bf Due to tree-level LFV:} In many radiative neutrino mass
  models the 4-lepton operators receive contributions at tree-level.
  The most prominent example of such models is the Zee-Babu model, in
  which the doubly-charged scalar $k^{++}$ mediates unsuppressed
  $\ell_\alpha \rightarrow \ell_\beta \overline{\ell_\beta}
  \ell_\beta$ decays. In such case one can easily find regions of
  parameter space where $\text{BR}\left(\ell_\alpha \rightarrow
  \ell_\beta \overline{\ell_\beta} \ell_\beta \right) \gg \text{BR}
  \left(\ell_\alpha \rightarrow \ell_\beta \gamma\right)$, see
  Ref.~\cite{Herrero-Garcia:2014hfa} for a recent study.

\item {\bf Due to loop-level LFV:} References~\cite{Kubo:2006yx,
  Sierra:2008wj, Suematsu:2009ww, Adulpravitchai:2009gi} explored the
  LFV phenomenology of the scotogenic model but only considered $\mu
  \to e \gamma$. However, this assumption has been shown to be valid
  only in some regions of the parameter space. In fact, box diagrams
  contributing to 4-lepton coefficients can actually dominate,
  dramatically affecting the phenomenology of the scotogenic model
  \cite{Toma:2013zsa,Vicente:2014wga}. Qualitatively similar results
  have been found in other variants of the scotogenic model
  \cite{Chowdhury:2015sla,Rocha-Moran:2016enp}.\footnote{Interestingly,
    the authors of \cite{Chowdhury:2015sla} have shown that in
    variants of the scotogenic model with higher $SU(2)$
    representations the LFV rates become larger due to additive
    effects from the components of the large multiplets.} In fact,
  this feature is not specific of the scotogenic model and its
  variants: one can find other radiative neutrino mass models with
  loop contributions dominating over the dipole. For instance,
  $Z$-penguin contributions have been found to be dominant in the angelic model~\cite{Angel:2012ug} and R$\nu$MDM models \cite{Cai:2016jrl}.

\end{itemize}

This clearly shows that radiative neutrino mass models typically have
a very rich LFV phenomenology with new (sometimes unexpected) patterns
and correlations.

\subsubsection{$\mu-e$ conversion}

The most spectacular improvements in the search for LFV are expected
in $\mu-e$ conversion experiments. Several projects will begin their
operation in the near future, with sensitivities that improve the
current bounds by several orders of magnitude. The experimental situation is shown in Tab.~\ref{tab:muec}.

\begin{table}[tb!]
\centering
\begin{tabular}{ccc}
\toprule
LFV Process CR & Present Bound & Future Sensitivity  \\
\midrule
    $\mu^-, \mathrm{Ti} \rightarrow e^-, \mathrm{Ti}$ &  $4.3\times 10^{-12}$~\cite{Dohmen:1993mp} & $\sim10^{-18}$~\cite{PRIME} \\
    $\mu^-, \mathrm{Au} \rightarrow e^-, \mathrm{Au}$ & $7\times 10^{-13}$~\cite{Bertl:2006up} & \\
    $\mu^-, \mathrm{Al} \rightarrow e^-, \mathrm{Al}$ &  & $10^{-15}-10^{-18}$~\cite{Pezzullo:2017iqq} \\
    $\mu^-, \mathrm{SiC} \rightarrow e^-, \mathrm{SiC}$ &  & $10^{-14}$~\cite{Natori:2014yba} \\
\bottomrule
\end{tabular}
\caption{Current experimental bounds and future sensitivities for $\mu-e$ conversion in nuclei.}
\label{tab:muec}
\end{table}

The conversion rate, normalized to the the muon capture rate
$\Gamma_{\rm capt}$, is given by~\cite{Kuno:1999jp,Arganda:2007jw}
\begin{align}
{\rm CR} (\mu- e, {\rm Nucleus}) &= 
\frac{p_e \, E_e \, m_\mu^3 \, G_F^2 \, \alpha^3 
\, Z_{\rm eff}^4 \, F_p^2}{8 \, \pi^2 \, Z \, \Gamma_{\rm capt}}  \nonumber \\
&\times \left\{ \left| (Z + N) \left( g_{LV}^{(0)} + g_{LS}^{(0)} \right) + 
(Z - N) \left( g_{LV}^{(1)} + g_{LS}^{(1)} \right) \right|^2 + 
\right. \nonumber \\
& \ \ \ 
 \ \left. \,\, \left| (Z + N) \left( g_{RV}^{(0)} + g_{RS}^{(0)} \right) + 
(Z - N) \left( g_{RV}^{(1)} + g_{RS}^{(1)} \right) \right|^2 \right\} \,.
\end{align}   
$Z$ and $N$ are the number of protons and neutrons in the nucleus and
$Z_{\rm eff}$ is the effective atomic charge~\cite{Chiang:1993xz}.
$G_F$ is the Fermi constant, $\alpha$ is the electromagnetic fine
structure constant, $p_e$ and $E_e$ are the momentum and energy of the
electron, $m_\mu$ is the muon mass and $F_p$ is the nuclear matrix
element. $g_{XK}^{(0)}$ and $g_{XK}^{(1)}$ (with $X = L, R$ and $K =
S, V$) are effective couplings at the nucleon level. They can be
written in terms of effective couplings at the quark level as
\begin{align}
g_{XK}^{(0)} &= \frac{1}{2} \sum_{q = u,d,s} \left( g_{XK(q)} G_K^{(q,p)} +
g_{XK(q)} G_K^{(q,n)} \right)\,, \nonumber \\
g_{XK}^{(1)} &= \frac{1}{2} \sum_{q = u,d,s} \left( g_{XK(q)} G_K^{(q,p)} - 
g_{XK(q)} G_K^{(q,n)} \right)\,.
\end{align}
The numerical values of the relevant $G_K$ factors can be found in Refs.~\cite{Kuno:1999jp,Kosmas:2001mv,Porod:2014xia}. For coherent $\mu-e$
conversion in nuclei, only scalar ($S$) and vector ($V$) couplings
contribute and sizable contributions are expected only from the
$u,d,s$ quark flavors.  The $g_{XK(q)}$ effective couplings can be
written in terms of the Wilson coefficients in Eqs.~\ref{eq:L-llg},
\ref{eq:L-2L2D} and \ref{eq:L-2L2U} as
\begin{eqnarray}
g_{LV(q)} &=& \frac{\sqrt{2}}{G_F} \left[ e^2 Q_q \left( K_1^L - K_2^R \right)- \frac{1}{2} \left( C_{\ell\ell qq}^{VLL} + C_{\ell\ell qq}^{VLR} \right) \right] \\
g_{RV(q)} &=& \left. g_{LV(q)} \right|_{L \to R} \\ 
g_{LS(q)} &=& - \frac{\sqrt{2}}{G_F} \frac{1}{2} \left( C_{\ell\ell qq}^{SLL} + C_{\ell\ell qq}^{SLR} \right) \\
g_{RS(q)} &=& \left. g_{LS(q)} \right|_{L \to R} \, ,
\end{eqnarray}
where $Q_q$ is the quark electric charge ($Q_d = -1/3$, $Q_u = 2/3$)
and $C_{\ell\ell qq}^{IXK} = B_{XY}^K \, \left( C_{XY}^K \right)$ for
d-quarks (u-quarks), with $X = \rm L, R$ and $K = \rm S, V$.

Radiative neutrino mass models can also be probed by looking for
$\mu-e$ conversion in nuclei. As already pointed out, the search for
this LFV process is going to be intensified in the next few years and,
in case no observation is made, it will soon become one of the most
constraining observables for this type of models. Similarly to the
leptonic LFV 3-body decays discussed above, the dipole coefficients
$K_2^{L,R}$ also enter the $\mu-e$ conversion rate, potentially
dominating it. In this case, one can derive a simple
relation~\cite{Albrecht:2013wet}
\begin{equation}
\label{eq:mueconv-dipole}
\frac{{\rm CR} (\mu- e, {\rm Nucleus})}{\text{BR}(\mu \to e \gamma)}
\approx \frac{f(Z,N)}{428} \, ,
\end{equation}
where $f(Z,N)$ is a function of the nucleus ranging from $1.1$ to
$2.2$ for the nuclei of interest. The reader is referred to Refs.~\cite{deGouvea:2013zba,Crivellin:2017rmk} for a discussion on the complementarity of $\mu\rightarrow e \gamma$ and $\mu-e$ conversion in nuclei. One can easily depart from this
\textit{dipole dominance scenario} in radiative neutrino mass models
due to the existence of sizable contributions to other LFV
operators. For instance, non-dipole contributions have been shown to
be potentially large in the scotogenic model in
Refs.~\cite{Toma:2013zsa,Vicente:2014wga}.  The dipole coefficients
may also be reduced due to partial cancellations in non-minimal
models, see for example
Refs.~\cite{Ahriche:2014cda,Ahriche:2014oda,Rocha-Moran:2016enp}. Finally,
as already pointed out in the case of $\ell_\alpha \rightarrow
\ell_\beta \ell_\delta \ell_\delta$ decays, some radiative neutrino
mass models contain new states that mediate LFV processes at tree
level. For instance, in R-parity violating models with trilinear terms
(discussed in Sec.~\ref{sec:RPV}), the superpotential terms $\lambda'
\widehat{L} \widehat{Q} \widehat{d}^c$ induce $\mu-e$ conversion at
tree level \cite{deGouvea:2000cf}. This easily breaks the expectation
in Eq.~\ref{eq:mueconv-dipole}.

Finally, we point out that the experiments looking for $\mu
\to eee$ and $\mu-e$ conversion in nuclei will soon take the lead in
the search for LFV. Therefore, even if dipole contributions turn out
to be dominant in a given model, $\mu \to eee$ and $\mu-e$ conversion
in nuclei might become the most constraining LFV processes in the near
future. Prospects illustrating this point for specific radiative
neutrino mass models have been presented in Refs.~\cite{Angel:2012ug,Vicente:2014wga,Klasen:2016vgl}.

\subsubsection{$h \rightarrow \overline{\ell_\alpha} \ell_\beta$}

In many radiative neutrino mass models, there can also be
contributions to lepton-flavor violating Higgs (HLFV) decays, like $h
\rightarrow \tau^- \mu^+,\tau^- e^+$ and their CP-conjugates. These
same interactions, however, also generate LFV processes such as $\tau
\rightarrow \mu (e)\, \gamma$, as no symmetry can prevent the
latter~\cite{Dorsner:2015mja}, which are subject to much stronger
constraints. In the effective field theory with just the 125 GeV Higgs
boson, HLFV decays involving the tau lepton can be sizable, and ATLAS
and CMS constraints on its flavor violating couplings (shown in
Tab.~\ref{tab:HLFVsignals}) are comparable or even stronger than those
coming from low-energy
observables~\cite{Blankenburg:2012ex,Harnik:2012pb,Herrero-Garcia:2016uab}.
However, in UV models, specially in radiative neutrino mass models,
the situation is generally the opposite.

\begin{table}[tb]
\begin{center}
  \begin{tabular}{  c c  c}
    \toprule
   HLFV Decay BR & ATLAS & CMS \\ 
   \midrule
    $h \to \tau \mu$ & $0.0143$ ~\cite{Aad:2016blu} & $0.0025$~\cite{CMS:2017onh}\\ 
    $h \to \tau e$ & $0.0104 $~\cite{Aad:2016blu} & $0.0061$~\cite{CMS:2017onh} \\ 
    \midrule
       \end{tabular}
       \begin{minipage}{0.9\linewidth}
	       \caption[Experimental upper bounds on Higgs LFV decays.]{\label{tab:HLFVsignals} Experimental $95~\%$ C.L. upper bounds on HLFV decays from ATLAS and CMS in the tau sector using the 13 TeV data sets.}
       \end{minipage}
\end{center}
\end{table}

The relevant gauge-invariant effective operators that generate HLFV are the \emph{Yukawa} operator:
\begin{equation} \label{eq:Yuk}
\mathcal{O}_Y = \overline{L}e_{\rm R}H(H^\dagger H)\,,
\end{equation}
and \emph{derivative} operators like
\begin{equation}\label{eq:DeR}
	\mathcal{O}_{D,\, e_{\rm R}} = (\overline{e_{\rm R}} H^\dagger) i\,\slashed D (e_{\rm R}\, H)\,,
\end{equation}
or
\begin{equation}\label{eq:DL}
\mathcal{O}_{D,\,L} =(\overline L H) i\,\slashed D (H^\dagger L)\,,
\end{equation}
plus their Hermitian conjugates. In Ref.~\cite{Herrero-Garcia:2016uab} all the possible tree-level realizations of these operators were outlined, some of which include particles that are present in radiative neutrino mass models, as we will see below. 
In Fig.~\ref{openoploop}, we show some possible UV completions of operators $\mathcal{O}_{Y}$, $\mathcal{O}_{D,\ L}$ and $\mathcal{O}_{D,\ e_{\rm R}}$.
The authors concluded that only $\mathcal{O}_Y$ can have sizable rates, and in particular only for UV completions that involve scalars, like in a type-III two-Higgs doublet model.

After electroweak symmetry breaking the \emph{Yukawa} operator gives rise to the interaction Lagrangian in Eq.~\ref{eq:L-llh}. For instance, the $S_{L,R}$ couplings are given by
\begin{equation}
S_L = \frac{v^2}{\sqrt{2} \Lambda^2} C_Y^\dagger + D_f \quad , \quad S_R = \frac{v^2}{\sqrt{2} \Lambda^2} C_Y + D_f \, ,
\end{equation}
where $D_f$ is the SM flavor-diagonal contribution, not relevant for
the present discussion, and $C_Y$ is the Wilson coefficient of the $\mathcal{O}_Y$ operator defined in Eq.~\ref{eq:Yuk}. Focusing on the contributions from the
\emph{Yukawa} operator, the branching ratio of the Higgs into a tau
and a muon reads:
\begin{equation}\label{eq:BRhtaumu}
\mathrm{BR}(h\rightarrow \tau\mu) = \frac{m_h}{8\pi \Gamma_h}\, \left(\frac{v^2}{\sqrt{2} \Lambda^2}\right)^2\,\left(|(C_Y)_{\tau\mu}|^2+|(C_Y)_{\mu\tau}|^2 \right)  \,.
\end{equation}

Most radiative neutrino mass models generate HLFV at 1-loop order~\cite{Herrero-Garcia:2016uab}.\footnote{Also in type-I seesaw (and inverse seesaw), and in the MSSM, HLFV is generated at 1-loop order~\cite{Pilaftsis:1992st,Arganda:2004bz,Arganda:2014dta,Thao:2017qtn,Arganda:2017vdb}.} For instance, the doubly-charged scalar singlet and the singly-charged scalar singlet of the Zee-Babu model (see Sec.~\ref{sec:modelleptons}) generate respectively the derivative operators $\mathcal{O}_{D,\,e_{\rm R}}$ and $\mathcal{O}_{D,\,L}$ at 1-loop order.
The scotogenic model (see Sec.~\ref{subsec:scotogenic}) also generates HLFV at 1-loop order ($\mathcal{O}_{D,\,L}$). 

\begin{figure}
\centering
\begin{subfigure}[t]{0.3\textwidth}
\centering
 \begin{tikzpicture}[node distance=1cm and 1cm]
	 \coordinate[label=left:$e_{\rm R}$] (nu1);
     \coordinate[vertex, right=of nu1] (v1);
     \coordinate[vertex, right=of v1] (lfv);
     \coordinate[vertex, above=of lfv] (v3);
     \coordinate[above left=of v3,  label=left:$H$] (h1);
     \coordinate[above right=of v3,  label=right:$H$] (h2);
     \coordinate[below=of lfv, label=below:$H$] (h3);
     \coordinate[vertex, right=of lfv] (v2);
     \coordinate[right=of v2, label=right:$L$] (nu2);
     \coordinate[above=of v1, label=left:$S_1$] (s1);
     \coordinate[above=of v2, label=right:$S_2$] (s2);

     \draw[fermion] (nu1)--(v1);
     \draw[fermion] (v1) -- node[below]{$F_1$} ++ (lfv);
     \draw[fermion] (lfv)-- node[below]{$F_2$} ++ (v2);
     \draw[fermion] (v2)--(nu2);
     \draw[scalar] (h1) -- (v3);
     \draw[scalar] (v3) -- (h2);
     \draw[scalar] (h3) -- (lfv);
     \draw[scalarnoarrow] (v3) to[out=180,in=90] (v1);
     \draw[scalarnoarrow] (v3) to[out=0,in=90] (v2);
   \end{tikzpicture}
\caption{$\mathcal{O}_{Y}$}
\end{subfigure}
\hspace{0.3cm}
\begin{subfigure}[t]{0.3\textwidth}
\centering
 \begin{tikzpicture}[node distance=1cm and 1cm]
	 \coordinate[label=left:$e_{\rm R}$] (nu1);
     \coordinate[vertex, right=of nu1] (v1);
     \coordinate[ right=of v1] (lfv);
     \coordinate[vertex, above=of lfv] (v3);
     \coordinate[above left=of v3,  label=left:$H$] (h1);
     \coordinate[above right=of v3,  label=right:$H$] (h2);
     \coordinate[vertex, right=of lfv] (v2);
     \coordinate[right=of v2, label=right:$e_{\rm R}$] (nu2);
     \coordinate[above=of v1, label=left:$S_1$] (s1);
     \coordinate[above=of v2, label=right:$S_2$] (s2);

     \draw[fermion] (nu1)--(v1);
     \draw[fermion] (v1) -- node[below]{$F$} ++ (v2);
     \draw[fermion] (v2)--(nu2);
     \draw[scalar] (h1) -- (v3);
     \draw[scalar] (v3) -- (h2);
     \draw[scalarnoarrow] (v3) to[out=180,in=90] (v1);
     \draw[scalarnoarrow] (v3) to[out=0,in=90] (v2);
   \end{tikzpicture}
   \caption{$\mathcal{O}_{D,\ e_{\rm R}}$}
\end{subfigure}
\hspace{0.4cm}
\begin{subfigure}[t]{0.3\textwidth}
\centering
 \begin{tikzpicture}[node distance=1cm and 1cm]
     \coordinate[label=left:$L$] (nu1);
     \coordinate[vertex, right=of nu1] (v1);
     \coordinate[ right=of v1] (lfv);
     \coordinate[vertex, above=of lfv] (v3);
     \coordinate[above left=of v3,  label=left:$H$] (h1);
     \coordinate[above right=of v3,  label=right:$H$] (h2);
     \coordinate[vertex, right=of lfv] (v2);
     \coordinate[right=of v2, label=right:$L$] (nu2);
     \coordinate[above=of v1, label=left:$S_1$] (s1);
     \coordinate[above=of v2, label=right:$S_2$] (s2);

     \draw[fermion] (nu1)--(v1);
     \draw[fermion] (v1) -- node[below]{$F$} ++ (v2);
     \draw[fermion] (v2)--(nu2);
     \draw[scalar] (h1) -- (v3);
     \draw[scalar] (v3) -- (h2);
     \draw[scalarnoarrow] (v3) to[out=180,in=90] (v1);
     \draw[scalarnoarrow] (v3) to[out=0,in=90] (v2);
   \end{tikzpicture}
\caption{$\mathcal{O}_{D,\ L}$}
\end{subfigure}
\caption[UV completions of operators contributing to Higgs LFV.]{Different 1-loop UV completions of the \emph{Yukawa} operator $\mathcal{O}_Y$ given in Eq.~\ref{eq:Yuk}, and the \emph{derivative} operators $\mathcal{O}_{D,\ e_{\rm R}}$ given in Eq.~\ref{eq:DeR} and $\mathcal{O}_{D,\ L}$ in Eq.~\ref{eq:DL}. $F$ and $F_{1,2}$ are fermion fields
 and $S_{1,2}$  scalar fields. The Zee-Babu and the scotogenic models are examples of radiative models with HLFV generated at 1-loop order. Figure reproduced from Ref.~\cite{Herrero-Garcia:2016uab}.}
\label{openoploop}
\end{figure}
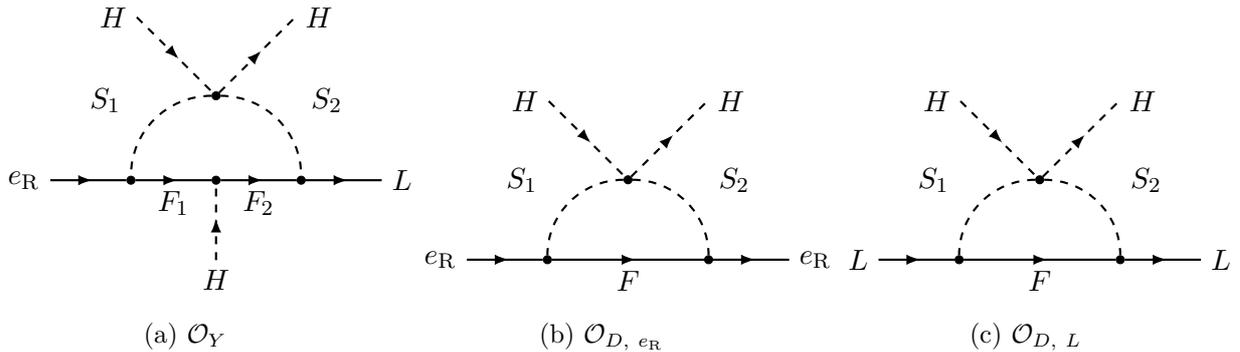

We can estimate the loop-induced HLFV in radiative neutrino mass models. Denoting a generic Yukawa coupling of the fermions and scalars with the SM leptons as $Y$, and a scalar quartic coupling with the Higgs as $\lambda_{i h}$, and taking into account that the amplitude of $h\rightarrow \mu \tau$ involves a tau mass, one can estimate the dominant contribution to be~\cite{Herrero-Garcia:2016uab}:
\begin{equation}
{\rm BR}(h\rightarrow \mu \tau) \sim {\rm BR}(h\rightarrow \tau \tau)\,\frac{ \lambda_{i h}^2}{(4\pi)^4} \left(\frac{v}{\rm TeV}\right)^4\, \left(\frac{Y}{M_i/\text{TeV}}\right)^4.
 \end{equation}
where $M$ is the largest mass in the loop. In all these models, in addition to the loop factor, there are in general limits from charged LFV processes, as usually all radiative neutrino mass models have charged particles that can generate $\ell_\alpha \rightarrow \ell_\beta \gamma$. As $\tau \rightarrow \mu \gamma$ typically gives the constraint $Y/(M/\text{TeV})^4\lesssim\mathcal{O}(0.01-1)$, we get:
\begin{equation}
{\rm BR}(h\rightarrow \mu \tau) \lesssim 10^{-8},\,
 \end{equation}
well below future experimental sensitivities. Thus, unless cancellations are invoked (which are difficult to achieve in all possible radiative decays), HLFV rates are very suppressed, well below future experimental sensitivities.

One class of models which can have large HLFV are those with another Higgs doublet such that both the SM and the new scalar doublet couple to the lepton doublets~\cite{Davidson:2010xv,Sierra:2014nqa,Dorsner:2015mja,Omura:2015nja, Botella:2015hoa}. In such scenarios, both Yukawa couplings cannot be diagonalized simultaneously, which leads to LFV Higgs interactions. One example is the Zee model discussed in Sec.~\ref{sec:modelleptons}, which can have ${\rm BR}(h\rightarrow \mu \tau)$ up to the percent level~\cite{Herrero-Garcia:2017xdu}.


\subsection{Anomalous magnetic moments and electric dipole moments}

The anomalous magnetic moments (AMMs) and electric dipole moments
(EDMs) of the SM leptons receive new contributions in radiative
neutrino mass models (see Ref.~\cite{Raidal:2008jk} for a review on the topic). These are contained in the dipole coefficients
that also contribute to the radiative $\ell_\alpha \to \ell_\beta
\gamma$ decays, typically leading to tight correlations between these
observables. Using the effective Lagrangian in Eq.~\ref{eq:L-llg},
the anomalous magnetic moment $a_\alpha$ and the electric dipole
moment $d_\alpha$ of the charged lepton $\ell_\alpha$ are given by~\cite{Raidal:2008jk}
\begin{align}
	a_{\alpha} & = m_{\ell_\alpha}^2 \, \text{Re} \left( K_2^L + K_2^R\right) \, , &
	\frac{d_{\alpha}}{e} & = \frac{1}{2} \, m_{\ell_\alpha} \, \text{Im} \left( K_2^R - K_2^L\right) \, .
\end{align}

The experimental values for the AMMs and EDMs of charged leptons are collected in Tab.~\ref{tab:ammedm}. In particular the muon AMM received a lot of intention in recent years due to the discrepancy between the experimentally measured value given in Tab.~\ref{tab:ammedm} and the SM prediction~\cite{Olive:2016xmw}
\begin{equation}
	a_\mu^{SM}= 116 591 803(1)(42)(26) \times 10^{-11}
\end{equation}
with the errors due to electroweak, lowest-order, and higher-order hadronic contributions.  
\begin{table}[bt]\centering
	\begin{tabular}{ccc}
		\toprule
		Lepton & AMM $a$ & EDM $d\;\; [e\,\mathrm{cm}]$ \\
		\midrule
		$e$ &  $(1159.65218091\pm0.00000026)\times 10^{-6}$ & $<0.87\times 10^{-28}$\\
		$\mu$ & $(11659208.9\pm5.4\pm3.3)\times 10^{-10}$ & $(-1\pm9 )\times 10^{-20} $\\
		$\tau$ & $[-0.52,0.013]$ & $[-2.20 ,4.5] \times 10^{-17} + i [-2.50,8.0] \times 10^{-19} $ \\
		\bottomrule
	\end{tabular}
	\caption[Experimental results for AMMs and EDMs.]{Experimental values for AMMs and EDMs~\cite{Olive:2016xmw}. Both statistical and systematic uncertainties are given for the muon AMM $a_\mu$.}
	\label{tab:ammedm}
\end{table}

There are many examples of radiative neutrino mass models leading to
sizable effects in these two observables. For some examples in the case of AMMs see for
instance Refs.~\cite{Dicus:2001ph,Babu:2010vp,Nomura:2016jnl,Nomura:2016ask,Chiang:2017tai,Lee:2017ekw}. In some cases, the new contributions
effects can help close the gap between the theory prediction and the
experimental measurement of the muon AMM, although in other cases they increase the disagreement, depending on their sign. We
refer to the recent review~\cite{Lindner:2016bgg} for a guide
regarding new physics contributions to the muon AMM.

Regarding lepton EDMs, some examples in
radiative neutrino models are given in
Refs.~\cite{Borah:2016zbd,Borah:2017leo,Chiang:2017tai}. In this case one
requires CP-violating new physics in the lepton sector, something that
is easily accommodated in new Yukawa couplings.


\subsection{Neutrinoless double beta decay} \label{sec:0nubb}

One of the main experimental probes to test the Majorana/Dirac nature of
neutrinos is neutrinoless double beta decay ($0\nu\beta\beta$), in which a
nucleus $(A,\,Z)$ decays into another nucleus $(A,\,Z+2)$ and two electrons~\cite{Furry:1939qr}. In order to have sizable $0\nu\beta\beta$ rates, the nuclei should not
have single beta decays. This is achieved with even-even nuclei which, thanks
to the nuclear pairing force, are lighter than the odd-odd nucleus, making
single beta decays kinematically forbidden. The current strongest experimental limits are obtained using $^{136}$Xe by EXO-200~\cite{Albert:2014awa} and KamLAND-Zen~\cite{Gando:2012zm,Asakura:2014lma} which yield lower bounds of the lifetime of $1.1\cdot 10^{25}$ y and $1.9\cdot 10^{25}$ y at 90~\% C.L, respectively. Uncertainties in the nuclear matrix elements translate into uncertainties in the extracted values of $|m_{ee}|$ (see eq.~\eqref{eq:Sec1:mbetabeta}), whose current strongest upper limits are in the ballpark of $\sim 0.15$ eV. For further details regarding the present and future experimental situations see Ref.~\cite{DellOro:2016tmg}.

The observation of $0\nu\beta\beta$ decay would imply that lepton
number is violated by two units ($\Delta L=2$), and therefore that neutrinos are Majorana
particles~\cite{Schechter:1981bd}. However, quantitatively, this contribution to neutrino masses occurs at 4-loop order and is therefore extremely suppressed, much lighter than
the observed neutrino masses (see Ref.~\cite{Duerr:2011zd} for a quantitative
study of this statement). So, even if it is true that neutrinos will necessarily be Majorana if $0\nu\beta\beta$ is observed, the main contribution to their masses may no be necessarily related to $0\nu\beta\beta$.

We will mainly focus in this section on radiative models which have new
\emph{direct} contributions to neutrinoless double beta decay beyond the
standard ones mediated by the light Majorana neutrinos, which are
\emph{indirect}, as they are generated by the new particles at higher-loop order 
(via light neutrino masses). For general reviews on the subject the interested
reader is referred to Refs.~\cite{Rodejohann:2012xd, Bilenky:2012qi,
DellOro:2016tmg}.

In Refs.~\cite{Pas:1999fc,Pas:2000vn} a general phenomenological formula for
the process including both long and short-range interactions was given. The
authors considered all possible Lorentz structures for the quarks involved in
the process and the outgoing electrons. In Ref.~\cite{delAguila:2012nu}
effective operators that involve gauge bosons were considered, such that there
are new effective vertices of the $W$-boson and the electrons. 

In Fig.~\ref{fig:conts_0nubb} (reproduced from Ref.~\cite{delAguila:2013zba}) all possible
contributions to $0\nu\beta\beta$ are shown, with the red dot representing
the $\Delta L=2$ vertex. Diagram \ref{fig:conts_0nubb-a} shows the light neutrino contribution, while
diagram \ref{fig:conts_0nubb-f} involves a dimension-9 effective operator. In
Ref.~\cite{Bonnet:2012kh} a systematic classification of possible UV models
stemming from the dimension 9 operator was performed (diagram \ref{fig:conts_0nubb-f}). See also
Ref.~\cite{Helo:2015fba} for scalar-mediated UV completions and its connection
to neutrino masses. Diagrams \ref{fig:conts_0nubb-d} and \ref{fig:conts_0nubb-e} involve new vertices between quarks,
leptons and gauge bosons.  Diagrams \ref{fig:conts_0nubb-b} and \ref{fig:conts_0nubb-c} involve new vertices with just
leptons and gauge bosons and no quarks. In Ref.~\cite{delAguila:2012nu}
operators that involve gauge bosons were considered, such that there are new
effective vertices of the $W$-boson and the electrons, as in diagrams \ref{fig:conts_0nubb-b} and \ref{fig:conts_0nubb-c}. See Sec.~\ref{sec:derivatives} for a discussion of the effective operators that generate the latter
diagrams and their connection to neutrino masses. A
systematic classification of UV models for all the dimension-7 operators was
given in Ref.~\cite{Helo:2016vsi}. Many (if not all) of these particles can be
present in radiative neutrino mass models.

We outline in the following two typical new contributions to $0\nu\beta\beta$
from radiative neutrino mass models: \begin{enumerate} \item New particles that
	couple to quarks. For instance, leptoquarks as in
	Refs.~\cite{Hirsch:1996ye, Hirsch:1996qy, Kohda:2012sr}. In R-parity
	violating SUSY (see Sec.~\ref{sec:RPV}) there can be new contributions
	to $0\nu\beta\beta $ from new states, see
	Refs.~\cite{Mohapatra:1986su,Babu:1995vh,Hirsch:1995ek,Hirsch:1995vr}.
	Another simple example due to exchange of color octet scalars and
	fermions that couple to quarks and leptons simultaneously is given in
	Ref.~\cite{Choubey:2012ux}. A model with two scalar diquarks, a
	dilepton and a second Higgs doublet is given in
	Ref.~\cite{Brahmachari:2002xc}. See other examples in
	Refs.~\cite{Gu:2011ak, Helo:2015fba}.  \item New particles that open
	operators that involve gauge bosons~\cite{delAguila:2012nu}, see
	discussion in Sec.~\ref{sec:derivatives}.  \end{enumerate}

Let us also mention that, in addition to $0\nu\beta\beta$, there are also
limits on other lepton number violating elements $m_{\alpha\beta}$ of the neutrino mass matrix in flavor basis (where the charged lepton mass matrix is diagonal),
different from the $m_{ee}$ (which equals $m_{\beta\beta}$) one, stemming from meson decays, tau decays, $e^+ p$
collider data among other processes~\cite{Quintero:2016iwi}. Also indirect bounds using
neutrino oscillations and the unitarity of the PMNS matrix can be
set~\cite{Rodejohann:2000th}. However, both the direct and indirect (even if
much stronger than the direct) bounds obtained are typically very
weak~\cite{Rodejohann:2000th,Quintero:2016iwi}. $\mu^-- e^+$ conversion also
offers a possibility to test the $m_{e\mu}$ element, however typically the
rates are not competitive with those of $0\nu\beta\beta$, although of course
they test a different element and flavor effects could be relevant. A study of
the contributions from effective operators was performed in
Ref.~\cite{Berryman:2016slh}, while a doubly-charged scalar was studied in
detail in Ref.~\cite{Geib:2016daa}.

Lepton number violation can also be searched for at colliders. This is
specially interesting for channels that do not involve electrons, as it is
necessarily the case for $0\nu\beta \beta$. Those will be discussed in
Sec.~\ref{sec:colliders}. Also the connection of lepton number violation to the
matter-antimatter asymmetry of the universe will be discussed in
Sec.~\ref{sec:matter}.


\subsection{Collider searches} \label{sec:colliders}

Radiative neutrino mass models generally have a much lower UV scale than the GUT scale,
which makes them testable at either current or future colliders.
The diversity of exotic particles and their interaction with the SM particles in 
radiative neutrino mass models leads to an extremely rich phenomenology at colliders.         
Processes pertaining to the Majorana nature of neutrino masses or 
LFV couplings between the exotic particles and the SM,  
i.e. processes violating lepton number and/or lepton flavor, 
are often chosen as signal regions in collider searches 
due to the low SM background.\footnote{Theoretically there is no SM background.
Realistically, however, object misidentification, undetected particles and fake objects 
can result in similar final states at the detector level.} Of course, there are searches for exotic particles in general if they are not too heavy and 
the couplings are sizable.
\footnote{ Some of
the exotic particles may also show up in tree-level neutrino mass models. 
The interested reader is referred to the recent review~\cite{Deppisch:2015qwa} for the collider tests of specific tree-level models.}
In the following, we sketch different search strategies at colliders, 
which often utilize the low SM background for LNV and LFV processes. 
We thus discuss LNV and LFV processes separately before discussing general searches for new particles, 
which rely on processes without any LNV/LFV.
 
\subsubsection{Lepton number violation}
\label{subsubsec:LNV}
At the LHC, the most sought-after channel of LNV\footnote{Strictly speaking the process is not necessarily LNV, because $X$ may carry lepton number as well, for example in form of neutrinos. Currently the searches are limited to electrons and muons. However $\tau$-leptons may also be used to search for LNV.} are same-sign leptons  
\begin{align}
pp\to \ell^\pm \ell^\pm X \; ,
\end{align} 
where $\ell$ denotes $e$ or $\mu$,
and $X$ can be any number of jets, $E_{\rm T}^{miss}$ or other SM objects.
The details of the production and the actual content of $X$ are very model-dependent:
typically heavy states are produced and decay to final states with same-sign dilepton due to their Majorana nature.      
We will take a doubly-charged scalar as a simple example to illustrate the basics of this search strategy.
A doubly-charged scalar $\phi^{++}$ is an $SU(2)_{\rm L}$ singlet with hypercharge $Y=2$. 
They can be pair-produced via Drell-Yan process and subsequently decay to 
two same-sign dileptons. For large masses the photon-initiated process becomes important and leads to an enhancement~\cite{Babu:2016rcr}.
Assuming the branching fraction of $\phi^{++}\to e^+ e^+$ is $100\%$, 
the signature for pair-produced doubly-charged scalars is four electrons and thus $ZZ$ production is the main SM background.    
To reduce the SM background, discriminating variables such as
the same-sign dilepton mass, the difference between the opposite sign dilepton mass and the $Z$ boson mass,
and the scalar sum of the lepton $p_T$ can be utilized.   
ATLAS~\cite{ATLAS:2016pbt} has 
excluded doubly-charged $SU(2)_{\rm L}$ singlet scalar with mass lower than 420 GeV at $95\%$ CL with LHC Run 2 data.
The improved limit can be extracted from the CMS search for doubly-charged component of an $SU(2)_{\rm L}$ triplet~\cite{CMS:2017pet}. 
In Refs.~\cite{Sugiyama:2012yw, delAguila:2013mia,delAguila:2013yaa, Kanemura:2013vxa} studies of doubly-charged scalars and how to discriminate the multiplet to which they belong were performed.

The sensitivities of $0\nu \beta\beta$ searches detailed in Sec.~\ref{sec:0nubb} and the same-sign
dilepton searches at the LHC can be compared in any specific model (see for 
example Refs.~\cite{Helo:2013ika,Peng:2015haa,Gonzales:2016krw}).
Specifically in Refs.~\cite{Helo:2013ika,Peng:2015haa} a simplified model with a scalar doublet $S\sim (1,2,1)$ and a 
Majorana fermion $F$, which has the same matter content as the scotogenic model, is adopted. 
In this model, the reach of tonne-scale $0\nu\beta\beta$ generally beats that of the LHC. 
In the parameter space region where the heavy particle masses are near the TeV scale, however, the two probes are complementary.  

\subsubsection{Lepton flavor violation}
As described in Sec.~\ref{subsec:LFV}, lepton flavor violating processes are commonly predicted in  
radiative neutrino mass models, which can also be probed at colliders.
The actual production topology of the LFV processes varies from model to model. 
For example, in models with the leptoquark $S_1\sim (\bar{3}, 1, 1/3)$, 
there are two possible decay channels, $S_1\to \bar \nu \bar b$ or $S_1\to \ell^+ \bar t$~\cite{Buchmuller:1986zs}.
The dilepton final states are produced from 
\begin{align}
pp\to S_1^* S_1\to b\nu \bar{t} \ell^+\to \ell^+\ell^{\prime -}b\bar{b} + X\,,
\qquad 
pp\to S_1^* S_1\to t\ell^{\prime -} \bar{t}\ell^+ \to \ell^+ \ell^{\prime -} b \bar{b} + X  \; , 
\end{align}
where $X$ can represent $E_T^{miss}$, multiple jets and leptons, and  the former contributes dominantly for normal ordering 
in the minimal model with two leptoquarks.
SUSY stop searches in the dilepton final states have the same signatures
and their collider bounds can be translated into that of the leptoquark.
This has been done for the LHC 8 TeV run~\cite{Aad:2014qaa} and the limit 
was $m_{S_1} \gtrsim 600$ GeV~\cite{Cai:2014kra}. 
Note that this limit in LFV channel is stronger than lepton flavor conserving ones ($m_{S_1}\gtrsim 500$ GeV)
as the SM background is lower.   
The stop search has been updated for LHC Run 2~\cite{CMS:2017qjo,CarraonbehalfoftheATLASCollaboration:2017ekz}, though 
a recast for leptoquarks in LFV dilepton final states still awaits further analysis.   

Alternatively, LFV processes can also be studied in an independent manner. 
In the framework of effective operators with two flavor-diagonal quarks and 
two flavor-off-diagonal leptons, constraints from LHC searches for LFV final states 
are interpreted as lower limits on the UV cut-off scale~\cite{Cai:2015poa}. 
Compared with the limits derived from low energy precision measurements~\cite{Carpentier:2010ue,Cai:2015poa}, LHC delivers less 
stringent limits for light quarks.
For heavier quarks, however, competitive limits of $\Lambda_{\rm UV}\gtrsim 600-800$ GeV
can already be set for operators with right-handed $\tau$ leptons using only LHC Run 1 data.

\subsubsection{Searches for new particles}
\label{subsubsec:newparticles}
Radiative neutrino mass models may contain exotic particles such as
vector-like quarks (VLQs), vector-like leptons (VLLs), scalar leptoquarks, singly- or doubly-charged scalars, 
colored octet fermions or scalars, and electroweak multiplets.
Note that the examples here are far from complete
and searches for each individual particle require their own
dedicated discussion.  
In Ref.~\cite{Cai:2014kra}, LHC searches for exotic particles 
in UV complete models based on $\Delta L=2$ dimension 7 operators are discussed systematically.        
Here we will only present a simple summary about a handful of new particles. 

{\bf Vector-like quarks:} We refer by VLQs to new $SU(3)_{\rm c}$ triplets 
which mix with the SM quarks and Higgs via Yukawa couplings~\cite{AguilarSaavedra:2009es}.    
The VLQs include different $SU(2)$ representations:  two singlets $T$ and $B$ with hypercharge $2/3$ and $-1/3$;
three doublets $(T, B)$, $(X, T)$, and $(B, Y)$ with hypercharge $1/6$, $7/6$ and $-5/6$; 
and two triplets $(X, T, B)$, and $(T, B, Y)$ with hypercharge $2/3$ and $-1/3$.
They can be pair produced at the LHC via gluon fusion and
quark-antiquark annihilations. Single production is model-dependent and can be dominant for large vector-like
quark masses and large mixings~\cite{AguilarSaavedra:2009es}. 
The mass splitting among the components of the fields is suppressed by the mixing angles 
between the SM quarks and the vector-like quarks, which in turn suppresses
the decays between the component fields.  
Therefore, VLQs will dominantly decay to either a gauge boson or a Higgs plus 
a SM quark.  
Both ATLAS and CMS have performed searches for VLQs 
and have set lower limits on the VLQs masses up to 990 GeV  
at the $95\%$ confidence level (CL) depending on the representations and 
the decay branching ratio~\cite{Aad:2015gdg, Aad:2014efa, Aad:2015kqa, Khachatryan:2015oba, Aaboud:2017qpr, Aaboud:2016lwz, ATLAS:2016sno, Sirunyan:2017usq, Chatrchyan:2013uxa, Aad:2015tba, Khachatryan:2015gza, Aad:2015mba}.

{\bf Vector-like leptons:} VLLs are the colorless version of VLQs.
Similar to VLQs, VLLs mix with the SM leptons via Yukawa couplings with Higgs.    
Due to the absence of right-handed neutrinos, there are less VLLs: two singlets $N$ and $E$
with hypercharge 0 and 1; 
two doublets $(N, E)$ and $(E, D)$ with hypercharge $3/2$ and $1/2$; and 
triplets $(P, N, E)$ and $(N, E, D)$ with hypercharge $0$ and $1$, respectively. 
Detailed studies have been performed in Refs.~\cite{Altmannshofer:2013zba,Falkowski:2013jya,Dermisek:2014qca,Kumar:2015tna}.
Contrary to the colored VLQs, VLLs are dominantly pair produced at the LHC via Drell-Yan process 
as the phase space suppression is less significant in the  parameter space of interest at the moment.
They can also be singly produced in association with $W$, $Z$ or $H$, which can be dominant if the pair
production channel is phase space suppressed and sizable mixing parameters are assumed.
Likewise VLLs decay either to a SM lepton and a boson, $W$ or $Z$, or Higgs.       
So far there is no dedicated search for VLLs at colliders, though SUSY searches for 
sleptons or charginos can be used to derive bounds on VLLs (see Ref.~\cite{Altmannshofer:2013zba, Hamada:2016vwk} for example).  
 
{\bf Leptoquarks:}
Leptoquarks appear frequently in theories beyond the SM such as grand unified theories~\cite{Pati:1974yy, Georgi:1974sy}.
As its name suggests, a leptoquark, which can be either a scalar or a vector~\cite{Buchmuller:1986zs}, possesses both nonzero lepton and baryon numbers. 
Here we will focus on scalar leptoquarks.  
At hadron colliders, leptoquarks are primarily produced in pairs via gluon fusion and quark-antiquark annihilation.
Each leptoquark subsequently decays to one quark and one charged or neutral lepton. 
Both ATLAS~\cite{Aaboud:2016qeg, Aad:2015caa} and CMS~\cite{CMS:2016imw, CMS:2016qhm, CMS:2016hsa} 
have performed searches for leptoquarks in final states with two charged leptons plus multiple jets.
Assuming $100\%$ branching fraction of the leptoquark decay into a charged lepton and a quark, 
current searches at the LHC Run 2 with 13 TeV center of mass energy 
have excluded leptoquarks with masses less than 1130 GeV~\cite{CMS:2016imw}, 
1165 GeV~\cite{CMS:2016qhm} and 900 GeV~\cite{CMS:2016hsa} at $95\%$ CL for leptoquark couplings to the first, second and third generations respectively. 

{\bf Charged scalars singlets:}
Singly- and doubly-charged scalars are introduced in various radiative neutrino mass models 
(see Refs.~\cite{Babu:2013pma, Zee:1985id, Babu:1988ki}, for instance). 
As singlets under $SU(3)_{\rm c}\times SU(2)_{\rm L}$, 
the singly (doubly) charged scalar can only couple to the lepton doublet (right-handed charged lepton) bilinear.
So the doubly-charged scalar can only decay to a pair of charged leptons, which 
leads to LNV signature at colliders (see discussion in Sec.~\ref{subsubsec:LNV} for details).      
As for the singly-charged scalar, it decays to a charged lepton and a neutrino whose LNV effects can 
not be detected at the LHC.   
Singly-charged scalars are mainly produced in pairs via the Drell-Yan pair process. 
They are searched for in final states with two leptons plus $E_T^{miss}$.\footnote{Long-lived charged particles have 
been searched at the LHC using anomalously high ionization signal~\cite{Aad:2015oga}, also in the context of dark matter~\cite{Khoze:2017ixx}. 
However, charged scalars in radiative neutrino
mass models usually have sizable couplings to SM leptons and decay promptly.}   
SUSY searches for sleptons and charginos at the LHC share the same signature as the singly-charged scalars. 
Thus we can in principle recast the slepton search in Ref.~\cite{ATLAS:2017uun} and 
extract the limit for our singly-charged scalars.  Note a slepton can also be produced via a $W$-boson, while 
singly-charged scalar only via a virtual photon.   
     
{\bf Higher-dimensional electroweak multiplet:} 
$SU(2)_{\rm L}$ higher-dimensional representations can also be incorporated in radiative neutrino mass 
theories~\cite{Cai:2011qr, Chen:2011bc, Law:2013saa, Chowdhury:2015sla, 
Cai:2016jrl, Ahriche:2016rgf, Sierra:2016qfa, Ahriche:2015wha}.
While the mass splittings among the component fields for scalar multiplets can be generally large due to couplings to the SM Higgs, those for fermion multiplets
are only generated radiatively and are typically $\sim\mathcal{O}(100)$ MeV, with the 
neutral component being the lightest.  This small mass splitting results in lifetimes $\sim \mathcal{O}(0.1)$ ns. 
At the LHC, charged component field can be produced in pair 
via electroweak interaction and decay to the neutral component plus a very soft pion, which leads to 
a disappearing track signature.   
For a triplet with a lifetime of about 0.2 ns, the current LHC searches set the lower mass limit to be 430 GeV
at $95\%$ CL~\cite{Aad:2013yna, CMS:2014gxa, ATLAS:2017bna}.


\subsection{Generation of the matter-antimatter asymmetry of the universe} \label{sec:matter}

The matter-antimatter asymmetry of the universe has been inferred independently (and consistently) by big bang nucleosynthesis (BBN) predictions of light elements, and by the temperature anisotropies of the cosmic microwave background. In order to generate it, the Sakharov conditions need to be fulfilled~\cite{Sakharov:1967dj}. There should be:
\begin{itemize} 
\item Processes that involve baryon number violation (BNV).
\item Processes in which both charge conjugation (C) and charge and parity conjugation (CP) are violated.
\item Departure from thermal equilibrium, so that (i) the number densities of
	particles and antiparticles can be different, and (ii) the generated
	baryon number is not erased.
\end{itemize}

In the standard model, it is well-known that due to the chiral nature of weak
interactions B+L is violated by sphaleron processes, while B$-$L is
preserved~\cite{Kuzmin:1985mm}. Also C and CP are violated in the quark sector
(in the CKM matrix), although the amount is too small to generate the required
CP asymmetry. In the lepton sector (with massive neutrinos) CP can be violated,
and there are in fact hints of $\delta \sim -\pi/2$~\cite{Esteban:2016qun}.
However, the measurement of the Higgs mass at $125$ GeV implies that the
phase transition is not strongly first-order, with no departure from thermal
equilibrium. Therefore, the SM has to be extended to explain the matter-antimatter asymmetry which raises the question whether this new physics is related to neutrino masses or not.

When sphalerons are active and in thermal equilibrium, roughly at temperatures
above the electroweak phase transition, B+L can be efficiently violated.
Therefore, one natural option in models of Majorana neutrinos is that an
asymmetry in lepton number is generated, which is converted by sphalerons into
a baryon asymmetry. This is known as leptogenesis~\cite{Fukugita:1986hr} (see
Ref.~\cite{Davidson:2008bu} for a review on the topic), the most popular
example being the case of type-I seesaw, where the out-of-equilibrium decays of
the lightest of the heavy right-handed neutrinos into lepton and Higgs doublets
and their conjugates, at a temperature equal or smaller than its mass, generate
the lepton asymmetry due to CP-violating interactions. 

The scotogenic model and its variants, see Sec.~\ref{subsec:scotogenic}, have been
studied in detail regarding the generation of the baryon asymmetry from
particle decays with TeV-scale masses. Reference \cite{Ma:2006fn} briefly discusses leptogenesis within the scotogenic model. This discussion is extended in Refs.~\cite{Kashiwase:2012xd,Kashiwase:2013uy,Racker:2013lua} to include resonant leptogenesis. Resonant leptogenesis has also been studied in a gauge extension of the scotogenic model~\cite{Chang:2011kv, Lindner:2013awa,Kownacki:2016hpm} in Ref.~\cite{Kashiwase:2015joo} and resonant baryogenesis in an extension with new colored states in Ref.~\cite{Dev:2015uca}. References~\cite{Hambye:2006zn,Babu:2007sm} consider extensions of the scotogenic model by an additional charged or neutral scalar to achieve viable non-resonant leptogenesis. The baryon asymmetry can similarly be enhanced by producing the SM singlet fermions in the scotogenic model non-thermally beyond the usual thermal abundance~\cite{Suematsu:2016vgz}. Leptogenesis via
decays of an inert Higgs doublet or a heavy Dirac fermion were studied in
Refs.~\cite{Lu:2016ucn,Lu:2016dbc} in scotogenic-like models,
respectively. In Ref.~\cite{Chen:2011bc} leptogenesis was studied in a scotogenic-like model with fermionic 5-plets and a scalar 6-plet, via the decays of the second-lightest fermionic 5-plet.  Reference~\cite{Baldes:2013eva} demonstrated the feasibility to generate the correct matter-antimatter asymmetry via leptogenesis in the model proposed in Ref.~\cite{Ma:2009gu}. 
It also showed that any pre-existing baryon asymmetry in the two models proposed in Refs.~\cite{Ma:2009gu,Law:2012mj} is washed out at temperatures above the mass of their heaviest fields.

In radiative models with extra scalars coupled to the Higgs field,
the phase transition can generally be stronger, as they contribute positively
to the beta function of the Higgs and therefore, they help to stabilize the Higgs
potential. Moreover in these models there are typically extra sources of CP
violation. These two ingredients allow the possibility of having electroweak
baryogenesis. In particular, the strong first-order phase transition has
been discussed using an effective potential in Ref.~\cite{Bertolini:2014aia},
and in Ref.~\cite{Aoki:2009vf} for the model of Ref.~\cite{Aoki:2008av}. Also
in the case of a supersymmetric radiative model in
Ref.~\cite{Kanemura:2014cka}.

However, in general the new states can also destroy a pre-existing asymmetry,
irrespective of their production mechanism, as they violate necessarily lepton
number by two units~\cite{Fischler:1990gn,Campbell:1990fa,Nelson:1990ir,Harvey:1990qw}. The new particles typically have gauge interactions, so
that they are in thermal equilibrium at lower temperatures than those at which
the asymmetry is generated (by high-scale baryogenesis or by leptogenesis, for
instance\footnote{In this last case, of course, the presence of low scale LNV
can be regarded as being less motivated, as in principle there would already be
an explanation for neutrino masses (at least for one neutrino).}) potentially
washing it out.

Some works have focused on the fact that if LNV is observed at the LHC, one
could falsify leptogenesis, as the wash-out processes would be too
large~\cite{Antaramian:1993nt,Frere:2008ct, Deppisch:2013jxa}. Similarly,
observations of $0\nu\beta\beta$ rates beyond the one generated by the light
neutrinos could impose constraints for the first
family~\cite{Deppisch:2015yqa}. LFV processes could be used to extend it to all
families. See Ref.~\cite{Deppisch:2015qwa} for further discussions about LNV processes in leptogenesis.

The limits on radiative models due to the requirement of not washing-out any
pre-existing asymmetry are model-dependent. A more systematic way to go is to
consider the LNV effective operators related to radiative
models~\cite{Weinberg:1979sa,Babu:2001ex,deGouvea:2007qla}. These operators
lead to wash-out processes if they are in thermal equilibrium above the electroweak
phase transition, and therefore their strength can be bounded by this
requirement.


\subsection{A possible connection to dark matter models}
\label{sec:DM}

In many radiative neutrino mass models the generation of neutrino
masses at tree-level is forbidden by a symmetry, $\mathcal{G}$. This
symmetry can be global or gauge, continuous or discrete (a typical
example is a $Z_2$ parity), imposed or accidental (a
by-product of other symmetries in the model). If $\mathcal{G}$ is
preserved after electroweak symmetry breaking, the lightest state
transforming non-trivially under it, the so-called lightest charged
particle (LCP), is completely stable and, in principle, could
constitute the dark matter (DM) of the universe. This opens up an
interesting connection between radiative neutrino masses and dark
matter. DM may be produced via its coupling to neutrinos and thus the
annihilation cross section is closely related to neutrino mass. This has been
studied using an effective Lagrangian for light, MeV-scale, scalar
DM~\cite{Boehm:2006mi} in a scotogenic-like model and for fermionic DM~\cite{Kubo:2006yx,Sierra:2008wj,Suematsu:2009ww,Adulpravitchai:2009gi,Toma:2013zsa,Vicente:2014wga}
in the scotogenic model.
A key signature of this close connection is a neutrino line from DM
annihilation. The constraints from neutrino mass generation on the
detectability of a neutrino line has been recently discussed in
Ref.~\cite{ElAisati:2017ppn}.

Based on the general classification of 1-loop models~\cite{Bonnet:2012kz}, the authors of Ref.~\cite{Restrepo:2013aga} performed a
systematic study for models compatible with DM stabilized by a discrete $Z_2$
symmetry. They focused on the topologies T1-x and T3. The topologies T4-2-i
and T4-3-i require an additional symmetry to forbid the tree-level contribution
and thus were not studied in Ref.~\cite{Restrepo:2013aga}. 
A similar classification for 2-loop models has been presented in
Ref.~\cite{Simoes:2017kqb} based on the possible 2-loop topologies discussed in
Ref.~\cite{Sierra:2014rxa}. Symmetries forbidding tree and lower-order loop diagrams have been discussed in Ref.~\cite{Farzan:2012ev}. In Sec.~\ref{subsec:scotogenic} we discuss the
prototype example of
such models: the scotogenic model.

Besides dark matter being stabilized by a fundamental symmetry, it may be
stable due to an accidental symmetry. 
For example, higher representations of $SU(2)_{\rm L}$ cannot couple to the SM 
in a renormalizable theory, which leads to an accidental $Z_2$ symmetry at the renormalizable level.
This has been dubbed minimal dark
matter~\cite{Cirelli:2005uq,Cirelli:2009uv}. After the initial proposal to
connect the minimal dark matter paradigm and radiative neutrino mass
generation~\cite{Cai:2011qr}, it has been conclusively demonstrated that the
minimal dark matter paradigm cannot be realized in 1-loop neutrino mass
models~\cite{Cai:2016jrl,Ahriche:2016rgf,Sierra:2016qfa}. However, there is a
viable variant of the KNT model at 3-loop order~\cite{Ahriche:2015wha},
which realizes the minimal dark matter paradigm without imposing any
additional symmetry beyond the SM gauge symmetry. 

Finally, the DM abundance in the universe may be explained by a light pseudo-Goldstone boson (pGB) associated with the spontaneous breaking of a global symmetry. It is commonly called Majoron in case the lepton number plays the role of the global symmetry. The possibility of pGB dark matter has been discussed in one of the models in Ref.~\cite{Dasgupta:2013cwa} which provides a pGB dark matter candidate after the breaking of a continuous U(1) symmetry to its $Z_2$ subgroup in addition to the LCP. Recently the authors of Ref.~\cite{Ma:2017vdv} proposed an extension of the Fileviez-Wise model~\cite{FileviezPerez:2009ud} to incorporate a Majoron DM candidate which simultaneously solves the strong CP problem.


\section{Selected examples of models}\label{sec:examples}
In the following subsections, we list and discuss different benchmark models
for neutrino mass that are qualitatively different. We start with the most well-studied models, which are the Zee model, discussed in
Sec.~\ref{sec:ZeeModel}, that is the first 1-loop model for Majorana
neutrino masses, and the Zee-Babu model, revisited in Sec.~\ref{sec:ZeeBabuModel}, which is the
first 2-loop model. 
In Sec.~\ref{sec:KNT} we discuss the first 3-loop model~\cite{Krauss:2002px},
which was proposed by Krauss, Nasri, and Trodden and is commonly called
KNT-model, and its variants. It is also the first model with a stable dark
matter candidate.  
The scotogenic model is discussed in Sec.~\ref{subsec:scotogenic}. It generates neutrino mass at 1-loop order and similarly to the KNT-model it features a stable dark matter candidate due to
the imposed $Z_2$ symmetry. These are the most well-studied models in the literature. However this preference is mostly due to the
historic development (and also simplicity) and we are proposing a few other interesting benchmark models in the following subsections.


\subsection{Models with leptophillic particles} \label{sec:modelleptons}

There are only three different structures which violate lepton number (LN) by two units that can be constructed with SM fields~\cite{Cheng:1980qt}:
\begin{align}
\overline{ \tilde L}\,  \vec{\tau}\, L \sim \left(1, 3, -1\right) \,,
\qquad
\overline{ \tilde L}\, L  \sim \left( 1, 1, -1 \right)\,,
\qquad
\overline{e^c_{\rm R}} \,e_{\rm R} \sim \left( 1, 1, -2\right)  \; .
\end{align}
The three different structures can couple respectively to a 
$SU(2)$ triplet scalar with $Y=1$ (we denote it by $\Delta$), a 
singly-charged $SU(2)$ singlet scalar (we call it $h^{+}$) and a 
doubly-charged SU(2) singlet scalar (we call it $k^{++}$). 

In all cases, we could assign LN equal to -2 to the new fields so that such interactions preserve it. However dimension-3 terms in the scalar potential will softly break LN, as there is no symmetry to prevent them. 
In the first case, the triplet can have in the potential the lepton-number violating term (with $\Delta L=2$) with the SM Higgs doublet H
\begin{equation}
V_\Delta \subset \mu_\Delta \tilde{H}^\dagger\Delta^\dagger H+{\rm H.c.}\,.
\end{equation}
Then, after electroweak symmetry breaking, the triplet gets an induced VEV $v_T
\simeq - \mu_\Delta v^2/m_\Delta^2$ (strongly bounded by the T parameter to be
$\lesssim O(1)$ GeV), and neutrino masses are generated at tree-level via
the type-II seesaw.

If only the singly-charged scalar $h^+$ is present, a $\Delta L=2$ term can be
constructed with two Higgs doublets, the SM Higgs H and an extra Higgs doublet
$\Phi$
\begin{equation}\label{eq:muZee}
	V_{\rm Zee} \subset \mu_{\rm Zee} \tilde{H}^\dagger \Phi (h^+)^* +{\rm H.c.}\,.
\end{equation}
In this case, however, neutrino masses are not induced by the Higgs VEV at tree-level, but they are generated at 1-loop order. This is known as the
Zee model~\cite{Zee:1980ai,Wolfenstein:1980sy}. 

For the case of the doubly-charged scalar, one can construct the $\Delta L=2$
term precisely with two singly-charged scalars $h^+$
\begin{equation}\label{eq:muZB}
	V_{\rm ZB} \subset \mu_{\rm ZB}\, h^+ h^+ (k^{++})^* +{\rm H.c.}\,.
\end{equation}
Notice that no other combination with SM fields exist, given the large electric
charge of $k^{++}$. In this case,  neutrino masses are generated at 2-loop order. This
is known as the Zee-Babu model~\cite{Cheng:1980qt,Babu:2002uu}. 

These are the simplest radiative models. By using particles that couple to a
lepton and a quark (leptoquarks), one can also have $\Delta L=2$ interactions
and generate neutrino masses at a different number of loops. In the following,
we will discuss the Zee and Zee-Babu models.

\subsubsection{The Zee model}
\label{sec:ZeeModel}
In addition to the SM content with a Higgs scalar doublet $H$, the Zee
model~\cite{Zee:1980ai,Wolfenstein:1980sy} contains an extra Higgs
scalar doublet $\Phi$ and a singly-charged scalar singlet $h^+$, which is shown in Tab.~\ref{tab:ZeeModel}. It is an
example of the operator $O_2 = L^i L^j L^k e^c H^l \epsilon_{ij}\epsilon_{kl}$.
Several aspects of the phenomenology of the model have been studied in
Refs.~\cite{Petcov:1982en,Zee:1985id,Bertolini:1987kz,Bertolini:1990vz,Smirnov:1994wj,Smirnov:1996hc,Frampton:1999yn,Jarlskog:1998uf,Ghosal:2001ep,Kanemura:2000bq,Balaji:2001ex,Koide:2002uk,Brahmachari:2001rn,Frampton:2001eu,Assamagan:2002kf,He:2003ih,Kanemura:2005hr,AristizabalSierra:2006ri}.
While the Zee-Wolfenstein version where just the SM Higgs doublet couples to the
leptons has been excluded by neutrino oscillation
data~\cite{Koide:2001xy,He:2003ih}, the most general version of the Zee model in
		which both couple remains allowed ~\cite{He:2011hs} and has been
		recently studied in Ref.~\cite{Herrero-Garcia:2017xdu} (see
		also Refs.~\cite{Babu:2013pma,Aranda:2011rt} for a variant with a flavor-dependent $Z_4$ symmetry).
		\begin{figure}[tb]
			\begin{minipage}[b]{0.45\linewidth}\centering
  \begin{tikzpicture}[node distance=1cm and 1cm]
     \coordinate[label=left:$L$] (nu1);
     \coordinate[vertex, right=of nu1] (v1);
     \coordinate[vertex, right=of v1] (lfv);
     \coordinate[vertex, above=of lfv] (v3);
     \coordinate[above=of v3,  label=above:$H/\Phi$] (h1);
     \coordinate[below=of lfv, label=below:$H/\Phi$] (h2);
     \coordinate[vertex, right=of lfv] (v2);
     \coordinate[right=of v2, label=right:$L$] (nu2);
     \coordinate[above=of v1, label=left:$h^+$] (s1);
     \coordinate[above=of v2, label=right:$\Phi/H$] (s2);

     \draw[fermion] (nu1)--(v1);
     \draw[fermion] (lfv) -- node[below]{$L$} ++ (v1);
     \draw[fermion] (lfv)-- node[below]{$\bar{e}$} ++ (v2);
     \draw[fermion] (nu2)--(v2);
     \draw[scalar] (h1) -- (v3);
     \draw[scalar] (h2) -- (lfv);
     \draw[scalar] (v3) to[out=180,in=90] (v1);
     \draw[scalar] (v2) to[out=90,in=0] (v3);
   \end{tikzpicture}
   \captionof{figure}{1-loop neutrino masses generated in the Zee model in the flavor basis. } \label{Zee}
   \end{minipage}%
   \hfill
   \begin{minipage}[b]{0.45\linewidth}\centering
			\begin{tabular}{ccc}
			\toprule
			Field &  Spin & $G_{\rm SM}$ \\
			\midrule
			$h^+$ & 0 & $(1,1,1)$\\
			$\Phi$ & 0 & $(1,2,\frac12)$\\
			\bottomrule
		\end{tabular}
		\captionof{table}{Quantum numbers for new particles in the Zee model.}
		\label{tab:ZeeModel}
	\end{minipage}
	\end{figure}

The Yukawa Lagrangian is
\begin{equation}  \label{eq:yuklep}
-\mathcal{L}_{L}=
\overline{L}\, (Y^\dagger_1 H + Y^\dagger_2 \Phi)e_{\rm R}  +  \overline{\tilde{L}}f\,L h^{+}+\mathrm{H.c.}  \,,
\end{equation}
where $L=(\nu_{\rm L},\,e_{\rm L})^T$ and $e_{\rm R}$ are the SU(2) lepton
doublets and singlets, respectively, and $\tilde{L} \equiv i \tau_2 L^c = i
\tau_2 C \overline{L}^T$ with $\tau_2$ being the second Pauli matrix. Due
to Fermi statistics, $f$ is an antisymmetric Yukawa matrix in flavor space,
while $Y_1$ and $Y_2$ are completely general complex Yukawa matrices.
Furthermore, the charged-lepton mass matrix is given by
\begin{equation} \label{mass_ch}
m_E=\frac{v}{\sqrt{2}}(c_\beta Y^\dagger_1+s_\beta Y^\dagger_2)\,,
\end{equation}
where $\tan\beta=s_\beta/c_\beta=v_2/v_1$ with $\langle H^0 \rangle=v_1$ and $\langle \Phi^0 \rangle=v_2$ and $v^2=v_1^2+v_2^2$. Without loss of generality, one can work in the basis where $m_E$ is diagonal.

Assuming CP-invariance there are two CP-even neutral scalars (one of which is the 125
GeV Higgs boson, with mass $m_h$, and the other is a heavy one with mass
$m_H$), one neutral CP-odd scalar with mass $m_A$, and two charged-scalars of
masses $m_{h^+_{1,2}}$, whose mixing due to the trilinear term in
Eq.~\ref{eq:muZee} is given by
\begin{equation} \label{eq:ch_mixing}
s_{2\varphi} = \dfrac{\sqrt{2} v \mu_{\rm Zee}}{m_{h^+_2}^2-m_{h^+_1}^2}\,.
\end{equation}
Interestingly, $\mu_{\rm Zee}$ cannot be arbitrarily large, as it contributes at 1-loop level to the mass of the light Higgs. Demanding no fine-tuning, we can estimate $|\mu_{\rm Zee}|\, \lesssim 4 \pi\,m_h \simeq1.5 \, {\rm TeV}$.

The Yukawa couplings of Eq.~\ref{eq:yuklep}, together with the term in the potential given in Eq.~\ref{eq:muZee}, imply that lepton number is violated by the product $m_E\,(Y_1v_2-Y_2 v_1) \,f\, \mu_{\rm Zee}$. Therefore, neutrino masses will be necessarily generated, in particular the lowest order contribution appears at 1-loop order, as shown diagram of Fig.~\ref{Zee}, where the charged scalars run in the loop. The neutrino mass matrix is given by:
\begin{equation}
\mathcal{M}_\nu= A\,\Big[f\,m_E^2+m_E^2f^T-\frac{v}{\sqrt{2}\,s_\beta}(f\,m_E\,Y_2+Y_2^T\,m_E\,f^T)\Big]\,\ln\frac{m^2_{h^+_2}}{m^2_{h^+_1}},\,\,\qquad A\equiv \frac{s_{2\varphi}\, t_\beta}{8 \sqrt{2}\pi^2\, v}\,,
\end{equation}
with $\varphi$ being the mixing angle for the charged scalars given in Eq.~\ref{eq:ch_mixing}. Therefore, in the Zee model, due to the loop and the chiral suppressions, the new physics scale can be light. From the form of the mass matrix it is clear that if one takes $Y_2 \rightarrow 0$ (Zee-Wolfenstein model), the diagonal elements vanish, yielding neutrino mixing angles that are not compatible with observations.

Neglecting $m_e \ll m_\mu,\,m_\tau$ and taking $f_{e\mu}=0$, the following Majorana mass matrix is obtained
\begin{equation}
	\mathcal{M}_\nu =A\, \frac{m_\tau v}{\sqrt{2}\,s_\beta}%
		\begin{pmatrix} 
-2 f^{e\tau} Y_2^{\tau e} & -f^{e\tau} Y_2^{\tau\mu} - f^{\mu\tau} Y_2^{\tau e} & \frac{\sqrt{2} s_\beta\,m_\tau}{v} f^{e\tau} - f^{e\tau} Y_2^{\tau\tau}\\
-f^{e\tau} Y_2^{\tau\mu} - f^{\mu\tau} Y_2^{\tau e} & -2 f^{\mu\tau} Y_2^{\tau\mu} & \frac{\sqrt{2} s_\beta m_\tau}{v} f^{\mu\tau} - f^{\mu\tau} Y_2^{\tau\tau}\\
\frac{\sqrt{2} s_\beta\,m_\tau}{v} f^{e\tau} - f^{e\tau} Y_2^{\tau\tau} & \frac{\sqrt{2} s_\beta m_\tau}{v} f^{\mu\tau} - f^{\mu\tau} Y_2^{\tau\tau} & 2 \frac{m_\mu}{m_\tau}f^{\mu\tau} Y_2^{\mu\tau}
\end{pmatrix}\,.
\label{Mnu}
\end{equation}
Notice that if the term proportional to the muon mass is neglected, one
neutrino remains massless. In order to obtain correct mixing angles, we need
both $Y_2^{\tau\mu}$ and $Y_2^{\tau e}$ different from zero
\cite{Herrero-Garcia:2016uab,Herrero-Garcia:2017xdu}, as they enter in the 1-2
submatrix of Eq.~\ref{Mnu}. This implies that LFV mediated by the scalars will
be induced. In fact, in the model large LFV signals are generated, like $\tau
\to \mu \gamma$ and $\mu-e$ conversion in nuclei. Moreover, also a full
numerical scan of the model performed in Ref.~\cite{Herrero-Garcia:2017xdu}
showed that large LFV Higgs decays are possible, in particular ${\rm BR} (h
\rightarrow \tau \mu)$ can reach the percent level. ${\rm BR} (h \rightarrow
\tau e)$ is roughly two-orders of magnitude smaller than ${\rm BR} (h
\rightarrow \tau \mu)$. The singly-charged h also generates violations of
universality, as it interferes constructively with the W boson, as well as
non-standard interactions, see Sec.~\ref{sec:univ_NSI}, which however are too
small to be observed~\cite{Herrero-Garcia:2017xdu}. 

In Ref.~\cite{Herrero-Garcia:2017xdu} it was also shown that the model is
testable in next-generation experiments. While normal mass ordering (NO) provided a
good fit, inverted mass ordering (IO) is disfavored, and if $\theta_{23}$ happens to be in the second
octant, then IO will be ruled-out. Notice also that the lightest neutrino is required to be massless for IO, as it has also been obtained in Ref.~\cite{He:2011hs}.
Furthermore, future $\tau \rightarrow \mu \gamma$ ($\mu-e$ conversion) will
test most regions of the parameter space in NO (IO). Regarding direct searches
at the LHC, the new scalars have to be below $\sim2$ TeV, which implies that
they can be searched for similarly as in a two-Higgs doublet model (with an
extra charged scalar that could be much heavier). Particularly, the charged
scalars are searched for at colliders. See the discussion in Sec.~\ref{sec:colliders}.

Let us mention that an interesting modification of the Zee model was proposed in Ref.~\cite{Babu:2013pma} (see also Ref.~\cite{Aranda:2011rt}), where a $Z_4$ symmetry was imposed, being able to reduce significantly the number of parameters. In that case, among the predictions of the model, is that the spectrum should be inverted.  
Other flavor symmetries beyond $Z_4$ in this framework have been studied in Refs.~\cite{Babu:1989wn,Babu:1990wv,Koide:2000jm,Kitabayashi:2001ex,Adhikary:2006wi,Fukuyama:2010ff,Aranda:2010im,Aranda:2011dx}.

\subsubsection{The Zee-Babu model}
\label{sec:ZeeBabuModel}
The Zee-Babu model contains, in addition to the SM, two $SU(2)$ singlet scalar
fields with electric charges one and two, denoted by $h^+$ and
$k^{++}$~\cite{Cheng:1980qt,Babu:2002uu} as shown in Tab.~\ref{tab:ZeeBabuModel}. It is a UV completion of the operator $O_9 =
L^i L^j L^k e^c L^l e^c \epsilon_{ij}\epsilon_{kl}$. Several studies of its
phenomenology exist in the
literature~\cite{AristizabalSierra:2006gb,Nebot:2007bc,Ohlsson:2009vk,Herrero-Garcia:2014hfa,Schmidt:2014zoa}. 

\begin{figure}[bt]\centering
	\begin{minipage}[b]{0.45\linewidth}\centering
		\begin{tikzpicture}[node distance=1cm and 1cm]
     \coordinate[label=left:$L$] (nu1);
     \coordinate[vertex, right=of nu1] (v1);
     \coordinate[vertex, right=0.75 of v1] (v1a);
     \coordinate[vertex, right=0.75 of v1a] (lfv);
     \coordinate[vertex, above=1.5 of lfv] (v3);
     \coordinate[vertex, right=0.75 of lfv] (v2a);
     \coordinate[vertex, right=0.75 of v2a] (v2);
     \coordinate[right=of v2, label=right:$L$] (nu2);
     \draw[scalar] (v3) to[out=0,in=90] node[label=above:$h^+$] {} (v2);
     \draw[scalar] (v3) to[out=180,in=90] node[label=above:$h^+$] {} (v1);
     \coordinate[below=of v2a, label=below:$H$] (h2);
     \coordinate[below=of v1a, label=below:$H$] (h1);
     \draw[fermion] (nu1)--(v1);
     \draw[fermion] (v1a)--  node[label=below:$L$] {} (v1);
     \draw[fermion] (v1a) -- node[label=below:$\bar{e}$] {} (lfv);
     \draw[fermion] (v2a)--  node[label=below:$\bar{e}$] {} (lfv) ;
     \draw[fermion] (v2a)--  node[label=below:$L$] {} (v2);
     \draw[fermion] (nu2)--(v2);
     \draw[scalar] (lfv) -- node[midway,label=right:$k^{++}$] {} (v3);
     \draw[scalar] (h1) -- (v1a);
     \draw[scalar] (h2) -- (v2a);
   \end{tikzpicture}
   \captionof{figure}{2-loop neutrino masses generated in the Zee-Babu model. }
\label{fig:babu}
\end{minipage}
\hfill
\begin{minipage}[b]{0.45\linewidth} \centering
			\begin{tabular}{ccc}
			\toprule
			Field &  Spin & $G_{\rm SM}$ \\
			\midrule
			$h^+$ & 0 & $(1,1,1)$\\
			$k^{++}$ & 0 & $(1,1,2)$\\
			\bottomrule
		\end{tabular}
		\captionof{table}{Quantum numbers for new particles in the Zee-Babu model.}
		\label{tab:ZeeBabuModel}
	\end{minipage}
\end{figure}
The leptonic Yukawa Lagrangian reads:
\begin{equation}
\mathcal{L}_{L}=
\overline{L}\, Y^\dagger \, e_{\rm R} H +  \overline{\tilde{L}}f L h^{+}+\overline{e_{\rm R}^{c}}g\, e_{\rm R}\, k^{++}+
\mathrm{H.c.}  \,, \label{eq:babuL}
\end{equation}
where like in the Zee model, due to Fermi statistics, $f$ is an antisymmetric matrix in flavor space. On the other hand, $g$ is symmetric. Charged lepton masses are given by $m_E=\frac{v}{\sqrt{2}}Y^\dagger$, which be take to be diagonal without loss of generality.

	Lepton number is violated by the simultaneous presence of the trilinear term
$\mu_{\rm ZB}$ in Eq.~\ref{eq:muZB}, together with $m_E,\,f,\,g$. Note that
the trilinear term cannot be arbitrarily large, as it contributes to the
charged scalar masses at loop level, and can also lead to charge-breaking
minima, if $|\mu_{\rm ZB}|$ is large compared to the charged scalar masses. For
naturalness considerations we demand $|\mu_{\rm ZB}|\ll 4\pi \,  {\rm
min}(m_{h},m_k)$. See Refs.~\cite{Nebot:2007bc,Herrero-Garcia:2014hfa} for 
detailed discussions.

As lepton number is not protected, neutrino masses are generated radiatively, in particular at 2-loop order, via the diagram of Fig.~\ref{fig:babu}.
The mass matrix is approximately given by (see for instance Refs.~\cite{McDonald:2003zj,Nebot:2007bc,Herrero-Garcia:2014hfa} for more details)
\begin{equation}
\mathcal{M}_{\nu} \simeq \frac{v^{2}\mu_{\rm ZB}}{96\pi^{2}M^{2}}\, f\, Y\, g^{\dagger}Y^{T}f^{T}\,,\label{eq:Mnu}
\end{equation}
where $M$ is the heaviest mass of the loop, either that of the singly-charged singlet $h^+$ or of the doubly-charged singlet $k^{++}$. A prediction of the model is that, since $f$ is a $3\times 3$ antisymmetric
matrix, $\det f=0$, and therefore $\det\mathcal{M}_{\nu}=0$. Thus,
at least one of the neutrinos is exactly massless at this order.

In the model, both NO and IO can be accommodated. The phenomenology of the singly-charged scalar is similar to that discussed in the Zee model, apart from the fact that in the Zee model the charged singlet mixes with the charged component of the doublet. Some of the most important predictions of the model are due to the presence of the doubly-charged scalar $k^{++}$. Firstly, $k^{++}$ mediates trilepton decays ($\ell_i \to \ell_j \overline{\ell_k} \ell_l$) at tree-level which unlike, in the Zee model, are not suppressed by the small charged lepton masses, as well as radiative decays ($\ell_i \to \ell_j \gamma$). Secondly, $k^{++}$ can be pair-produced at the LHC via Drell-Yan, decaying among other final states into same-sign leptons which yields a clean experimental signature. See the discussion in Sec.~\ref{sec:colliders}.


\subsection{KNT-models}
\label{sec:KNT}
The first radiative neutrino mass model at 3-loop order is the KNT model~\cite{Krauss:2002px} which  
has one fermionic singlet $N$ and two singly-charged scalars $S_{1,2}$ in addition 
to the SM particles. A discrete $Z_2$ symmetry is imposed, under which only $S_2$ and $N$ are odd. 
We list the quantum numbers of the exotic particles in Tab.~\ref{tab:KNT}. 

The $Z_2$ symmetry forbids the usual type-I seesaw contribution at tree-level. 
The relevant Lagrangian is expressed as 
\begin{align}
	\mathcal{L}&= f \; L^T C i\tau_2  L S_1^* + g\; \overline{N^c} e_{\rm R} S_2^* + \frac{1}{2}M_N N^T C N + {\rm H.c.} \\
           & + M_{S_1} S_1 S_1^* + M_{S_2} S_2 S_2^* + \frac{1}{4}\lambda_S (S_1 S_2^*)^2    \; ,  
\end{align}    
where the flavor indices of $f$ and $g$ are all suppressed. 
With this setup, neutrino masses are generated first at 3-loop order as shown in Fig.~\ref{fig:KNTmodel}.
The neutrino mass matrix is then
\begin{align}
\left(\mathcal{M}_\nu\right)_{ij} = \sum_{\alpha\beta}\frac{\lambda_S}{(4\pi^2)^3}\frac{m_\alpha m_\beta}{M_{S_2}}
                    f_{i\alpha} f_{j\beta} g_\alpha^* g_\beta^* F\left( \frac{M_N^2}{M_{S_2}^2}, \frac{M_{S_1}^2}{M_{S_2}^2}\right)\,,
\end{align}
where the function $F$ is defined in Ref.~\cite{Ahriche:2013zwa}.
This matrix is, however, only rank one and thus can give exactly one nonzero neutrino mass.
Adding more copies of $N$ can increase the rank of the matrix.
The phenomenology of this model including flavor physics, dark matter, Higgs decay, electroweak phase transition 
and collider searches is discussed in detail in Ref.~\cite{Ahriche:2013zwa}.

\begin{figure}[tb]\centering
	\begin{minipage}[b]{0.45\linewidth}
  \begin{tikzpicture}[node distance=1cm and 1cm]
   \coordinate[label=left:$L$] (nu1);
   \coordinate[vertex, right of=nu1] (v1);
   \coordinate[vertex, right of=v1, xshift=1cm] (v03);
   \coordinate[vertex, right of=v03, xshift=1cm] (v06);
   \coordinate[vertex, right of=v06, xshift=1cm] (v2);
   \coordinate[right of=v2, label=right:$L$] (nu2);
   \coordinate[] (v5) at ($(v1)!0.5!(v2)$);
   \coordinate[vertex, above of=v5, yshift=1cm] (v05);

    \coordinate[vertex] (ve1) at ($(v1)!0.5!(v03)$);
    \coordinate[vertex] (ve2) at ($(v2)!0.5!(v06)$);
    \node [below=of ve1] (h1) {$H$};
    \node [below=of ve2] (h2) {$H$};
    \draw[fermion] (ve1)--node[below]{$L$} ++ (v1);
    \draw[fermion] (ve1)--node[below]{$\bar e$} ++ (v03);
    \draw[fermion] (ve2)--node[below]{$L$} ++ (v2);
    \draw[fermion] (ve2)--node[below]{$\bar e$} ++ (v06);
    \draw[fermionnoarrow] (v03)-- node[midway,cross] {} node[below]{$N$} ++ (v06);
    \draw[fermion] (nu1)--(v1);
    \draw[fermion] (nu2)--(v2);
    \draw[scalar] (v05) --node[left, xshift=-0.2cm]{$S_1$} ++ (v1);
    \draw[scalar] (v05) --node[right, xshift=0.2cm]{$S_1$} ++ (v2);
    \draw[scalar] (v03) --node[left]{$S_2$} ++ (v05);
    \draw[scalar] (v06) --node[right]{$S_2$} ++ (v05);
    \draw[scalar] (h1) -- (ve1);
    \draw[scalar] (h2) -- (ve2);
  \end{tikzpicture}
  \captionof{figure}{3-loop neutrino masses generated in the KNT model.} 
  \label{fig:KNTmodel}
\end{minipage}
\hfill
\begin{minipage}[b]{0.45\linewidth}\centering
   \begin{tabular}{ccccc}
   \toprule
   Field & Spin & $G_{\rm SM}$ & $Z_2$ \\
   \midrule
   $S_1$ & 0  & $(1, 1, -1$) & $+$\\
   $S_2$ &  0 & $(1, 1, -1)$ & $-$ \\ 
   $N$ & $\frac{1}{2}$  & $(1,1,0)$ & $-$ \\ 
   \bottomrule
   \end{tabular}
   \captionof{table}{Quantum numbers for new particles in the original KNT model.}
   \label{tab:KNT}
   \end{minipage}
   \end{figure}
       
This model is subject to constraints from LFV experiments such as $\mu\to e\gamma$ 
which requires three copies of $N$ for the neutrino mixing to be in agreement with the observations.\footnote{
Less copies of $N$ means less contribution to the neutrino mass matrix, which in turn generally leads to 
larger Yukawa couplings to generate the same neutrino mass scale and thus more likely to violate 
constraints from LFV processes. }
Meanwhile in order to be consistent with the measurements of muon anomalous magnetic moment 
and the $0\nu \beta\beta$ decay,
strong constraints are imposed.
For $M_{S_1, S_2}\geqslant 100$ GeV, $10^{-5}\lesssim\left|g_{i1}g_{i2}\right|\lesssim 10$ and 
$10^{-5}\lesssim \left|f_{13} f_{23}\right|\lesssim 1 $, it can satisfy all flavor constraints while reproducing 
the neutrino mixing data.        

Assuming a mass hierarchy $M_N< M_{S_2}$, the lightest fermion singlet is stable and serves as a good DM candidate.   
This is also the first radiative neutrino mass theory with a stable DM candidate running in the loop.
If the DM relic density is saturated and all previously discussed constraints are satisfied, 
the DM mass cannot exceed 225 GeV while the lighter charged scalar $S_2$ cannot be heavier than 245 GeV.
If the fermion singlets have very small mass splitting, DM coannihilation effects should be taken into account.
With about $5\%$ mass splitting, the DM relic density increases by $50\%$.

As discussed in Sec.~\ref{subsubsec:newparticles}, 
the singly-charged scalars can be pair-produced at the LHC and subsequently decay to a pair of charged leptons and 
the fermion singlets which appear as missing transverse energy. 
This signature is exactly the same as the direct slepton pair production in SUSY theories. 
ATLAS has performed the search for sleptons in this channel with 36.1 fb$^{-1}$ data of $\sqrt{s}=13$ TeV~\cite{ATLAS:2017uun} and has ruled out slepton masses below $\sim 500$ GeV in the non-compressed region. 
The actual constraint on $M_{S_2}$ depends on the decay branching ratio of $S_2$ to different leptons
and in principle will be substantially relaxed compared to the ATLAS search.

With the same topology, a lot of variations of the KNT model can be constructed.
Reference~\cite{Chen:2014ska} discusses several possibilities to
replace the electron with other SM fermions\footnote{The authors of Ref.~\cite{Chen:2014ska} also point out that up-quarks are not feasible due to gauge invariance.} or
vector-like fermions. 
A similar model in which the electron is replaced by a fermion doublet with hypercharge
$5/2$ and $S_{1,2}$ with doubly-charged scalar 
is discussed in Ref.~\cite{Okada:2016rav}.
The $Z_2$-odd particles in this model form instead the outer loop.


\subsection{The scotogenic model}
\label{subsec:scotogenic}

The most popular model linking dark matter to the radiative generation
of neutrino masses is the one proposed by E. Ma in 2006. We will refer to it as \textit{scotogenic
  model}~\cite{Ma:2006km}.\footnote{The scotogenic model has been
  extensively studied, sometimes referring to it with different
  names. For instance, some authors prefer the denomination
  \textit{radiative seesaw}. In this review we will stick to the more
  popular name \textit{scotogenic model}, which comes from the Greek
  word \textit{skotos} (\textsigma \textkappa \textomikron \texttau
  \textomikron \textvarsigma), \textit{darkness}. \textit{scotogenic}
  would then mean \textit{created from darkness}.} In the scotogenic
model, the SM particle content is extended with three
singlet fermions, $N_i$ ($i=1,2,3$), and one $SU(2)_{\rm L}$ doublet,
$\eta$, with hypercharge $\frac{1}{2}$,
\begin{equation}
\eta = \left( \begin{array}{c}
\eta^+ \\
\eta^0
\end{array} \right) \, .
\end{equation}
This setup is supplemented with a $Z_2$ parity, under which
the new states are odd and all the SM particles are
even.\footnote{The $Z_2$ symmetry can obtained from the
  spontaneous breaking of an Abelian $U(1)$ factor, see for instance
 Refs.~\cite{Sierra:2014kua}.} The newly-introduced particles with their respective charges of the scotogenic model are shown in Tab.~\ref{tab:scotogenic}.
\begin{figure}[tb]\centering
	\begin{minipage}[b]{0.45\linewidth}\centering
		\begin{tikzpicture}[node distance=1cm and 1cm]
     \coordinate[label=left:$L$] (nu1);
     \coordinate[vertex, right=of nu1] (v1);
     \coordinate[cross, right=of v1] (lfv);
     \coordinate[vertex, above=of lfv] (v3);
     \coordinate[above left=of v3,  label=left:$H$] (h1);
     \coordinate[above right=of v3,  label=right:$H$] (h2);
     \coordinate[vertex, right=of lfv] (v2);
     \coordinate[right=of v2, label=right:$L$] (nu2);
     \coordinate[above=of v1, label=left:$\eta$] (s1);
     \coordinate[above=of v2, label=right:$\eta$] (s2);

     \draw[fermion] (nu1)--(v1);
     \draw[fermion] (lfv) -- node[below]{$N$} ++ (v1);
     \draw[fermion] (lfv)-- node[below]{$N$} ++ (v2);
     \draw[fermion] (nu2)--(v2);
     \draw[scalar] (h1) -- (v3);
     \draw[scalar] (h2) -- (v3);
     \draw[scalar] (v3) to[out=180,in=90] (v1);
     \draw[scalar] (v3) to[out=0,in=90] (v2);
   \end{tikzpicture}
   \captionof{figure}{1-loop neutrino masses generated in the scotogenic model.}
	\label{fig:scotogenic}
\end{minipage}
\hfill
\begin{minipage}[b]{0.45\linewidth}\centering
\begin{tabular}{cccccc} 
\toprule 
Field & Spin  & Generations & $G_{\rm SM}$ & $Z_2$ \\ 
\midrule
\(\eta\) & \(0\)  & 1 & \( (1, 2, \frac{1}{2}) \) & $-$ \\ 
\(N\) & \(\frac{1}{2}\)  & 3 & \(( 1, 1,0) \) & $-$ \\
\bottomrule
\end{tabular}
\captionof{table}{Quantum numbers of new particles in the scotogenic model.}
\label{tab:scotogenic}
\end{minipage}
\end{figure}
The gauge and discrete symmetries of the model allow us to write
the Lagrangian terms involving the fermion singlets
\begin{equation}
\mathcal{L}_N=
\frac{M_N}{2} \, \overline{N^c} \, N+
Y_N \, \eta \, \overline{N} \, L + {\rm H.c.} \, .
\end{equation}
We do not write the kinetic term for the fermion singlet as it takes
the standard canonical form. $Y_N$ is an arbitrary $3 \times 3$
complex matrix, whereas the $3 \times 3$ Majorana mass matrix $M_N$
can be taken to be diagonal without loss of generality. We highlight
that the usual neutrino Yukawa couplings with the SM Higgs doublet are
not allowed due to the $Z_2$ symmetry. This is what prevents
the light neutrinos from getting a nonzero mass at tree-level. The
scalar potential of the model is given by
\begin{align}\label{eq:scotogenicScalarPotential}
\mathcal{V} =& \:
-m_{H}^2 H^\dagger H+m_\eta^2\eta^\dagger\eta+
\frac{\lambda_1}{2}\left( H^\dagger H\right)^2+
\frac{\lambda_2}{2}\left(\eta^\dagger\eta\right)^2+
\lambda_3\left( H^\dagger H\right)\left(\eta^\dagger\eta\right)\nonumber\\
& +
\lambda_4\left( H^\dagger\eta\right)\left(\eta^\dagger H\right)+
\frac{\lambda_5}{2}\left[\left( H^\dagger\eta\right)^2+
\left(\eta^\dagger H\right)^2\right] \, .
\end{align}
Neutrino masses are induced at the 1-loop level via the diagram in Fig.~\ref{fig:scotogenic} 
\begin{equation}
	\left(\mathcal{M}_\nu\right)_{ij} = \sum_{k=1}^3 \frac{Y_{Nki}Y_{Nkj}}{32\pi^2} M_{Nk} \left[ \frac{m_R^2}{m_R^2-M_{Nk}^2} \ln\left(\frac{m_R^2}{M_{Nk}^2}\right) - \frac{m_I^2}{m_I^2-M_{Nk}^2} \ln\left(\frac{m_I^2}{M_{Nk}^2}\right)\right]\;,
\end{equation}
where the masses of the scalar $\eta_R$ and pseudo-scalar part $\eta_I$ of the neutral scalar $\eta^0=(\eta_R+i\eta_I)/\sqrt{2}$ are given by
\begin{align}
	m_{R,I}^2 & = m_\eta^2 + \frac12 \left(\lambda_3+\lambda_4 \pm\lambda_5\right) v^2
\end{align}
with the electroweak VEV $v=\sqrt{2}\braket{H^0}\simeq 246 \mathrm{GeV}$. Neutrino mass vanishes in the limit of $\lambda_5=0$ and thus degenerate masses for the neutral scalars $\eta_{R,I}$, because it is possible to define a generalized lepton number which forbids a Majorana mass term.

In the scotogenic model, the $Z_2$ parity is assumed to be
preserved after electroweak symmetry breaking. This will be so if
$\langle \eta \rangle = 0$. In this case, the lightest
$Z_2$-odd state (to be identified with the LCP defined in Sec.~\ref{sec:DM})
will be stable and, if neutral, will constitute a potentially good DM
candidate. The LCP in the scotogenic model can be either a fermion or
a scalar: the lightest singlet fermion $N_1$ or the lightest neutral
$\eta$ scalar ($\eta_R$ or $\eta_I$). As the neutrino Yukawa couplings are generally required to be small to satisfy LFV constraints, the DM phenomenology for a scalar
LCP is generally the same as in the inert doublet
model~\cite{Deshpande:1977rw,LopezHonorez:2006gr}. Recently it has been pointed
out~\cite{Borah:2017dfn} that late decay of the lightest SM singlet fermion
$N_1$ may repopulate the dark matter abundance and thus resurrect the
intermediate dark matter mass window between $m_W$, the mass of the $W$ boson, and $550$ GeV. In the case of a fermionic LCP, for which the annihilation cross section is governed by the neutrino Yukawa couplings, the connection of the dark matter abundance 
with neutrino masses leads to a very
constrained scenario due to the bounds from lepton flavor violation
\cite{Kubo:2006yx,Sierra:2008wj,Suematsu:2009ww,Adulpravitchai:2009gi,Toma:2013zsa,Vicente:2014wga}. 

Many \textit{scotogenic variations} have been proposed since the
publication of the minimal model described above. All these models are
characterized by neutrino masses being induced by new dark sector
particles running in a
loop~\cite{Ma:2008cu,Chao:2012sz,vonderPahlen:2016cbw,Hirsch:2013ola,Farzan:2009ji,Hehn:2012kz,Lu:2016dbc,FileviezPerez:2009ud,Liao:2009fm,Farzan:2010mr,Okada:2015vwh,Brdar:2013iea,Aoki:2011yk,Cai:2011qr,Chen:2011bc,Law:2013saa,Chowdhury:2015sla,Chen:2009gd,Patra:2014sua,Ferreira:2016sbb,Fortes:2017ndr,Ma:2008ym,Bhattacharya:2013mpa,Ma:2013xqa,Ma:2014eka,Adulpravitchai:2009re,Parida:2011wh,Ma:2013nga,Fraser:2014yha,Ma:2012ez}. One
of them involves a global continuous dark symmetry, instead of a discrete
dark symmetry~\cite{Ma:2013yga,generalized}. A gauge dark symmetry was
considered in Ref.~\cite{Yu:2016lof} and a scale-invariant version
presented in Ref.~\cite{Ahriche:2016cio}. The collider
\cite{Ho:2013hia,Ho:2013spa,Hessler:2016kwm,Diaz:2016udz} and dark
matter
\cite{Molinaro:2014lfa,Faisel:2014gda,Chowdhury:2016mtl}
phenomenologies of different scotogenic variants have also been
discussed in detail. Finally, we point out that the authors of
Ref.~\cite{Merle:2015gea} identified a potential problem in this
family of models, since some parameter regions lead to the breaking of
the $Z_2$ parity at high energies. This problem, how it can
be escaped and its phenomenological implications have been explored in Refs.~\cite{Merle:2015ica,Merle:2016scw,Lindner:2016kqk}.


\subsection{Models with leptoquarks}
\label{sec:leptoquarks}
Leptoquarks are common ingredients of radiative neutrino mass models. For
example neutrino mass can be generated at loop level by two leptoquarks which
mix via a trilinear coupling to the SM Higgs
boson~\cite{Nieves:1981tv,Chua:1999si,Mahanta:1999xd,AristizabalSierra:2007nf,Helo:2015fba,Pas:2015hca,Cheung:2016fjo,Dorsner:2017wwn}.
Neutrino mass generation at 1-loop order with all possible leptoquarks has
been systematically studied in Ref.~\cite{AristizabalSierra:2007nf}.
At 1-loop order and especially at a higher-loop order, leptoquarks usually appear together with other exotic particles such as 
vector-like quarks and leptons, charged scalar singlets and electroweak multiplets~\cite{Cai:2014kra}. 
We will review two models here, one at 1-loop and one at 2-loop order. 

\subsubsection{A 1-loop model}
\label{sec:leptoquarks-1}
\begin{figure}[bt]\centering
	\begin{minipage}[b]{0.45\linewidth}\centering
	\begin{tikzpicture}[node distance=1cm and 1cm]
     \coordinate[label=left:$L$] (nu1);
     \coordinate[vertex, right=of nu1] (v1);
     \coordinate[vertex, right=of v1] (lfv);
     \coordinate[vertex, above=of lfv] (v3);
     \coordinate[above=of v3,  label=above:$H$] (h1);
     \coordinate[below=of lfv, label=below:$H$] (h2);
     \coordinate[vertex, right=of lfv] (v2);
     \coordinate[right=of v2, label=right:$L$] (nu2);
     \coordinate[above=of v1, label=left:$S_1/S_3$] (s1);
     \coordinate[above=of v2, label=right:$\tilde{R}_2$] (s2);

     \draw[fermion] (nu1)--(v1);
     \draw[fermion] (lfv) -- node[below]{$Q$} ++ (v1);
     \draw[fermion] (lfv)-- node[below]{$\bar{d}$} ++ (v2);
     \draw[fermion] (nu2)--(v2);
     \draw[scalar] (h1) -- (v3);
     \draw[scalar] (h2) -- (lfv);
     \draw[scalar] (v3) to[out=180,in=90] (v1);
     \draw[scalar] (v3) to[out=0,in=90] (v2);
   \end{tikzpicture}
   \captionof{figure}{1-loop neutrino masses generated via leptoquark mixing.}
   \label{fig:1loop-LQ}
   \end{minipage}
   \hfill
   \begin{minipage}[b]{0.45\linewidth}\centering
\begin{tabular}{ccc}
\toprule
Field & Spin & $G_{\rm SM}$  \\ 
\midrule 
$S_1$ & $0$  & $(\bar{3},1, \frac{1}{3})$\\
$S_3$ & $0$ & $(\bar{3},3, \frac{1}{3})$ \\
$\tilde{R}_2$  & 0 & $\left(3, 2, \frac{1}{6}\right)$\\
\bottomrule
\end{tabular}
\captionof{table}{Quantum numbers of leptoquarks.}
\label{tab:oneloopleptoquark}
\end{minipage}
\end{figure}
Without introducing exotic fermions, 
the only possible topology that can contribute at 1-loop order to the Weinberg operator is 
T1-ii shown in Fig.~\ref{fig:WeinbergOneLoop} as we need the fermion arrow to flip only once.         
With this topology and leptoquarks as the only exotic particles,
the only UV completion we can realize is depicted in Fig.~\ref{fig:1loop-LQ}.
The relevant scalar leptoquarks\footnote{We follow the nomenclature in Ref.~\cite{Dorsner:2016wpm, Buchmuller:1986zs} for the names of the leptoquarks, where subscripts indicate dimension of the $SU(2)_{\rm L}$ representations.} are $S_1$, $S_3$ and $\tilde{R}_2$ with quantum numbers detailed in Tab.~\ref{tab:oneloopleptoquark}.
The relevant Lagrangian reads
\begin{align}
	\Delta\mathcal{L}&= y_1 \overline{Q^c} L S_1 + y_3 \overline{Q^c} S_3 L + \tilde{y}_2 \bar{d} L \tilde{R}_2 +\lambda_1 S_1^* \tilde{R}_2^\dagger H
 +\lambda_3 \tilde{R}_2^\dagger S_3^\dagger H + {\rm H.c. } \; ,
\label{eqn:leptoquark1loop}
\end{align}
following the convention in Ref.~\cite{Dorsner:2017wwn} with all generation indices suppressed.
Apparently only the leptoquark component fields with electric charge $Q=-\frac{1}{3}$ can contribute.
These leptoquarks, in the interaction basis $(S_1 , S_3^{\frac{1}{3}}, \tilde{R}_2^{-\frac{1}{3} *})$,  
will mix with each other through the $\lambda_{1,3}$ terms 
in Eq.~\ref{eqn:leptoquark1loop}.\footnote{  
Reference~\cite{AristizabalSierra:2007nf} considered the most general interactions with all possible leptoquarks and found
in total four mass matrices for leptoquarks with electric charges $Q = -\frac{1}{3}$, $-\frac{2}{3}$, 
$-\frac{4}{3}$ and $-\frac{5}{3}$. }
We will consider simplified scenarios where either $S_1$ or $S_3$ appears together with $\tilde{R}_2$. 
For the model with $S_{1,3}$, the squared-mass matrix will be diagonalized with angle $\theta_{1,3}$ 
and the mass eigenvalues are $m_{1}$ and $m_{2}$.  
So the neutrino mass matrix is expressed as~\cite{Dorsner:2017wwn, AristizabalSierra:2007nf} 
\begin{align}
\mathcal{M}_\nu & \simeq \frac{3 \sin 2\theta_{1,3} }{32\pi^2} \ln \frac{m_{2}^2}{m_{1}^2} 
 \left(\tilde{y}_2^T \  M_d \ y_{1,3} +  y_{1,3}^T \ M_d \ \tilde{y}_2      \right) \; ,
\end{align} 
where $M_d=\mathrm{diag}(m_d, m_s, m_b)$ with $m_{d,s,b}$ being the down, strange and bottom quark masses.
Due to the hierarchy of down-type quark masses,
the neutrino mass matrix will be approximately rank-2 with one nearly massless neutrino.
Current neutrino oscillation data put lower bounds on the product of Yukawa couplings ranging from $10^{-12}$ to $10^{-7}$
for leptoquarks with TeV scale masses~\cite{AristizabalSierra:2007nf}.
On the other hand, low energy precision experiments constrain the Yukawa couplings from above.
For example, $\mu-e$ conversion in titanium bounds the first generation Yukawa couplings with
\begin{align}
\left(\tilde{y}_2\right)_{11} \left(\tilde{y}_2\right)_{21} & < 2.6\times 10^{-3}\,, & 
\qquad 
\left(y_3\right)_{11} \left(y_3\right)_{21} &< 1.7\times 10^{-3} \; , & 
\end{align}    
for 1 TeV leptoquark masses. 
Their decay branching fractions 
are dictated by the same couplings that determine the neutrino masses and mixings, 
which leads to a specific connection between the decay channels of the leptoquark and the neutrino mixings.
Generally LFV decays with similar branching ratios to final states with muon and tau are expected 
in some leptoquark decays.   
This neutrino mass model can also be tested at colliders. 
The leptoquarks running in the loop can be created in pairs and decay to final states containing leptons plus jets 
with predicted branching ratios. 
We refer to Sec.~\ref{sec:colliders} for further details on searches of leptoquarks at colliders.
\begin{figure}[bt]\centering
	\begin{minipage}[b]{0.45\linewidth}\centering
\begin{tikzpicture}[node distance=1cm and 1cm]
     \coordinate[label=left:$L$] (nu1);
     \coordinate[vertex, right=of nu1] (v1);
     \coordinate[right=0.75 of v1] (v1a);
     \coordinate[vertex, right=0.75 of v1a] (lfv);
     \coordinate[vertex, above=1.5 of lfv] (v3);
     \coordinate[vertex, right=0.75 of lfv] (v2a);
     \coordinate[vertex, right=0.75 of v2a] (v2);
     \coordinate[right=of v2, label=right:$L$] (nu2);
     \path (v3) to[out=180,in=90] node[vertex,midway] (v13) {} (v1);
     \draw[fermion] (v13) to[in=180,out=45] node[label=above:$\bar d$] {} (v3) ;
     \draw[fermion] (v13) to[out=225,in=90] node[label=left:$Q$] {} (v1) ;
     \draw[scalar] (v3) to[out=0,in=90] node[label=above:$S_1$] {} (v2);
     \coordinate[above left=0.7 of v13, label=above left:$H$] (h1);
     \coordinate[below=of v2a, label=below:$H$] (h2);
     \draw[fermion] (nu1)--(v1);
     \draw[scalar] (lfv) -- node[label=below:$S_1$] {} (v1);
     \draw[fermion] (v2a)--  node[label=below:$\bar d$] {} (lfv) ;
     \draw[fermion] (v2a)--  node[label=below:$Q$] {} (v2);
     \draw[fermion] (nu2)--(v2);
     \draw (lfv) -- node[cross,midway,black,label=right:$f$] {} (v3);
     \draw[scalar] (h1) -- (v13);
     \draw[scalar] (h2) -- (v2a);
   \end{tikzpicture}
   \captionof{figure}{2-loop neutrino masses generated in the Angelic model.}
   \label{fig:2loop-LQ}
\end{minipage}
\hfill
\begin{minipage}[b]{0.48\linewidth}\centering
\begin{tabular}{ccccc}
\toprule
Field &Spin & Generation & $G_{\rm SM}$ & B\\
\midrule
$S_1$ & 0 & 2&  $(\bar{3}, 1, \frac{1}{3})$ & -1 \\
$f$ & $\frac{1}{2}$ & 1 & $(8, 1, 0)$ & 0\\
\bottomrule
\end{tabular}
\captionof{table}{Quantum numbers of new particles in the Angelic model.}
\label{tab:o11b}
\end{minipage}
\end{figure}

Reference~\cite{Pas:2015hca} explored the possibility to explain the anomalous $b\to s ll$ transitions with $S_3$ and $\tilde{R}_2$.
Different texture of the Yukawa coupling matrices $y_3$ and $\tilde{y}_2$ were considered and leptoquark masses in the 
the range of 1 to 50 TeV can reproduce the neutrino masses and mixings in addition to $R_K$~\cite{Aaij:2014ora}.       

\subsubsection{A 2-loop model}
\label{sec:leptoquarks-2}
Based on the gauge-invariant effective operator $\mathcal{O}_{11b}=LLQd^c Q d^c$, which violates lepton number by two units,
a UV complete radiative neutrino mass model at 2-loop order containing leptoquark $S_1$ and fermion color octet $f$ 
can be constructed~\cite{Angel:2013hla}. 
We list their quantum numbers in Tab.~\ref{tab:o11b} for the convenience of the readers.

The general gauge invariant Lagrangian for the exotic particles is then expressed as 
\begin{align}
\Delta \mathcal{L} =  &\left( \lambda^{LQ} \overline{L}^c \, Q \, S_1 
+ \lambda^{df}\, \overline{d} \, f \, S_1^*
+	\lambda^{eu}\, \overline{e}^c \, u \, S_1 
+ {\rm H.c.} \right)
 - \frac{1}{2}\, m_f \, \overline{f}^c f 
\ , \label{eq:leptoYuk}
\end{align}
where generation indices for all parameters and fields are suppressed.
We demand baryon number conservation to forbid the terms $\bar{Q}QS_1$ and $\bar{u} d^c S_1$  
which induce proton decay.  
With this setup, Majorana neutrino mass will be generated at 2-loop order as shown in Fig.~\ref{fig:2loop-LQ}.
Generally the contribution to the neutrino mass matrix is proportional to the down-type Yukawa coupling squared
which is dominated by the third generation unless strong hierarchy in $\lambda^{LQ} \lambda^{df}$ exists.   
As a result, we can simplify the formula for the neutrino mass matrix to 
\begin{align}
	(M_\nu)_{ij}  \simeq 4 \frac{m_f m_b^2 V_{tb}^2}{(2\pi)^8} \sum_{\alpha,\beta=1}^{N_{S_1}}
\left(\lambda_{i3\alpha}^{LQ} \lambda_{3\alpha}^{df} \right)
\left( I_{\alpha\beta} \right)
\left(\lambda_{j3\beta}^{LQ} \lambda_{3\beta}^{df}  \right) \; ,  
\label{eqn:paul_nu_mass}
\end{align}
with the CKM-matrix element $V_{tb}$ and 
$I_{\alpha\beta}$ as a function of $m_f$ and $m_{S_1}$ whose exact form can be read from Ref.~\cite{Angel:2013hla}.
The indices $\alpha$ and $\beta$ label the leptoquark copies. 
This neutrino mass matrix is only rank one if there is only one leptoquark flavor assuming the dominance of the bottom-quark loop.\footnote{The contributions of the strange and down quarks are suppressed by $m_{s,d}^2/m_b^2$ and thus have been neglected in the discussion of Ref.~\cite{Angel:2013hla}.} 
At least two leptoquarks are needed to fit to the current neutrino oscillation data in this model, where
one neutrino mass eigenvalue is nearly vanishing.  
Among all flavor processes, $\mu-e$ conversion in nuclei, $\mu\to e\gamma$ and $\mu\to eee$ give    
the most stringent constraints.

The leptoquark $S_1$ can explain the recent anomalies observed in semileptonic $B$ decays,
i.e. the violation of lepton flavor universality (LFU) of $R_{K^{(*)}}$~\cite{Aaij:2014ora} 
and $R_{D^{(*)}}$~\cite{Lees:2012xj, Lees:2013uzd, Huschle:2015rga, Sato:2016svk, Hirose:2016wfn, Aaij:2015yra}.
In the parameter space with relatively large $\lambda^{eu}_{32}$, the combination of left- and 
right-handed couplings induces scalar and tensor operators, which lift the chirality suppression of the semi-leptonic $B$-decay $B\to D^{(*)} \ell\nu$ and produce sizable effects in the LFU observables $R_{D^{(*)}}$~\cite{Cai:2017wry}.  


\subsection{Supersymmetric models with R-parity violation}
\label{sec:RPV}

Supersymmetric models with R-parity violation naturally lead to
nonzero neutrino masses and mixings. These models have been regarded
as very economical, since no new superfields besides those already
present in the MSSM are
required. Moreover, their phenomenology clearly departs from the
standard phenomenology in the usual SUSY models, typically providing
new experimental probes.

With the MSSM particle content, one can write the following
superpotential, invariant under supersymmetry, as well as the gauge
and Lorentz symmetries,
\begin{equation}
\mathcal{W} = \mathcal{W}^{MSSM} + \mathcal{W}^{\textnormal{\rpv}} \, .
\end{equation}
Here $\mathcal{W}^{MSSM}$ is the MSSM superpotential, whereas
\begin{equation}\label{rpv-superpotential}
\mathcal{W}^{\textnormal{\rpv}} = \frac{1}{2} \lambda_{ijk} \widehat{L}_i \widehat{L}_j \widehat{e}^c_k  + \lambda'_{ijk} \widehat{L}_i \widehat{Q}_j \widehat{d}^c_k + \epsilon_i \widehat{L}_i \widehat{H}_u + \frac{1}{2} \lambda''_{ijk} \widehat{u}^c_i \widehat{d}^c_j \widehat{d}^c_k \, .
\end{equation}
The $\epsilon$ coupling has dimensions of mass, $\{i,j,k\}$ denote
flavor indices and gauge indices have been omitted for the sake of
clarity. The first three terms in $\mathcal{W}^{\textnormal{\rpv}}$
break lepton number (L) whereas the last one breaks baryon number
(B). The non-observation of processes violating these symmetries
impose strong constraints on these parameters, which are required to
be rather small \cite{Barbier:2004ez}. Also importantly, their
simultaneous presence would lead to proton decay, a process that has
never been observed and whose rate has been constrained to
increasingly small numbers along the years. For this reason, it is
common to forbid the couplings in Eq.~\ref{rpv-superpotential} by
introducing a discrete symmetry called R-parity. The R-parity of a
particle is defined as
\begin{equation}\label{Rp}
	R_p = (-1)^{3({\rm B}-{\rm L})+2s}\,,
\end{equation}
where $s$ is the spin of the particle. With this definition, all SM
particles have $R_p=+1$ while their superpartners have $R_p=-1$, and
the four terms in $\mathcal{W}^{\textnormal{\rpv}}$ are
forbidden. Furthermore, as a side effect, the lightest supersymmetric
particle (LSP) becomes stable and can be a dark matter candidate.

However, there is no fundamental reason to forbid all four couplings
in $\mathcal{W}^{\textnormal{\rpv}}$. When R-parity is conserved both
lepton and baryon numbers are conserved, but in order to prevent
proton decay just one these two symmetries suffices. Furthermore, the
breaking of R-parity by L-violating couplings generates nonzero
neutrino masses, and thus constitutes a well-motivated scenario beyond
the standard SUSY models. This scenario (with only L-violating
couplings) can be theoretically justified by replacing R-parity by a
less restrictive symmetry, such as baryon triality
\cite{Ibanez:1991pr}.

We can distinguish two types of R-parity violating (RPV) neutrino mass
models:

\begin{itemize}
\item {\bf Bilinear R-parity violation (b-\rpv)}: In this case the
  only RPV term in the superpotential is the bilinear
  $\mathcal{W}^{\textnormal{b-\rpv}} = \epsilon_i \widehat{L}_i
  \widehat{H}_u$, which breaks lepton number by one unit. This leads
  to the generation of one mass scale for the light neutrinos at
  tree-level via a low-scale seesaw mechanism with the neutralinos
  playing the role of the right-handed neutrinos. The second
  (necessary) mass scale is induced at the 1-loop level. Therefore,
  this can be regarded as a hybrid radiative neutrino mass model.
\item {\bf Trilinear R-parity violation (t-\rpv)}: When one allows for
  the violation of R-parity with the trilinear superpotential terms
  $\mathcal{W}^{\textnormal{t-\rpv}} = \frac{1}{2} \lambda_{ijk}
  \widehat{L}_i \widehat{L}_j \widehat{e}^c_k + \lambda'_{ijk}
  \widehat{L}_i \widehat{Q}_j \widehat{d}^c_k$, lepton number is also
  broken by one unit and Majorana neutrino masses are generated at the
  1-loop level. Therefore, this setup constitutes a pure radiative
  neutrino mass scenario.
\end{itemize}

We now proceed to discuss some of the central features of these two
types of leptonic RPV models, highlighting the most remarkable
experimental predictions. Although in general one can have both types
of leptonic RPV simultaneously, we will discuss them separately for
the sake of clarity.

\subsubsection*{Neutrino masses with b-\rpv}

Bilinear R-parity violation \cite{Hall:1983id} is arguably the most
economical supersymmetric scenario for neutrino masses. The bilinear
$\epsilon_i = \left( \epsilon_e , \epsilon_\mu , \epsilon_\tau
\right)$ terms in the superpotential come along with new $B_\epsilon^i
= \left( B_\epsilon^e , B_\epsilon^\mu , B_\epsilon^\tau \right)$
terms in the soft SUSY breaking potential. Therefore, the number of
new parameters in b-\rpv with respect to the MSSM is $6$, without
modifying its particle content, and they suffice to accommodate all
neutrino oscillation data. For a comprehensive review on b-\rpv see Ref.~\cite{Hirsch:2004he}.

\begin{figure}[bt]
\centering
\begin{subfigure}[t]{0.45\linewidth}\centering
	\begin{tikzpicture}[node distance=1cm and 1cm]
		\begin{scope}
			\clip (-1.5,0) rectangle (1.5,1.5);
			\draw [thick,dashed] (0,0) circle [radius=1];
		\end{scope}
			\begin{scope}
			\clip (-1.5,0) rectangle (1.5,-1.5);
			\draw [thick] (0,0) circle [radius=1];
		\end{scope}
		\coordinate  (l2) at (0:1) ;
		\node[left] at (l2.west) {$g^\prime$};
		\coordinate  (l1) at (180:1);
		\node[right] at (l1.east) {$g,g^{\prime}$};
		\coordinate[cross] (x1) at (270:1);
		\node[below] at (x1.south) {$b$};
		\coordinate[fillvertex] (x2) at (60:1);
		\node[above] at (x2.north) {$s_{\tilde b}$};
		\coordinate[fillvertex] (x3)at (120:1);
		\node[above] at (x3.north) {$c_{\tilde b}$};
		\node at (90:1.3) {$\tilde b_1$};
		\node at (30:1.3) {$\tilde b_{\rm R}$};
		\node at (140:1.3) {$\tilde b_{\rm L}$};

		\coordinate[openvertex, left=of l1] (v1) ;
		\node[below] at (v1.south) {$(a_2,a_1)|\vec\Lambda|\delta_{j3}$};
		\node[left=of v1] (nu1) {$\nu_j$};
		\draw[fermion] (nu1) -- (v1);
		\draw[fermionnoarrow] (v1) -- node[midway,above] {$\tilde W,\tilde B$} (l1);
		
		\coordinate[openvertex, right=of l2 ] (v2) ;
		\node[below] at (v2.south) {$a_1|\vec\Lambda|\delta_{i3}$};
		\node[right=of v2] (nu2) {$\nu_i$} ;
		\draw[fermion] (nu2) -- (v2);
		\draw[fermionnoarrow] (v2) -- node[midway,above] {$\tilde B$}(l2);
	\end{tikzpicture}
\end{subfigure}
\hfill
\begin{subfigure}[t]{0.45\linewidth}\centering
	\begin{tikzpicture}[node distance=1cm and 1cm]
		\begin{scope}
			\clip (-1.5,0) rectangle (1.5,1.5);
			\draw [thick,dashed] (0,0) circle [radius=1];
		\end{scope}
			\begin{scope}
			\clip (-1.5,0) rectangle (1.5,-1.5);
			\draw [thick] (0,0) circle [radius=1];
		\end{scope}
		\coordinate  (l2) at (0:1) ;
		\node[left] at (l2.west) {$h_b$};
		\coordinate  (l1) at (180:1);
		\node[right] at (l1.east) {$h_b$};
		\coordinate[cross] (x1) at (270:1);
		\node[below] at (x1.south) {$b$};
		\coordinate[fillvertex] (x2) at (60:1);
		\node[above] at (x2.north) {$c_{\tilde b}$};
		\coordinate[fillvertex] (x3)at (120:1);
		\node[above] at (x3.north) {$s_{\tilde b}$};
		\node at (90:1.3) {$\tilde b_1$};
		\node at (30:1.3) {$\tilde b_{\rm L}$};
		\node at (140:1.3) {$\tilde b_{\rm R}$};

		\coordinate[openvertex, left=of l1] (v1) ;
		\node[below] at (v1.south) {$a_3|\vec\Lambda|\delta_{j3}+b \tilde\epsilon_j$};
		\node[left=of v1] (nu1) {$\nu_j$};
		\draw[fermion] (nu1) -- (v1);
		\draw[fermionnoarrow] (v1) -- node[midway,above] {$\tilde H$} (l1);
		
		\coordinate[openvertex, right=of l2 ] (v2) ;
		\node[below] at (v2.south) {$a_3|\vec\Lambda|\delta_{i3}+b\tilde \epsilon_i$};
		\node[right=of v2] (nu2) {$\nu_i$} ;
		\draw[fermion] (nu2) -- (v2);
		\draw[fermionnoarrow] (v2) -- node[midway,above] {$\tilde H$}(l2);
	\end{tikzpicture}
\end{subfigure}
\caption[Neutrino masses in the b-\rpv model.]{Bottom--Sbottom diagrams for solar neutrino mass in the
  b-\rpv model. Open circles correspond to small R-parity violating
  projections, full circles correspond to R-parity conserving
  projections and crosses 
  indicate genuine
  mass insertions which flip chirality. $h_b \equiv Y_b$ is the bottom
  quark Yukawa coupling. Figure reproduced from Ref.~\cite{Diaz:2003as}.}
\label{fig:brpv-loop}
\end{figure}
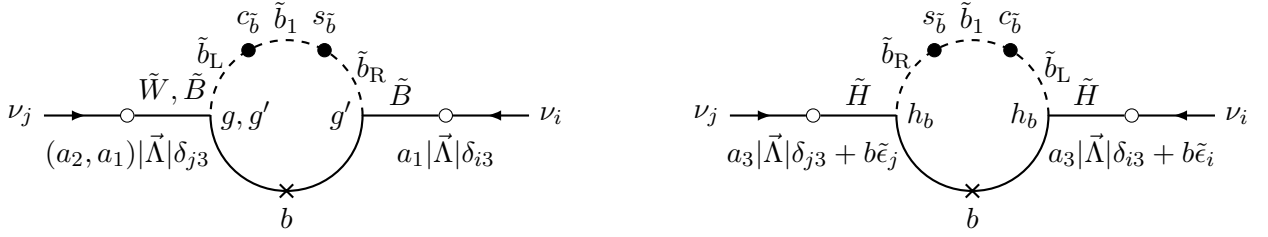

The $\epsilon_i$ couplings induce mixing between the neutrinos and the
MSSM neutralinos. In the basis $(\psi^0)^T=
(-i\tilde{B}^0,-i\tilde{W}_3^0,\widetilde{H}_d^0,\widetilde{H}_u^0,
\nu_{e}, \nu_{\mu}, \nu_{\tau} )$, the neutral fermion mass matrix
$\mathcal{M}_N$ is given by
\begin{equation} \label{brpv-massF0}
\mathcal{M}_N=\left(
\begin{array}{cc}
\mathcal{M}_{\chi^0}& m^T \cr \vb{20} m & 0 \cr
\end{array}
\right) \, .
\end{equation}
Here $\mathcal{M}_{\chi^0}$ is the standard MSSM neutralino mass matrix
and $m \propto \epsilon$ is the matrix containing the
neutrino-neutralino mixing. Assuming the hierarchy $m \ll \mathcal{M}_{\chi^0}$ (naturally fulfilled if $\epsilon \ll m_W$), one can
diagonalize the mass matrix in Eq.~\ref{brpv-massF0} in the seesaw
approximation, $m_\nu = - m \cdot \mathcal{M}_{\chi^0}^{-1} m^T$,
obtaining
\begin{equation} \label{meff-lambda-brpv}
m_{\nu} = \frac{M_1 g^2 \!+\! M_2 {g^\prime}^2}{4\, {\rm Det}(\mathcal{M}_{\chi^0})}\left( \begin{array}{ccc}
\Lambda_e^2
\hskip -1pt&\hskip -1pt
\Lambda_e \Lambda_\mu
\hskip -1pt&\hskip -1pt
\Lambda_e \Lambda_\tau \\
\Lambda_e \Lambda_\mu
\hskip -1pt&\hskip -1pt
\Lambda_\mu^2
\hskip -1pt&\hskip -1pt
\Lambda_\mu \Lambda_\tau \\
\Lambda_e \Lambda_\tau
\hskip -1pt&\hskip -1pt
\Lambda_\mu \Lambda_\tau
\hskip -1pt&\hskip -1pt
\Lambda_\tau^2
\end{array} \right)
\end{equation}
where $\Lambda_i = \mu v_i + v_d \epsilon_i$ are the so-called
\emph{alignment parameters}. Here $M_{1,2}$ are the usual gaugino soft
mass terms, $\mu$ is the Higgsino superpotential mass term,
$v_d/\sqrt{2}$ is the $H_d^0$ VEV and $v_i/\sqrt{2}$ are the sneutrino
VEVs (induced by $\epsilon_i \ne 0$). The special (projective) form of
$m_{\nu}$ implies that it is a rank $1$ matrix, with only one nonzero
eigenvalue, identified with the atmospheric mass scale. Furthermore,
one can obtain two leptonic mixing angles in terms of the alignment
parameters,
\begin{equation}
\tan\theta_{13} = - \frac{\Lambda_e}{(\Lambda_{\mu}^2+\Lambda_{\tau}^2)^{\frac{1}{2}}} \quad , \quad
\tan\theta_{23} = - \frac{\Lambda_{\mu}}{\Lambda_{\tau}} \, .
\end{equation}
The generation of the solar mass scale, which is much smaller ($\Delta
m_{sol}^2 \ll \Delta m_{atm}^2$), requires one to go beyond the
tree-level approximation. This makes b-\rpv a hybrid radiative
neutrino mass model, since loop corrections are necessary in order to
reconcile the model with the observations in neutrino oscillation
experiments. An example of such loops is shown in
Fig.~\ref{fig:brpv-loop}, where the bottom--sbottom diagrams are
displayed. These are found to be the dominant contributions to the
solar mass scale generation in most parts of the parameter space of
the model. Other relevant contributions are given by the tau-stau and
neutrino-sneutrino loops~\cite{Hirsch:2000ef,Diaz:2003as,Grossman:2003gq}. In all cases two
\rpv projections are required, hence leading to the generation of
$\Delta L = 2$ Majorana masses for the light neutrinos.

The most important consequence of the breaking of R-parity at the LHC
is that the LSP is no longer stable and decays. In fact, this is the
only relevant change with respect to the standard MSSM
phenomenology. Since the \rpv couplings are constrained to be small,
they do not affect the production cross-sections or the intermediate
steps of the decay chains, and hence only the LSP decay is altered in
an observable way. For instance, the smallness of the \rpv couplings
typically imply observable displaced vertices at the LHC, see for
instance Ref.~\cite{deCampos:2012pf}.  Furthermore, in b-\rpv there is a
sharp correlation between the LSP decay and the mixing angles measured
in neutrino oscillation experiments~\cite{Mukhopadhyaya:1998xj,Choi:1999tq,Romao:1999up,Porod:2000hv}. This
connection allows to test the model at colliders. For instance, for a
neutralino LSP one finds
\begin{equation}
\frac{\text{BR}(\tilde{\chi}_1^0 \to W \mu)}{\text{BR}(\tilde{\chi}_1^0 \to W \tau)} \simeq \left( \frac{\Lambda_\mu}{\Lambda_\tau} \right)^2 = \tan^2 \theta_{23} \simeq 1 \, .
\end{equation}
A departure from this value would rule out the model
completely. Interestingly, these correlations are also found in
extended models which effectively lead to bilinear \rpv
\cite{Hirsch:2008ur,Bartl:2009an,Liebler:2011tp}.

\subsubsection*{Neutrino masses with t-\rpv}

Supersymmetry with trilinear \rpv has many similarities with
leptoquark models. Once the trilinear RPV interactions are allowed in
the superpotential, the sfermions become scalar fields with lepton
and/or baryon number violating interactions, defining properties of a
leptoquark. For instance, the right sbottom $\tilde{b}_{\rm R}$ has the same
quantum numbers as the leptoquark $S_1$ discussed in
Sec.~\ref{sec:leptoquarks-2} and the $\lambda'$ coupling in
Eq.~\ref{rpv-superpotential} originates a Yukawa interaction exactly
like $\lambda^{LQ}$ in Eq.~\ref{eq:leptoYuk}\footnote{There are,
  however, additional couplings that supersymmetry forbids but would
  be allowed for general leptoquarks. Therefore, t-\rpv can then be
  regarded as a constrained leptoquark scenario. See
  Ref.~\cite{Deshpand:2016cpw} for a paper on t-\rpv as a possible
  explanation for the B-meson anomalies that highlights the
  similarities between this setup and leptoquark models.}. For this
reason, neutrino mass generation takes place in analogous ways, t-\rpv
being a pure radiative model.

As already discussed, the breaking of R-parity leads to the decay of
the LSP. This is the most distinctive signature of this family of
models. However, in contrast to b-\rpv, the large number of free
parameters in t-\rpv exclude the possibility of making definite
predictions for the LSP decay. Nevertheless, one expects novel
signatures at the LHC, typically with many leptons in the final states
\cite{Dreiner:2012wm}. Other signatures, already mentioned in
Sec.~\ref{subsec:LFV}, include LFV observables, see for instance Ref.~\cite{deGouvea:2000cf}.


\section{Conclusions and outlook} \label{sec:conc}

The discovery of neutrino oscillations and its explanation in terms of massive
neutrinos has been one of the most exciting discoveries in particle physics
in recent years and a clear sign of lepton flavor violation and physics beyond the SM. Neutrino masses
being the first discovery of physics beyond the SM may be related to the fact
that the lowest-order effective operator, the Weinberg operator, generates
Majorana neutrino masses. This may point to Majorana neutrinos and consequently
lepton number violation introducing a new scale beyond the SM.
The magnitude of this scale, and that of  lepton flavor violation,
are unknown.

The sensitivity to many lepton flavor violating processes will be increased by
2-4 orders of magnitude in the next decade and thus test lepton flavor
violation at scales of $\mathcal{O}(1-1000)$ TeV. In particular the expected improvement of up to 4 orders of magnitude for $\mu-e$ conversion and the decay $\mu\to eee$, but also other processes,  will yield strong constraints on the parameter space of currently allowed
models or even more excitingly lead to a discovery.
Moreover the LHC is directly probing the TeV-scale and several possible
options for colliders are discussed to probe even higher scales. These exciting experimental prospects, together with the simplicity of the
explanation for the smallness of neutrino mass, are the main
motivations to study radiative neutrino mass models. 

Radiative neutrino mass models explain the lightness of neutrinos without
introducing heavy scales. The main idea is that neutrino masses are absent at
tree-level, being generated radiatively at 1- or higher-loop orders. This,
together with the suppressions due to the possible presence of SM masses and/or
extra Yukawa and quartic couplings, implies that the scale of these models
may be in the range of $\mathcal{O}(1-100)$ TeV. This is also
theoretically desirable, because all new particles are light and no hierarchy problem is introduced.

The plethora of neutrino mass models studied in the last decades is
overwhelming, reaching the hundreds. We believe that at this point an ordering principle for the theory space is necessary to (i) help scientists outside the
field to acquire an overview of the topic, (ii) cover the theory space and spot
possible holes, (iii) try to draw generic phenomenological conclusions that
can be looked for experimentally, and last but not least (iv) serve as
reference for model-builders and phenomenologists.

One can choose to systematically classify the different possibilities and
models in different complementary ways: in terms of (i) the effective operators they
generate after integrating out the heavy particles at tree-level, (ii)
the number of loops at which the Weinberg operator is generated, and (iii) the
possible topologies within a particular loop order.\footnote{A fourth
complementary classification in terms of particles can be done, which will
appear in a future publication~\cite{particles}.} In the first case, the
contribution of the matching to the Weinberg operator can be easily estimated,
and possible UV completions can be outlined. The second option also sheds light
on the scale of the new particles. Finally, the study of possible topologies,
which have been analyzed up to 2-loop order, helps to
systematically pin down neutrino mass models.

We presented selected examples of radiative neutrino mass models in
Sec.~\ref{sec:examples} which serve as benchmark models and discussed their
main phenomenological implications such as lepton flavor-violating processes
and direct
production of the heavy particles at colliders. The phenomenology is
generally very rich and quite model-dependent including extra contributions to
neutrinoless double beta decay, electric dipole moments, anomalous magnetic moments, and meson decays. Furthermore, radiative neutrino mass models may solve the dark
matter problem with a weakly-interacting massive particle running in
the loop generating neutrino mass. Also, the new states can play a crucial role
for the matter-antimatter asymmetry, although not necessarily in a positive way,
and therefore extra bounds can be set on the lepton number violating
interactions.

From our work, we have found that there are several interesting avenues that
can be pursued in the future:
\begin{itemize}
\item If anomalies in B-physics~\cite{Lees:2012xj, Lees:2013uzd,
	Huschle:2015rga, Sato:2016svk, Hirose:2016wfn, Aaij:2015yra,
Aaij:2014ora}, or in the muon anomalous magnetic moment~\cite{Bennett:2006fi},
persist, their connection to radiative models should be further pursued. 
\item There are only a few studies of the matter anti-matter asymmetry in radiative neutrino mass models and more detailed studies are required.
\item A systematic classification of models generated from effective
	operators with covariant derivatives\footnote{All possible dimension-7 operators with SM fields and right-handed neutrinos have been listed in Ref.~\cite{Bhattacharya:2015vja}.} would help to pin down the
	possible models involving gauge bosons. 
\item Further studies of the symmetries that allow the generation of Dirac masses at  loop level.
\item Beyond the LHC, radiative neutrino mass models can be further tested
	specially if a future collider has initial leptonic states. If those
	are same sign, one could directly test the neutrino mass matrix by
	producing for instance the doubly-charged scalar of the Zee-Babu
	model~\cite{Schmidt:2014zoa}.
\end{itemize}

To conclude, it is interesting that there are many combinations of what one may call ``aesthetically reasonable'' particles -- those that have SM multiplet assignments and hypercharges that are not too high -- that couple to SM particles in such a way as to realize neutrino mass generation at loop level.  Radiative mass generation, as well as being a reasonable hypothesis for explaining the smallness of neutrino masses, also provides many phenomenological signatures at relatively low new-physics scales.
So, even if nature realizes the seesaw mechanism with heavy right-handed
neutrinos, given the difficulty of testing such a paradigm, falsifying radiative
models by means of studying in detail their phenomenology and actively
searching for their signals seems the only way to strengthen the case of the
former by reducing as much as possible the theory space. Not to mention
all the useful insights learned on such a journey.

\section*{Acknowledgments}
We acknowledge the use of the Ti\textit{k}Z-Feynman
package~\cite{Ellis:2016jkw}. J.H-G. acknowledges discussions with
Arcadi Santamaria and Nuria Rius. 
This work was supported in part
by the Australian Research Council through the
ARC Centre of Excellence for Particle Physics at the Terascale (CoEPP)
(CE110001104). A.V. acknowledges financial support from the ``Juan de
la Cierva'' program (27-13-463B-731) funded by the Spanish MINECO as
well as from the Spanish grants FPA2014-58183-P, Multidark
CSD2009-00064, SEV-2014-0398 and PROMETEOII/ 2014/084 (Generalitat
Valenciana).

\appendix
	\section{On the relative contribution of operators}
\label{app:PowerCounting}
Oftentimes the effective $\Delta L=2$ operators are discussed using a cutoff regularization scheme. 
In the following, however, we outline the relative contribution of the different $\Delta L=2$ operators to neutrino mass
using dimensional regularization with a momentum-independent renormalization scheme 
such as $\overline{\textrm{MS}}$ renormalization. 
Power counting in the SM effective theory establishes that the dominant
contributions to neutrino mass are given by (i) the lowest-dimensional Weinberg-like
operator $O_1^{(n)}\equiv LLHH(H^\dagger H)^n$ which is induced via matching at the new physics scale $\Lambda$
and (ii)
the contributions induced by mixing via renormalization group running of the
operator $O_1^{(n)}$ into the Weinberg operator or other lower-dimensional
Weinberg-like operators.

Using naive dimensional analysis we discuss in more detail the relative
contribution to neutrino mass from each operator in the
SM effective field theory. 
Note that here we follow the matching and running from low energy scale to high energy scale. 
Below the electroweak scale 
effective operators that can contribute to neutrino masses should contain
two neutrinos and possibly additional fields.
Those additional fields have to be closed off
and their contribution to neutrino masses vary:
for photons and gluons, the contribution from the tadpole diagram vanishes;   
for fermions $f$, the  contribution is proportional to a factor ${m^3}/{16\pi^2 \Lambda^3}$ per fermion loop. 
Thus the contribution of operators with additional fields to neutrino mass either vanishes or is generally suppressed. 
Matching at the electroweak
scale may similarly include loops with electroweak
gauge bosons or the top quark and lead to a suppression
of the respective operator. Additional Higgs fields
yield a factor ${v}/{\Lambda}$ each. Above the
electroweak scale the operators generally mix. 
Higher-dimensional operators also mix into lower
dimensional ones. 
For example although the operator
$O_1^\prime$ mixes into the operator $O_1$ via
renormalization group running and thus it is an operator of
lower dimension, its contribution to the Wilson
coefficient is suppressed by a factor of order
${m_H^2}/{16\pi^2 \Lambda^2}$ and therefore it is of the same
order as the operator $O_1^\prime$. At the new
physics scale the relative size of the Wilson
coefficients is determined by the couplings and the
loop level at which they are generated. The Wilson
coefficient of the Weinberg-like operators at the new
physics scale may be suppressed by a loop factor
compared to other operators, but the other operators
receive a further loop-factor suppression when matching
onto the effective interactions at the electroweak
scale or finally onto the neutrino mass term at a lower
scale. The contributions of all operators to neutrino mass has at least the
same loop-factor suppression as the leading Weinberg-like operator which is induced by matching at the new physics scale.
Higher-dimensional Weinberg-like operators will induce
the lower-dimensional ones via mixing when running the
Wilson coefficients to the low scale, but the
contribution of the induced operator is still of the
same order as the original higher-dimensional operator.
In summary, an order of magnitude estimate of neutrino
mass can be obtained from the leading Weinberg-like operator which is induced from matching at the new physics scale keeping in mind that its
contribution to lower-dimensional Weinberg-like
operators will be of a similar order of magnitude.

\footnotesize
\scriptsize
\setlength{\bibsep}{0pt}
\bibliography{refs}

\end{document}